\documentclass[aps,prx,twocolumn,longbibliography]{revtex4-2}
\usepackage{times,bm,soul,ulem,color,physics}
\usepackage{amssymb,amsmath,amsfonts,mathrsfs}
\usepackage{graphicx,dcolumn,multirow}
\usepackage[svgnames]{xcolor}
\usepackage[colorlinks,bookmarks=false,citecolor=blue,linkcolor=blue,urlcolor=blue]{hyperref}
\renewcommand{\Re}{\operatorname{Re}}
\renewcommand{\Im}{\operatorname{Im}}

\newcommand{\tbf}[1]{\textbf{#1}}

\begin{document}

\title{Algebraic non-Hermitian skin effect and generalized Fermi surface formula in arbitrary dimensions}

\author{Kai Zhang}
\affiliation{Department of Physics, University of Michigan Ann Arbor, Ann Arbor, Michigan, 48109, United States}
\author{Chang Shu}
\affiliation{Department of Physics, University of Michigan Ann Arbor, Ann Arbor, Michigan, 48109, United States}
\author{Kai Sun}
\email{sunkai@umich.edu}
\affiliation{Department of Physics, University of Michigan Ann Arbor, Ann Arbor, Michigan, 48109, United States}


\begin{abstract}
The non-Hermitian skin effect, characterized by a proliferation of exponentially localized edge modes in open-boundary systems, has led to the discovery of numerous novel physical phenomena that challenge the limits of conventional band theory. In sharp contrast to this familiar exponential localization, we report a distinct phenomenon---the algebraic non-Hermitian skin effect---which arises generically in non-Hermitian systems with two or more spatial dimensions. In such cases, the amplitude of skin modes typically decays from the boundary following a power law, rather than an exponential form---a behavior not captured by existing theoretical frameworks. To bridge this gap and describe the transition in localization from one to higher dimensions, we develop a generalized Fermi surface framework applicable to open-boundary systems in arbitrary dimensions. This framework not only reproduces known results for the exponential skin effect in 1D, but also predicts a new class of skin effects with algebraic decay in 2D and above. We demonstrate this framework in both tight-binding and continuum models in two and three dimensions. This investigation not only unveils a novel category of the non-Hermitian skin effect but also offers a comprehensive theoretical structure that describes skin effects in any non-Hermitian system, irrespective of its spatial dimensionality.
\end{abstract}

\maketitle

\section{Introduction}
For systems that interact with external environments, the utilization of non-Hermitian Hamiltonians has proven effective in encapsulating core physical properties of these systems~\cite{Ashida2020,Bergholtz2021RMP,Ding2022NRP}. These non-Hermitian frameworks have been increasingly recognized for their utility in open classical-wave systems ---including photonic~\cite{ZhenBo2015,Zhou2018,Cerjan2019NP}, acoustic~\cite{Huang2024NRP,Zhang2021NC}, and mechanical metamaterials~\cite{Vincenzo2020NP,Vincenzo2020,ZhouDi2020PRR,Banerjee2021PRL,ShankarNature}--- as well as in open quantum systems~\cite{SongFei2019,CHLiu2020PRR,Ueda2021PRL,WenTan2022PRL,Kawabata2023PRX}. In recent years, research interest in non-Hermitian systems has surged as a result of their unique and distinct properties: non-Hermitian Hamiltonians allow complex-valued eigenvalues and relieve the requirement for eigenstates to be orthogonal to each other~\cite{Brody2014}. This deviation from the conventional norms enriched our understanding and led to the emergence of new insights and principles in physics beyond the scope of traditional band theory~\cite{Hatano1996,Sato2011,FuLiang2018PRL,GongPRX2018,FLPRL2020}. 

One striking phenomenon of non-Hermitian band systems is the non-Hermitian skin effect~\cite{Yao2018,Kunst2018PRL,Murakami2019PRL,ChingHua2019,Slager2020PRL,Kai2020,Okuma2020_PRL,ZSaGBZPRL,KawabataPRB2020,Longhi2019PRR,DengTS2019,XuePeng2020,Thomale2020,Ghatak2020,LiLH2020NC,LLHScaleFree2021,TaylorH2021PRB,LuMing2021,Lee2022Sci,Lee2022electrostatics,SBZhang2023PRL,TransientSE2022NC,Nan2024PRL,Fulga2024NP,OkumaSatoReview,LeeCHReview,YFChen2022Review}, which is characterized by the localization of almost all eigenstates at the open boundaries~\cite{LeeModel2016,Torres2018}, and a notable distinction between spectra under periodic and open boundary conditions (OBCs)~\cite{Xiong2018}. This effect defies traditional band theory and compels a reevaluation of well-established principles, such as the bulk-boundary correspondence that characterizes Hermitian systems~\cite{ChingKai2016RMP,SatoPRX2019,XuePeng2020,Thomale2020,Ghatak2020}. Furthermore, the non-Hermitian skin effect spawns a variety of fascinating phenomena unique to non-Hermitian systems, including non-reciprocal responses~\cite{YYFPRL2020,XueWT2021PRB,LiLH2021NC}, directional wavepacket invisibility~\cite{Longhi2015SR,McDonaldPRX2018,Ashvin2019PRL,Sato2021PRL,LQPRL2022,Kai2023PRL}, and improved sensitivity in sensors that scales with the size of the system~\cite{Budich2020PRL,Clerk2020NC,Dong2022PRA}, among others~\cite{Longhi2022PRL,Gong2022PRL,Kai2023PRB}. Such advancements hold promise for robust and adaptable control over wave and signal transmission~\cite{Sebastian2020Science,MaGC2022Nature}. To fully comprehend the non-Hermitian skin effect, it is necessary to consider an extension of Bloch wavevectors into the complex plane, thereby advancing Bloch band theory into the more encompassing non-Bloch band paradigm~\cite{Yao2018,Murakami2019PRL,ZSaGBZPRL,SongFei2019,DengTS2019,Kawabata2020,Thomale2020,Ghatak2020,XuePeng2020}. This evolution of the theory accounts for the emergent complexities and novel behaviors witnessed in non-Hermitian systems. 

The non-Hermitian skin effect in one dimension has been well-established through the generalized Brillouin zone (GBZ) framework. By mapping the Bloch wavevector $k$ to $\beta=e^{ik}$, the Brillouin zone (BZ) is represented as the unit circle in the complex-$\beta$ plane as $k$ varies from $-\pi$ to $\pi$. To characterize the non-Hermitian skin effect, the wavevectors $k$ are extended to complex values, and accordingly, the BZ is generalized to the GBZ, represented as piecewise analytic closed loops of $\beta$ in the complex-$\beta$ plane~\cite{Yao2018,Murakami2019PRL,Kai2020,ZSaGBZPRL}. The GBZ accurately captures the localization of the skin effect, with its radius indicating the localization length and direction. Within this framework, the 1D non-Hermitian skin modes can be interpreted as Bloch waves modulated by an exponentially localized envelope. A local imaginary gauge transformation can be applied to transform these skin modes back into conventional Bloch waves by removing the exponential prefactors~\cite{Hatano1996,Yao2018,ChingHua2019}. This illustrates that the 1D non-Bloch band paradigm is not far from the traditional Bloch band framework, owing to the exponential localization law of the 1D non-Hermitian skin effect. 

\begin{figure}[b]
    \begin{center}
    \includegraphics[width=1\linewidth]{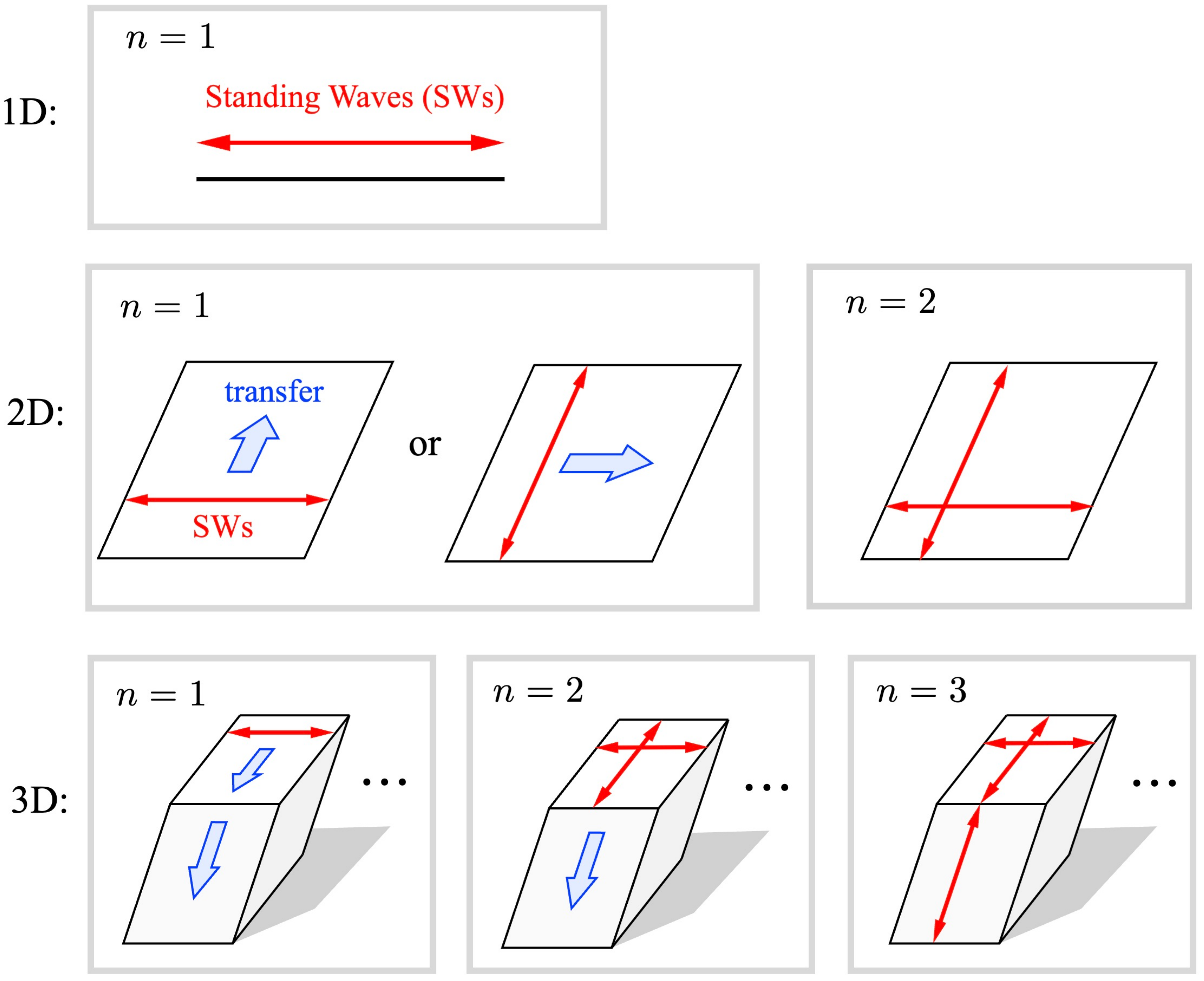}
    \par\end{center}
    \protect\caption{\label{fig:1}~Illustration of the standing wave formation in one, two, and $d$ dimensions. Here, $n$ denotes the number of enforced standing-wave conditions, with red double arrows illustrating the standing-wave directions and the blue wide arrows indicating the transfer directions. 
    In one dimension, standing wave condition can only be enforced along the single spatial direction. In contrast, in two and higher dimensions, standing waves can form along one, multiple, or all spatial directions. 
    In the case of $n=d$ (where $d$ is the spatial dimension), standing wave conditions fully constrain the system, a bulk formula independent of boundary conditions can be established. However, in more general cases where $n < d$, such as $n=1$ in two dimensions, the boundary conditions become significant.}
\end{figure}

However, in two and higher dimensions, the exponential localization characteristic of the non-Hermitian skin effect warrants reevaluation, due to the absence of an accurate high-dimensional GBZ description for wavefunctions and the following considerations. A straightforward extension of the 1D skin effect to higher dimensions would suggest that skin modes also exhibit exponential localization. As a result, these modes would accumulate at the corners of open-boundary systems in the thermodynamic limit~\cite{WangZhong2018,LeeCH2019_PRL,Nori2019,LiLinhuPRL2020,XDZhang2021NC,CYF2021NC}. However, recent numerical and experimental observations have indicated that skin modes universally accumulate on edges rather than at corners of open-boundary geometries~\cite{Kai2022NC,Wang2022NC,KunDing2022NanoP,QYZhou2023NC,DingKun2023PRL,WanTuoSciB,QinYi2023PRA,KaiEdge2023,XJLiu2023PRB,GBJo2025Nature}, representing a phenomenon without a direct 1D counterpart and lacking a comprehensive theoretical framework for accurately describing higher-dimensional OBC wavefunctions. 
Additionally, the GBZ framework generally implies the formation of standing waves under OBCs. For instance, the 1D GBZ framework suggests that two non-Bloch waves with the same localization length can combine to form standing waves under OBCs, as illustrated by the 1D case in Fig.~\ref{fig:1}. However, extending this concept of standing wave formation to higher dimensions remains challenging~\cite{Murakami2022} and is not yet fully understood, posing a significant barrier to understand and develop a comprehensive non-Bloch framework in higher dimensions.
Given these considerations, we pose the following key questions: 
(i) Are high-dimensional skin effects generally characterized by short-range exponential localization? If not, what forms of localization behavior can they exhibit? 
(ii) Is it possible to develop a systematic theoretical framework that accurately captures higher-dimensional non-Hermitian skin effects in the thermodynamic limit? 
Although several attempts have been made to extend the GBZ formalism to higher dimensions~\cite{WangZhong2018,Murakami2022,HuiJiang2022,HYWang2022,HuHP2025SciB,KaiEdge2023,Zhesen2023}, establishing a theoretical framework that can account for diverse localization behaviors, such as edge-localized skin effects, and accurately construct OBC wavefunctions remains a significant challenge and is still under active debate. 

In this paper, we report the algebraic non-Hermitian skin effect in two and higher dimensions. In contrast to the exponential localization of skin modes commonly observed in 1D systems, we find that the amplitudes of skin modes in higher dimensions exhibit power-law localization under OBCs. When the system's geometry is modified into a cylindrical shape, the decay of the skin modes transitions to exponential. This power-law decay and the sensitivity of the decay pattern to boundary conditions are highly unconventional. In Hermitian systems, comparable behavior is only observed for very special systems under specific and fine-tuned conditions: for instance, in 2D elasticity with zero bulk modulus~\cite{Sun2012PNAS}. In these Hermitian systems, power-law and exponential edge modes arise due to a special characteristic of the underlying system---an emergent conformal symmetry which supports indefinitely many conserved quantities. Given that 2D conformal mappings are highly sensitive to the topology of the underlying geometry, they naturally give rise to exponential functions for a cylindrical geometry and power-law functions for a disk or annulus geometry~\cite{Sun2012PNAS}. Any deviation from the conformal limit eliminates this effect in these Hermitian systems. In contrast, in non-Hermitian systems, similar phenomena occur from a totally different origin, without the need for specialized conditions or fine-tuning, and there is no requirement for conformal symmetry to observe such phenomena. The presence of generic and robust algebraic edge modes represents a unique feature of non-Hermitian systems in 2D and above, and stands in stark contrast to what can be achieved in Hermitian setups and/or 1D non-Hermitian systems. 

This finding of a robust algebraic non-Hermitian skin effect is supported by the establishment of a generalized Fermi surface (GFS) framework, which applies to generic non-Hermitian systems with regular open-boundary geometries. While the OBC in one dimension is unambiguous, corresponding to the ends of an open chain, the notion of OBC geometry becomes more diverse in higher dimensions. Here, regular boundary geometries specifically refer to $d$-dimensional parallelotopes, such as parallelograms in 2D and parallelepipeds in 3D.
The GFS framework is established based on the standing wave formation, as illustrated in Fig.~\ref{fig:1}. In one dimension, the standing waves can form along the only one spatial dimension and make the $\dim \text{GBZ} = {d} = 1$ under the existing non-Bloch framework~\cite{Yao2018,Murakami2019PRL}. 
While in two or higher $ d$-dimensional systems, there are $d$ distinct scenarios in terms of the number of standing wave conditions as shown in Fig.~\ref{fig:1}.
In general higher-dimensional cases, the GBZ constraint can be fewer than the space dimensionality, meaning that standing waves can be formed in either one direction or multiple directions, without necessarily forming along all space dimensions. 
We find that in $d$-dimensional lattice systems, the dimension of GBZ manifold ranges from $d$ to ${2d-1}$. 
A non-Bloch bulk formula, independent of sub-extensive boundary details, can be established only when $\text{dim GBZ}=d$; otherwise, when $\text{dim GBZ}>d$, boundary conditions become crucial for characterizing the system in the thermodynamic limit.
We have also implemented this framework on tight-binding lattice Hamiltonians in two and three dimensions and applied it to 2D non-Hermitian reciprocal continuum systems, successfully demonstrating these new phenomena. 
The outcomes are in complete accord with numerical simulations, exhibiting consistency both quantitatively and qualitatively. 

The remainder of the paper is organized as follows. 
In Sec.~\ref{SecII}, we summarize key concepts, focusing on dimensional counting to explain the emergence of algebraic skin modes in $d>1$ systems and their transition to exponential decay in cylindrical geometries, without delving into detailed theory analysis. 
In Sec.~\ref{SecIII}, we introduce rigorous methodologies to derive the GFS framework by employing the standing-wave formation as illustrated in Fig.~\ref{fig:1}. With this exact theoretical method, we construct the GBZ, compute the open boundary spectrum/eigenstates, and prove the dimension analysis and theoretical framework outlined in Sec.~\ref{SecII}. 
In Sec.~\ref{SecIV}, utilizing this framework, we demonstrate that the non-Hermitian reciprocal systems exhibit the quasi-long-range algebraic skin effect, in which the amplitudes of skin modes follow a power-law decay, instead of the conventionally expected exponential law, when away from the open boundaries. 
Notably, we emphasize that this algebraic localization is not exclusive to reciprocal skin effects; it can also arise in non-reciprocal skin effects, which are only required to occur in systems with spatial dimension $d\geq 2$.
In Sec.~\ref{SecV}, we show that the algebraic non-Hermitian skin effect is robust even with the presence of random boundary disorders.
In Sec.~\ref{SecVI}, we generalize the GFS formulation into arbitrary $d$ dimensions, and provide examples in three dimensions. 
In Sec.~\ref{SecVII}, we extend the above framework from lattice systems to continuum systems. 

\section{Dimensionality analysis for the generalized Brillouin zone and summary of key concepts}~\label{SecII}

Before delving into the universal GFS formula in arbitrary spatial dimensions, we commence with an analysis of dimensionality regarding BZ and GBZ. In this section, our discussion pivots solely on qualitative aspects, leaving the meticulous formulation of the GFS theory to be explored in the subsequent section. 

Without loss of generality, here we examine a lattice system in a $d$-dimensional space with finite-range couplings and a finite number of degrees of freedom per unit cell ($d\geq 1$). For Hermitian systems, the dimensions of the BZ and the energy spectrum are
\begin{align}
    &\dim{\text{BZ}}=d, \nonumber \\
    &\dim{E}=1. 
    \label{eq:dim_BZ}
\end{align}
Here, each point in the BZ corresponds to a Bloch wave with a real wavevector $k$, and $\dim{E}=1$ implies that all eigenenergies of the Hamiltonian form some 1D line segments in the complex energy plane, along the real energy axis. 

For non-Hermitian systems in 1D with open boundary conditions, we can substitute the BZ with the GBZ, and the aforementioned dimension counting remains
\begin{align}
    &\dim{\text{GBZ}}=d=1,  \nonumber \\
    &\dim{E}=1. 
\end{align}
Here, each point in the GBZ represents a complex wavevector. Although eigenenergy may deviate from the real energy axis, they still form 1D line sections in  the complex energy plane.

In contrast, open-boundary non-Hermitian systems with $d>1$ exhibit a distinct dimensional hierarchy, which is a key result from this study: 
\begin{align}
    &d \le \dim{\text{GBZ}}\le 2d-1,  \nonumber \\
    &\dim{E}=1 \;\; \textrm{or} \;\; 2.  
\end{align}
It is crucial to underline that in spatial dimensions $d>1$, the GBZ's dimension can surpass that of the spatial dimension $d$. This scenario is permissible exclusively in spatial dimensions beyond 1D, paving the way for higher-dimensional systems to manifest novel physical phenomena that are unattainable in 1D, such as the algebraic non-Hermitian skin effect. In addition, OBC eigenenergies could now occupy 1D lines or 2D regions on the complex plane, i.e., $\dim{E}=1$ or $2$. 

In the construction of eigenstates for a given eigenenergy $E$, one needs to consider only a specific subset of the BZ in Hermitian setups or the GBZ in non-Hermitian systems. This relevant subset will be termed the Fermi surface (FS) in the case of the BZ or the generalized Fermi surface (GFS) in connection with the GBZ. The dimensionality of the FS or GFS can be determined through the following relation:
\begin{align}
    \dim{\text{FS}}&=\dim{\text{BZ}}-\dim{E}, 
    \label{eq:FSdimrelation}\\
    \dim{\text{GFS}}&=\dim{\text{GBZ}}-\dim{E}.
    \label{eq:GFSdimrelation}
\end{align}
$\dim{\text{GFS}}$ is a key factor in determining the characteristics of skin modes. When $\dim\text{GFS} = 0$, skin modes must decay exponentially with the increase in distance from the open boundaries. On the other hand, if $\dim\text{GFS} > 0$, the decay of edge modes may follow different functional forms, such as power-law. Using the dimension counting provided above, it becomes evident that $\dim\text{GFS} > 0$ can only occur in dimensions greater than one, $d > 1$. 
It implies that the algebraic non-Hermitian skin effect derived from the nonzero dimension of GFS is exclusive to two and higher dimensional systems. 
A power-law skin effect at a critical phase transition point in one dimension was reported in Ref.~\cite{Kawabata2023PRX}, which arises from a distinct mechanism compared to the algebraic skin effect proposed in this work. 

\subsection{Hermitian systems}

For a Hermitian system in $d$ dimensions, the eigenstates of the Hamiltonian can be expressed as a superposition of Bloch waves: 
\begin{equation}\label{eq:BZ}
    \Psi=\sum_{\mathbf{k}\in \text{BZ}} A_{k_1,k_2, \ldots, k_d} e^{i (k_1 x_1+k_2 x_2 + \ldots + k_d x_d)}
\end{equation}
Here, $\mathbf{k} = (k_1, \ldots, k_d)$ represents the wavevector, and the collection of all possible $\mathbf{k}$ defines a $d$-dimensional space, i.e., the Brillouin zone. 

In the thermodynamic limit, the energy eigenvalue $E$ becomes a function of the wavevector $\mathbf{k}$, leading to the dispersion relation $E=\epsilon_n(k_1, \ldots, k_d)$, with the integer $n$ denoting the band index. If we mark all possible energy eigenvalues on the complex energy plane, they will span a set of 1D line sections along the real axis, indicating that the energy spectrum is 1D, $\dim{E}=1$. 

For a specific eigenenergy $E$, the dispersion relation $\epsilon_n(k_1, \ldots, k_d) = E$ demarcates a $(d-1)$-dimensional subspace within the $d$-dimensional BZ, namely the equal-energy manifold. In this paper, we adopt the terminology from Fermi liquid theory and refer to this equal-energy manifold as the Fermi surface, which inherently possesses a dimensionality $\dim{\text{FS}}=\dim{\text{BZ}}-\dim{E}=d-1$. As will be shown below, this concept of Fermi surface plays an important role in the construction of eigenstates of $H$. This is because for each given eigenenergy, the corresponding eigenstate can be written as the superposition of Bloch waves in the corresponding Fermi surface
\begin{equation}\label{eq:BZ_FS}
    \Psi_E=\sum_{\mathbf{k}\in \text{FS}} A_{k_1,k_2, \ldots, k_d} e^{i (k_1 x_1+k_2 x_2 + \ldots + k_d x_d)}
\end{equation}
where the sum over $\mathbf{k}$ is performed over the Fermi surface for eigenenergy $E$.

To summarize, the dimensions of the BZ, energy spectrum, and Fermi surfaces for Hermitian systems are $\dim{\text{BZ}} = d$, $\dim{E} = 1$ and $\dim{\text{FS}} = d - 1$, respectively, as shown in Eqs.~\eqref{eq:dim_BZ} and \eqref{eq:FSdimrelation}. 

\subsection{1D Non-Hermitian Systems}

In this section, we briefly review the fundamental concepts and framework of 1D GBZ theory, emphasizing that 1D non-Hermitian systems maintain the same dimensional structure as their Hermitian counterparts. 

Under open boundary conditions, though we can still compose eigenstates as superpositions of exponential functions, the wavevector $k$ is no longer restricted to being real. In 1D, an OBC eigenstate may be expressed as
\begin{equation}
	\Psi = \sum_{k \in \text{GBZ}} A_{k} (e^{i k})^x = \sum_{\beta \in \text{GBZ}} A_{\beta} \beta^x,
 \label{eq:GBZ:states}
\end{equation}
where $k$ is the complex wavevector and, for the sake of simplicity, it can be represented by $\beta = e^{i k}$. The collection of allowed $k$ values forms the GBZ, extending the concept of the BZ to non-Hermitian systems with open boundaries. With $k$ manifesting complex values, the GBZ naturally embeds within a two-dimensional space composed of $(\Re k, \Im k)$, denoting the real and imaginary parts of $k$, respectively.

Although embedded in a 2D momentum space, the actual dimension of GBZ is in fact less than two. This reduction results from the open boundary conditions, where the standing-wave condition imposes extra constraints on the permitted $\beta$ values, thereby diminishing the GBZ's dimensionality. The construction of the GBZ for 1D systems is already well-established, so we only provide a succinct example here. For simplicity, consider a standard 1D single-band tight-binding model, in which the Bloch Hamiltonian $\mathcal{H}(k)$ is analytically continued into its non-Bloch counterpart with the substitution $e^{ik} \to \beta$:
\begin{equation}
    \mathcal{H}(\beta) = t_0 \beta^{-M} + t_1 \beta^{-M+1} + \dots + t_{M+N} \beta^{N},
\end{equation}
where $t_i$ denotes generally complex coefficients, with $t_0, t_{M+N} \neq 0$. The non-Bloch Hamiltonian $\mathcal{H}(\beta)$ thus becomes a Laurent polynomial in the complex variable $\beta$. For a given complex eigenenergy $E$, by solving the characteristic equation $\det[\mathcal{H}(\beta) - E] = 0$, typically order-1 non-Bloch wave solutions $\beta_i(E)$ will be obtained, which can be arranged by their magnitudes as $|\beta_{i}(E)|\leq |\beta_{i+1}(E)|$. Apart from the characteristic equation, boundary conditions at the two ends of the 1D open chain impose another important constraint, known as the standing wave condition
\begin{equation}\label{eq:1DGBZCond}
    |\beta_{M}(E)| = |\beta_{M+1}(E)|,
\end{equation}
which stands as a cornerstone of 1D GBZ theory~\cite{Yao2018,Murakami2019PRL,Kai2020,ZSaGBZPRL,KawabataPRB2020}. 
For further conclusions on generic multi-band models, see Refs.\cite{Murakami2019PRL,KawabataPRB2020}. 
This condition ensures the formation of standing waves, which can simultaneously satisfy the OBCs at both ends, and thus allows the emergence of an extensive number of edge states, each of which is composed of two plane waves of exponential form with identical localization length, i.e., the non-Hermitian skin effect. 
Since the standing wave condition introduces one restriction on the values of $\beta$, the space encompassed by the allowable $\beta$ values experiences a one-dimensional reduction, and thus $\dim{\text{GBZ}}=2d-1=1$. 

Once the 1D GBZ is established, for each point within it (i.e., each permissible value of $\beta$), the corresponding eigenenergies $E$ can be inferred from the characteristic equation $\det[\mathcal{H}(\beta) - E] = 0$. As the allowable $\beta$ values constitute a one-dimensional continuum, it follows that the admissible $E$ values also create a one-dimensional continuum, corresponding to a set of 1D segments in the complex energy plane, that is, $\dim{E} = 1$, as illustrated in Fig.~\ref{fig:2}. 

For eigenstates at a specific eigenenergy, $\Psi_E$, not all possible complex wavevectors in the GBZ are needed. Instead, it just requires a relevant subset of the GBZ, which will be called the generalized Fermi Surface (GFS)
\begin{align}
    \Psi_E = \sum_{\beta \in \text{GFS}} A_{\beta} \beta^x.
\end{align}
In 1D, the dimensionality of the GFS is determined by  $\dim{\text{GFS}}=\dim{\text{GBZ}}-\dim{E}=0$. This zero dimensionality implies that the GFS consists of only a discrete set of $\beta$. This set can be calculated by solving the characteristic equation while satisfying the standing wave condition in Eq.~(\ref{eq:1DGBZCond}) simultaneously. In such cases, the wavefunction $\Psi_E$ becomes the sum of a finite number of exponential functions in terms of $x$. Consequently, $\Psi_E$ must decay exponentially at large $x$, if it is not an extended Bloch wave. 
The existing GBZ framework implies that for a given OBC eigenvalue, the localization length is uniquely determined, resulting in the exponential localization characteristic of the 1D non-Hermitian skin effect.

\subsection{Non-Hermitian Systems in $d>1$}

For $d$-dimensional non-Hermitian systems under open boundary conditions, we can represent eigenstates as sums of exponential functions with complex wavevectors: 
\begin{align}
    \Psi &= \sum_{\mathbf{k}\in \text{GBZ}} A_{k_1, \ldots, k_d} e^{i (k_1 x_1 + \ldots + k_d x_d)} \nonumber \\
    &= \sum_{\mathbf{\beta}\in \text{GBZ}}A_{\beta_1, \ldots, \beta_d} \beta_1^{x_1} \beta_2^{x_2} \ldots \beta_d^{x_d},
\end{align}
where $\mathbf{k} = (k_1, \ldots, k_d)$ and $\beta_i = e^{i k_i}$. Because each $k_i$ is a complex number, the GBZ is embedded in a $2d$-dimensional space of $\{\Re k_1, \ldots, \Re k_d, \Im k_1, \ldots, \Im k_d\}$, or equivalently, $\{\Re \beta_1,  \ldots, \Re \beta_d, \Im \beta_1, \ldots, \Im \beta_d\}$, where $\Re$ and $\Im$ represent real and imaginary parts respectively. 

While the GBZ resides in a higher-dimensional space of $2d$, OBCs and resultant eigenvalue relations place constraints on the permissible complex wavevectors $k_i$. These constraints effectively confine the dimensionality of the GBZ to less than $2d$, captured by the following dimensional equation:
\begin{align}
    \dim \text{GBZ} = 2d - \text{number of GBZ constraints},
\label{eq:GBZ_cont}
\end{align}
in which ``$\text{number of GBZ constraints}$" signifies the count of constraints influencing the eigenstates within the GBZ. 

In analogy to 1D systems, for each spatial direction, one standing wave condition could be introduced
\begin{equation}\label{eq:dDim_GBZCond}
    |\beta_{i,M_i}(E)|=|\beta_{i,M_i+1}(E)|,
\end{equation}
where $i=1,2,\ldots d$. However, one principal finding of this study is that the enforcement of separate standing wave conditions for each spatial direction might not be compatible. For example, imposing simultaneous standing wave conditions for $\beta_i$ and $\beta_j$ ($i \neq j$) can lead to contradictions, rendering it impossible to form eigenstates. Generally, the number of possible standing wave conditions one can enforce is constrained between $1$ and $d$:
\begin{align}
    1 \le  \text{ number of GBZ constraints} \le d.
\end{align}
The lower limit, $\text{number of GBZ constraints} = 1$, indicates that standing wave conditions for any two distinct spatial directions are not reconcilable, thereby allowing for just one such condition for a single spatial direction. The upper limit, $\text{number of GBZ constraints} = d$, suggests that standing wave conditions for all distinct directions are congruent, permitting their simultaneous enforcement. Drawing from Eq.~\eqref{eq:GBZ_cont}, at the lower limit of standing wave conditions ($\text{number of GBZ constraints} = 1$), the GBZ dimension reaches $2d - 1$, while at the upper limit ($\text{number of GBZ constraints} = d$), the GBZ dimension is reduced to $d$. Typically, a non-Hermitian system will fall between these extremes, hence the dimension of the GBZ lies between $d$ and $2d - 1$:
\begin{align}
    d \le \dim \text{GBZ} \le 2d - 1.
\end{align}
This insight is a crucial outcome of our study, suggesting that a comprehensive characterization of the non-Hermitian skin effect in higher dimensions might necessitate a GBZ with a dimension exceeding that of the real-space lattice.

Regarding the OBC eigenenergies, non-Hermitian systems admit complex energy eigenvalues. Plotting all permissible energy eigenvalues on the complex energy plane, two scenarios may emerge. In the most generic scenario, these OBC eigenenergies will occupy certain 2D areas of the complex energy plane, i.e., the energy spectrum is two-dimensional ($\dim{E} = 2$), as shown in Fig.~\ref{fig:2}. This represents a marked departure from Hermitian systems or 1D non-Hermitian systems, where $\dim{E} = 1$. In an alternative scenario, which may arise from symmetry, fine-tuning, or other factors, the allowed OBC energy values fail to span any 2D regions on the complex energy plane. Instead, they are confined to 1D curves, and hence $\dim{E} = 1$. 

For a given complex eigenenergy $E$, the allowed $\beta$ values constitute a subspace within the GBZ, i.e., the generalized Fermi surface for $E$. Correspondingly, the eigenstate associated with this complex energy can be delineated as
\begin{align}
    \Psi_E = \sum_{\beta \in \text{GFS}}A_{\beta_1, \ldots, \beta_d} \beta_1^{x_1} \beta_2^{x_2} \ldots \beta_d^{x_d},
\label{eq:GBZ_FS_high_d}
\end{align}
where the summation of $\beta$ extends over all possible values within the generalized Fermi surface of $E$. The dimension of the generalized Fermi surface is determined by the relation $\dim{\text{GFS}} = \dim{\text{GBZ}} - \dim{E}$. Therefore, the dimension of the generalized Fermi surface may vary from $0$ up to $2d - 2$. 
Unlike 1D non-Hermitian systems, which are characterized by a 0-dimensional generalized Fermi surface, systems with $d>1$ may exhibit $\dim{\text{GFS}}>0$. The presence of a nonzero $\dim{\text{GFS}}$ indicates that the summation in Eq.~\eqref{eq:GBZ_FS_high_d} becomes an integration under the thermodynamic limit with OBCs. Because integrals of exponential functions can engender nearly any function form (for instance, through the Laplace transform), even though the eigenstate is built from exponential functions of the coordinates $(x_1,\ldots,x_d)$ [Eq.~\eqref{eq:GBZ_FS_high_d}], it can manifest extremely diverse behaviors in non-Hermitian systems where $d>1$. For example, the eigenstate may turn into a power-law function of the coordinates, as demonstrated in Sec.~\ref{SecIV}. It is this richness that underpins the potential for the algebraic skin effect exclusive to non-Hermitian systems in dimensions higher than one. 

Here we emphasize that in an open-boundary system with $\dim\text{GFS}> 0$, power-law decaying skin modes are allowed by Eq.~\eqref{eq:GBZ_FS_high_d}. However, if we modify the geometry of the same system to a cylinder, imposing OBCs along the $x_1$ direction and PBCs along $x_2, x_3, \ldots, x_d$, the skin modes will become exponential. The underlying reason for this behavior is that within the cylindrical configuration, the wave vectors $k_2, k_3, \ldots, k_d$ are real and conserved due to PBCs in the corresponding directions. This effectively reduces the problem to a series of 1D systems, each characterized by a specific set of $k_2, k_3, \ldots, k_d$. In such reduced dimensions, the skin modes are constrained to exhibit exponential decay. This phenomenon is a key characteristic of the algebraic skin effect: the decay characteristics of the skin modes are profoundly influenced by the topology of the system's underlying manifold.

\subsection{Strategy to formulate the universal GFS formula}

Here, we succinctly outline our approach for deriving the GFS formula for non-Hermitian systems in arbitrary dimensions, with comprehensive formalism to be detailed in the subsequent sections. 
The GFS formula is used for constructing OBC wavefunctions in the thermodynamic limit, based on a standing-wave ansatz illustrated in Fig.~\ref{fig:1}.
For finite-size lattice systems, we employ the layer transfer matrix method as a practical and efficient tool for computing the discretized GFS that matches the boundary degrees of freedom, with the standing-wave conditions automatically satisfied.
In the thermodynamic limit, we derive continuous GFS curves analytically using the mathematical resultant method~\cite{ZSaGBZPRL}, as detailed in Appendix~\ref{App_ResGFS}, which agrees well with the finite-size results. 
To further demonstrate the applicability of the GFS framework, we apply it to a two-dimensional continuum model in Section~\ref{SecVII}, where an explicit analytic expression for the GFS is obtained without relying on the transfer matrix method. 
Below, we outline the layer transfer matrix approach for computing the GFS in finite-size lattice systems, while deferring the analytic treatment to Section~\ref{SecVII} and Appendix~\ref{App_ResGFS}.

Consider a system of dimension $d$, with the system size $L_1 \times L_2 \times \ldots \times L_d$. 
We can label the unit cells using integer coordinates $(x_1,x_2,\ldots,x_3)$ with $1 \le x_i \le L_i$.
The transfer matrix method is employed to construct the eigenstates of the system --- an exact analytic technique that systematically yields all eigenenergies and eigenstates for a given Hamiltonian. This method has been widely used for Hermitian systems~\cite{Lee1981PRB,Hatsugai1993PRB,Dwivedi2016PRB} under both open and periodic boundary conditions, and it is equally effective for 1D non-Hermitian systems~\cite{KunstPRB2019}. 

In the transfer matrix approach, we first select an arbitrary direction in real space, say the $x_d$ direction, and transform the bulk eigenequation of the Hamiltonian into a recursion for the coordinate $x_d$,
\begin{align}
    \psi(x_1, \ldots ,x_d) = \mathbb{T}_d \; \psi(x_1, \ldots, x_d - 1)
\end{align}
where $\psi$ denotes the eigenstate of the Hamiltonian with eigenenergy $E$. The operator $\mathbb{T}_d$ connects the $\psi$ values at $x_d$ and $x_d - 1$. This operator, known as the transfer matrix, is determined solely by the bulk Hamiltonian and the eigenenergy $E$, regardless of boundary conditions. The eigenvalues of the transfer matrix $\mathbb{T}_d$ yield complex wavevectors $\beta_d$, as outlined in Eq.~\eqref{eq:GBZ_FS_high_d}. 

Using this transfer matrix, we can thereby express the eigenstate as
\begin{align}
    \psi(x_1, \ldots , x_{d-1},x_d) = \mathbb{T}_d^{x_d-1} \; \psi(x_1, \ldots, x_{d-1} ,1).
\end{align}
Here the bulk Hamiltonian is encoded in $\mathbb{T}$. 
For a fixed $\beta_d$, the reduced subsystem can be treated as a $(d-1)$-dimensional system, $\psi_{\beta_d}(x_1, \ldots x_{d-1})$, and thus we can solve this $(d-1)$-dimensional problem using the transfer matrix method once again, defining a transfer matrix for $x_{d-1}$:
\begin{align}
\psi_{\beta_d}&(x_1, \ldots x_{d-2},x_{d-1})= \mathbb{T}_{d-1}^{x_{d-1}-1} \psi_{\beta_d}(x_1, \ldots, x_{d-2} ,1). 
\end{align}
Iterating this process ultimately simplifies the issue to a one-dimensional setup along the direction of $x_1$, $\psi_{\beta_d \beta_{d-1} \dots \beta_2}(x_1)$.
In this way, the $d$-dimensional system is reduced into a 1D system of $\psi_{\beta_d \beta_{d-1} \dots \beta_2}(x_1)$, supplemented by transfer matrices for the other $d-1$ directions which only rely on bulk properties.
We can then treat this 1D problem as a 1D non-Hermitian system, where the GBZ theory is well understood: the non-Hermitian skin effect necessitates a standing wave condition for the direction of $x_1$. This condition ensures that standing waves can be formed to obey boundary conditions at $x_1=1$ and $x_1=L_1$, allowing extensive number of skin modes. 
However, once this condition is enforced along $x_1$, imposing additional standing-wave constraints in other directions is generally untenable, which will be discussed below. Thus, for other directions, we must allow all possible eigenvalues of the transfer matrices $\mathbb{T}_2, \ldots, \mathbb{T}_d$ without further stipulations. 

Since the standing wave condition can typically be imposed in just one spatial direction, as previously discussed, the GBZ's complex wavevector space $(\beta_1, \ldots, \beta_d)$ characteristically spans a space of dimension $2d-1$, surpassing real space dimensions $d$ if $d>1$. This distinctive characteristic underpins why non-Hermitian systems with dimensions greater than 1D can exhibit unprecedented phenomena. 

\begin{figure}[t]
    \begin{center}
    \includegraphics[width=1\linewidth]{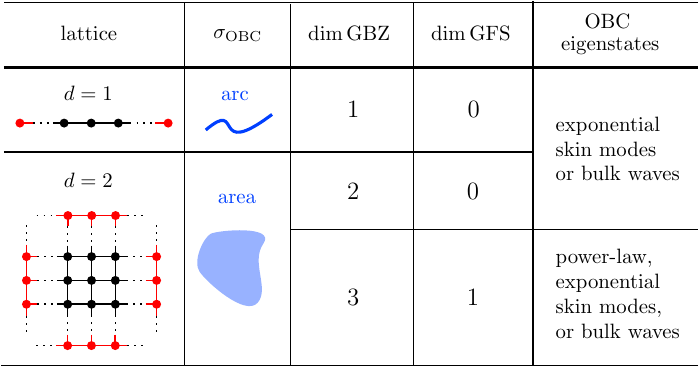}
    \par\end{center}
    \protect\caption{\label{fig:2}~Dimension analysis for GBZ and GFS in 1D and 2D systems. The GBZ dimension in $d$-dimensional systems ranges from $d$ to $2d-1$. In 1D systems, $\dim{\text{GBZ}}=1$, and the OBC spectrum (denoted by $\sigma_{\text{OBC}}$) is limited to forming arcs, meaning $\dim{\sigma_{\text{OBC}}}=1$. This leads to $\dim{\text{GFS}} = 0$ based on Eq.~(\ref{eq:GFSdimrelation}), with OBC eigenstates either being bulk or exponentially localized wavefunctions. In 2D systems, $\dim{\text{GBZ}}$ can be $2$ or $3$, corresponding to the standing wave conditions that are enforced in two or one spatial dimensions, respectively. The OBC spectrum in 2D generally spans a finite area in the complex energy plane, $\dim{\sigma_{\text{OBC}}}=2$. Consequently, $\dim{\text{GBZ}}=0$ or $1$. When $\dim{\text{GFS}} = 0$, the OBC eigenstates behave as bulk or exponentially localized wavefunctions. When $\dim{\text{GFS}} > 0$, the OBC eigenstates can exhibit power-law decaying behavior, a phenomenon unique to high dimensions. }
\end{figure}

As we conclude this section, it is important to draw attention to two key points. The first point is regarding the application of the standing wave condition, which can be imposed in an arbitrary direction. This flexibility implies that, depending on the choice of standing-wave-condition directions, $d$ distinct GBZs can be defined, and each GBZ corresponds to a unique set of basis functions. Because standing wave conditions in different directions in general produces different basis functions, they are typically not compatible with each other and thus cannot be simultaneously enforced. However, it is essential to recognize that while different bases associated with different GBZs can be used, the resulting eigenvalue spectrum and eigenstates---constructed as superpositions of these basis functions---remain invariant with respect to the choice of GBZ. Put simply, although multiple GBZ or GFS bases can be defined, they all ultimately describe the same physical reality.
Secondly, although typically only one direction can accommodate the standing wave condition, special scenarios may allow the standing-wave conditions across multiple spatial directions, e.g., decoupled 1D chains populating a $d$-dimensional space. In such systems, GBZ dimension is further reduced to below $2d-1$. If standing wave conditions for all directions are compatible with each other, the dimension of the GBZ reaches its minimum possible value, which coincides with the real-space dimensions $d$. 
This framework for one and higher dimensional systems is encapsulated in Fig.~\ref{fig:1} and Fig.~\ref{fig:2}. 

\section{Generalized fermi surface formula in two dimensions based on the layer transfer matrix method}~\label{SecIII}

In this section, we offer a general methodology for deriving the GFS formula of 2D lattice systems using a layer transfer matrix approach. 
The extension of the GFS formula to higher dimensions ($d>2$) is detailed in Section~\ref{SecV}; 
The application to a 2D continuum system is discussed in Section~\ref{SecVII};
The analytic approach to derive the GFS in the thermodynamic limit is provided in Appendix~\ref{App_ResGFS}.

\begin{figure}[b]
    \begin{center}
    \includegraphics[width=1\linewidth]{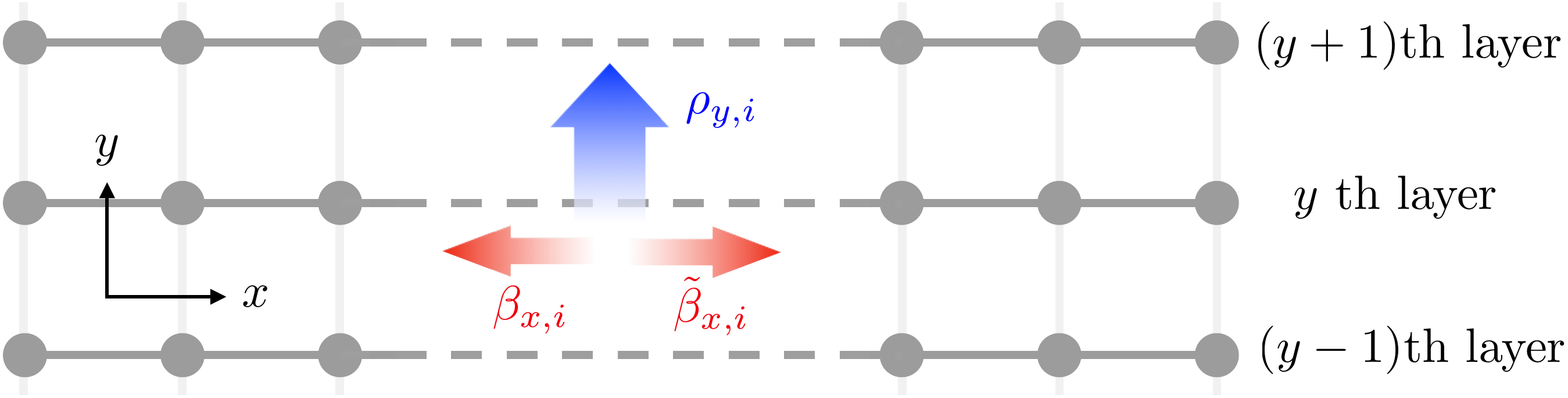}
    \par\end{center}
    \protect\caption{\label{fig:3}~Schematic of $i$-th standing-wave component of an OBC eigenstate under the layer-$y$ transfer matrix basis. Here, the $i$-th component is a standing wave $\varphi_i(x)$ in the $x$ direction, formed by two equal amplitude and counter-propagating non-Bloch waves $\beta_{x,i}$ and $\widetilde{\beta}_{x,i}$, coupled with the associated $\rho_{y,i}$ in the $y$ direction. }
\end{figure}

To construct OBC eigenstates, we begin with the layer transfer matrix for a given tight-binding Hamiltonian. The eigenvalues of the transfer matrix dictate the propagation of the wavefunction's layer components along the transfer matrix direction. 
We demonstrate that a generic 2D OBC eigenstate can be represented by a superposition of standing wave components along one spatial direction (e.g., the $x$ direction), each coupled with the permissible non-Bloch wave $\beta_y$ in the $y$ direction. 
It is expressed as: $\psi(x,y) = \sum_{i}A_i \, \rho^y_{y,i} \varphi_i(x)$, where $\varphi_i(x)$ represents $i$-th standing-wave component along the $x$ direction and $\rho_{y,i}$ is the corresponding $i$-th transfer matrix eigenvalue. 
One of standing-wave components in a 2D OBC wavefunction is illustrated in Fig.~\ref{fig:3}. 
Here, two equal-amplitude, counter-propagating complex waves, $\beta_{x,i}$ and $\widetilde{\beta}_{x,i}$, form a standing wave $\varphi_i(x)$ along the $x$ direction and couple with the non-Bloch wave $\rho_{y,i}$ in the $y$ direction. 

As stated above, a set of basis functions to construct the OBC wavefunction can be determined by imposing the standing-wave condition along either the $x$ or $y$ spatial direction. This leads to distinct GFS or GBZ manifolds, depending on the chosen standing-wave direction. 
For example, in the cases of $n=1$ in two dimensions, as illustrated in Fig.~\ref{fig:1}, when applying standing-wave conditions in different directions, distinct GFS curves are obtained. A concrete example demonstrating this is provided in Appendix~\ref{App_DiffGFS}.
In contrast, when standing-wave conditions can be imposed along all directions, such as 1D systems and certain 2D cases, the GFS or GBZ is uniquely determined. 
Following this principle, the GFS formula applies to 2D systems with parallelogram-shaped open boundaries, where the entire lattice can be generated by translating a one-dimensional line of sites along a specific spatial direction.

In the following subsections, we elaborate on the main findings presented above and illustrate them through concrete examples. This section is organized as follows: Subsections~\ref{SubSecA}$-$\ref{SubSecC} focus on the GFS formalism for the most general cases in two dimensions, where standing waves can be imposed along only one direction and $\dim{\text{GBZ}}=3$, as illustrated by the 2D case with $n=1$ in Fig.~\ref{fig:1}. In Subsection~\ref{SubSecD}, we comment on the non-uniqueness of higher-dimensional GBZ and GFS manifolds and the associated standing-wave conditions. Finally, subsection~\ref{SubSecE} discusses the other possible 2D scenarios. 

\subsection{Model and non-Bloch layer transfer matrix}~\label{SubSecA}

We start with a concrete 2D tight-binding model to present our key findings. 
The model Hamiltonian in real space is given by 
\begin{equation}\label{EQ_2DTBModel}
    \begin{split}
        H & = \sum_{x,y} t_x \, (c^{\dagger}_{x+1,y} c_{x,y} + \text{h.c.}) + t_y \, (c^{\dagger}_{x,y+1} c_{x,y} + \text{h.c.}) \\
        & + t_{xy} \, (c^{\dagger}_{x+1,y+1} c_{x,y} + \text{h.c.}) + u \, c^{\dagger}_{x,y} c_{x,y},
    \end{split}
\end{equation}
where $c^{\dagger}_{x,y}$ denotes the creation operator at lattice site $(x,y)$. The Hamiltonian parameters $t_x,t_{y},t_{xy}$ indicate the hopping strengths along the $x$, $y$, and $x{+}y$ directions, respectively, and $u$ is the on-site potential. These Hamiltonian parameters are generally complex values and thus the Hamiltonian is non-Hermitian and reciprocal. This non-Hermitian Hamiltonian can be implemented in dissipative, reciprocal acoustic and mechanical metamaterials~\cite{QYZhou2023NC,DingKun2023PRL}, as well as in ultracold atom systems~\cite{GBJo2025Nature}. 
Here, we utilize this reciprocal tight-binding model to demonstrate the effectiveness of our GFS formula, though this method is equally applicable to non-reciprocal non-Hermitian systems. 
A non-reciprocal example is provided in Appendix~\ref{App_NRSEModel}. 

The reciprocal Hamiltonian in Eq.~\eqref{EQ_2DTBModel} can be reexpressed in the layer-$y$ basis as follows
\begin{equation}\label{EQ_NNLayerModel}
    H = \sum\nolimits_{y}\sum\nolimits_{l=0,\pm 1} \tbf{c}_{y}^{\dagger} \, \tbf{h}_{l} \, \tbf{c}_{y+l}
\end{equation}
with lattice size $L_x\times L_y$, where $\tbf{c}_{y}^{\dagger}$ is a vector of creation operators at the $y$-th lattice layer, with $L_x$ components, expressed as $\tbf{c}_{y}^{\dagger} = (c^{\dagger}_{x=1,y},\dots , c^{\dagger}_{x=L_x,y})$. 
Under this layer basis, the Hamiltonian $H$ can be viewed as a band Toeplitz matrix~\cite{Bottcher2012} of dimension $L_y$ with matrices $\textbf{h}_{l=0,\pm 1}$ as its entries.
The matrix elements $\textbf{h}_{l=0,\pm 1}$ themselves are matrices of dimension $L_x$ and they represent the couplings between different $y$ layers separated by distance $l$.
In total, the Hamiltonian is represented as a matrix of dimension $L_y\times L_x$. 
The eigenequation of the Hamiltonian is $H\Psi_E = E \Psi_E$, and $\Psi_E$ is the eigenstate of eigenenergy $E$, represented under the layer basis as $\Psi_E = (\psi_{y=1},\dots,\psi_{y=L_y})^T$. 
We can cast the bulk eigenequation into a recursion equation along the $y$ direction
\begin{equation}\label{EQ_Recursion}
    \tbf{h}_{-1} \, \psi_{y-1} + (\textbf{h}_0- E \, \mathbb{I}_{L_x}) \, \psi_{y} + \tbf{h}_{1} \, \psi_{y+1} = 0,
\end{equation}
where $\mathbb{I}_{L_x}$ denotes an identity matrix of dimension $L_x$ and $\psi_y$ is a column with $L_x$ components. Here, $\psi_y$ can be further represented as $\psi_{y} = (\psi_{x=1,y},\dots, \psi_{x=L_x,y})^T$, where $\psi_{x,y}$ indicates the wavefunction component at lattice site ($x,y$). 
The recursion equation can be written in a matrix form 
\begin{align}\label{EQ_TransMat}
    \begin{pmatrix}
        \psi_{y+1} \\ \psi_{y}
    \end{pmatrix} &= \mathbb{T}(E) \begin{pmatrix}
        \psi_{y} \\ \psi_{y-1}
    \end{pmatrix} \\
    &\nonumber=
    \begin{pmatrix}
        \textbf{h}_1^{-1}(E\,\mathbb{I}_{L_x}-\tbf{h}_0) & -\tbf{h}_1^{-1}\tbf{h}_{-1}\\ 
        \mathbb{I}_{L_x} & 0 
    \end{pmatrix}
    \begin{pmatrix}
        \psi_{y} \\ \psi_{y-1}
    \end{pmatrix},
\end{align}
where $\textbf{h}_1^{-1}$ is the inverse of matrix $\textbf{h}_{1}$. 
This $2L_x$-dimensional square matrix $\mathbb{T}(E)$, termed the layer-$y$ transfer matrix, determines the propagation of the wavefunction's layer components along the $y$ direction. 

The transfer matrix usually takes the form of $\mathbb{T}(k_x,E)$ when periodic boundary condition (PBC) in $x$ direction is applied. Here, we adopt the OBCs for both the $x$ and $y$ directions. It is natural to extend the transfer matrix into a non-Bloch form by substituting $e^{ik_x}$ with complex variable $\beta_x$. As such, the matrix elements $h_{0,\pm 1}(k_x)$ are extended into Laurent polynomials of $\beta_x$, that is, $h_{0,\pm 1}(\beta_x)$. 
As an example given by Eq.~(\ref{EQ_2DTBModel}), its non-Bloch transfer matrix under layer-$y$ basis is obtained as 
\begin{equation}\label{EQ_NBTransMat}
    \mathbb{T}(\beta_x,E) = 
    \begin{pmatrix}
    \frac{E-h_0(\beta_x)}{h_1(\beta_x)} & \frac{-h_{-1}(\beta_x)}{h_1(\beta_x)} \\ 1 & 0
    \end{pmatrix},
\end{equation}
where $h_{\pm 1}(\beta_x) = t_y + t_{xy}\beta_x^{\pm}$ and $h_0(\beta_x) = t_x (\beta_x+\beta_x^{-1})+u$. 
The eigenvalues of the transfer matrix $\rho_y$ can be obtained by solving the bulk equation
\begin{equation}\label{EQ_NBChE}
    \begin{split}
        &f(\beta_x,\rho_y,E) = \det{[\mathbb{T}(\beta_x,E)-\rho_y \, \mathbb{I}_2]} \\
        & = \rho_y^2 + \rho_y \, \frac{h_0(\beta_x)-E}{h_1(\beta_x)} + \frac{h_{-1}(\beta_x)}{h_1(\beta_x)} = 0. 
    \end{split}
\end{equation}
We use the symbol $\rho$ to represent non-Bloch waves in the transferring direction, and $\beta$ for those in the standing-wave direction. 
It is worth noting that the bulk equation Eq.~(\ref{EQ_NBChE}) provides constraints on $\beta_x$ and $\rho_y$, which are fully equivalent to the constraints from the bulk characteristic equation $\det[\mathcal{H}(\beta_x,\rho_y)-E]=0$. 
Here, we introduce the transfer matrix to better solve and understand the wavefunction for a finite-size lattice system. The GFS and OBC wavefunctions in the thermodynamic limit can be obtained via an analytic method, which is presented in Appendix~\ref{App_ResGFS}. 
From Eq.~\eqref{EQ_NBChE}, there are two branches of $\rho_y$ for given $\beta_x$ and energy $E$. They characterize the propagation of the non-Bloch wave $\beta_x$ along the $y$ direction in the bulk. 
However, the standing-wave constraints from OBCs have not yet been applied at this stage. 
We will in the following section demonstrate that not all $\rho_y$ branches contribute to the open-boundary eigenstates; instead, only the branches satisfying the standing-wave condition are relevant. 

\subsection{GFS formula and standing-wave construction in two dimensions}\label{SubSecB}

Here, we focus on generic two-dimensional non-Hermitian systems and demonstrate that $\dim{\sigma_{\text{OBC}}} = 2$, $\dim{\text{GBZ}} = 3$, and $\dim{\text{GFS}} = 1$. 
According to Eq.~(\ref{eq:GBZ_cont}), constructing the three-dimensional GBZ manifold requires imposing a single constraint, specifically, a standing-wave condition along one spatial direction. 
We present this standing-wave condition and subsequently derive the resulting GFS at a fixed energy. 
Based on this, we propose an ansatz for exactly constructing the 2D OBC wavefunction using the GFS basis.

For a generic 2D lattice Hamiltonian, one can enforce the standing wave condition along at least one spatial direction. 
Consider a 2D non-Bloch Hamiltonian $\mathcal{H}(\beta_x,\beta_y)$. Without loss of generality, fixing $\beta_y$ reduces the 2D Hamiltonian to a 1D subsystem characterized by this fixed $\beta_y$. For a given energy $E$, generally, there exist multiple $\beta_x$ solutions to the characteristic equation $f(\beta_x,\beta_y,E)=0$. These solutions can be organized by their amplitudes as $|\beta_{x,1}(\beta_y,E)|\leq \dots \leq |\beta_{x,M_x+N_x}(\beta_y,E)|$, where $M_x$ and $N_x$ denote the maximum leftward and rightward hopping ranges, respectively, in the 1D $\beta_y$-subsystem. For example, in the model given by Eq.~(\ref{EQ_2DTBModel}), both $M_x$ and $N_x$ equal 1, indicating that the model only incorporates nearest-neighbor hopping along the $x$ direction for a fixed $\beta_y$. 
The implementation of open-boundary conditions on the 2D system implies that OBCs each 1D subsystem with fixed $\beta_y$ must also satisfy OBCs in the $x$ direction. To reproduce the continuum spectrum for each 1D subsystem under OBCs, the following condition inherited from the 1D GBZ condition is required: 
\begin{equation}\label{EQ_GBZCondition}
    |\beta_{x,M_x}(\beta_y,E)| = |\beta_{x,M_x+1}(\beta_y,E)|,
\end{equation}
which imposes a real constraint on the complex momentum space of ($\beta_x,\beta_y$) for a given energy $E$. 
Since this condition involves a fixed energy for the 2D system and determines the GFS curves, we also refer to it as the GFS condition.
Therefore, at least one standing-wave condition can be imposed in the most generic systems. According to Eq.~(\ref{eq:GBZ_cont}), $\dim{\text{GBZ}}=3$ in this most generic 2D cases. 

From the perspective of wavefunctions, a generic 2D OBC eigenstate can be represented as a superposition of wave components, formally written as $\Psi_E(x,y) = \sum_{(\beta_x,\beta_y)} A_{\beta_x\beta_y} \beta_x^x \beta_y^y$. The number of such components must match the number of constraints imposed by the open boundary conditions in order to construct the full OBC eigenstate. Thus, the number of components scales with the system length $L$. In the thermodynamic limit, these wave components collectively trace out 1D curves in the 4D space spanned by $(\Re \beta_x, \Im \beta_x, \Re \beta_y, \Im \beta_y)$. 
In Hermitian systems, this 1D curve is the Fermi surface that provides the suitable Bloch wave basis for forming the open-boundary eigenstates. 
However, in 2D non-Hermitian systems, for a fixed energy $E_0$, with only the characteristic equation, $\Re{f}(\beta_x,\beta_y,E)= \Im{f}(\beta_x,\beta_y,E) = 0$, the solution domain of ($\beta_x,\beta_y$) is confined to a 2D surface. 
Therefore, an additional constraint---the standing wave condition---is necessary to further reduce this 2D surface to 1D curves, namely GFS curves. With the GFS basis, the OBC eigenstates can then be constructed exactly. 
The standing wave condition can be applied along either one spatial direction, e.g., the standing wave condition enforced in the $x$ direction as given by Eq.~(\ref{EQ_GBZCondition}). 
From dimensionality relation, in general, $\dim{\text{GFS}}=1$ and OBC spectrum occupies a finite region, i.e., $\dim{\sigma_{\text{OBC}}}=2$. Thus, $\dim{\text{GBZ}}=\dim{\text{GFS}}+\dim{\sigma_{\text{OBC}}}=3$. 
If we impose one more standing wave condition, it would reduce the GBZ dimension to $2$ and the GFS dimension to 0, leading to contradictions that fails to satisfy the constraints imposed by the open-boundary conditions. 
To summarize, in the most generic 2D setting, it is both permissible and necessary to impose only a single standing-wave condition along one spatial direction. The requirement for two standing wave conditions needs fine-tuning, which we address in subsection~\ref{SubSecE}. 

The GBZ condition in Eq.~(\ref{EQ_GBZCondition}) suggests the following general procedure for constructing the OBC eigenstate at a given energy $E$:
First, form standing waves in the $x$-direction by superposing non-Bloch wave components $\beta_x$ that satisfy the open-boundary conditions along $x$. 
Next, propagate these standing waves along the $y$-direction using the non-Bloch transfer matrix defined in Eq.~(\ref{EQ_NBTransMat}), where each x-direction standing wave is associated with a transfer matrix eigenvalue $\rho_y$.
Finally, superimpose these $y$-propagating components to satisfy the open-boundary conditions in the $y$-direction. The corresponding superposition coefficients are determined by these boundary constraints, yielding the full open-boundary wavefunction.
Accordingly, we propose the following ansatz for the 2D OBC wavefunction at energy $E$:
\begin{equation}\label{EQ_WavFucRecipe}
    \begin{split}
    \Psi_E(x,y) = \sum\nolimits_{\rho_y\in \text{GFS}} A(\rho_y) \, \rho^y_{y} \, \varphi_{\rho_y}(x),
    \end{split}
\end{equation}
where $\varphi_{\rho_y}(x)$ represents the standing-wave component in the $x$ direction and $A(\rho_y)$ represents the superposition coefficient. 
The standing-wave component can be further expressed as:
\begin{equation}\label{EQ_GStandingWave}
    \varphi_{\rho_y}(x)=\sum\nolimits_{j=1}^{M_x+N_x} B_j(\rho_y) \, \beta^x_{x,j}(\rho_y,E), 
\end{equation}
where $B_j(\rho_y)$ represents the superposition coefficient that is determined by the OBCs in the $x$ direction, and $\beta^x_{x,j}(\rho_y,E)$ satisfies the GFS condition given by Eq.~(\ref{EQ_GBZCondition}). 

\begin{figure*}[t]
    \begin{center}
    \includegraphics[width=1\linewidth]{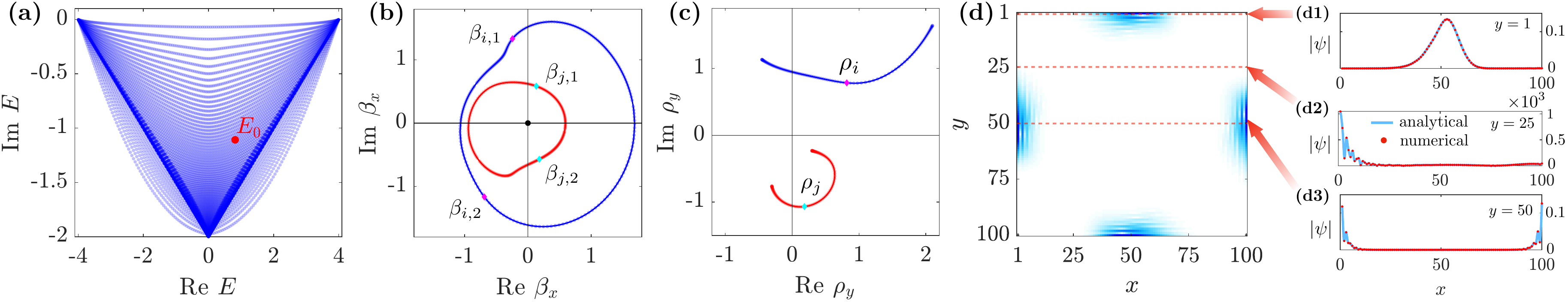}
    \par\end{center}
    \protect\caption{\label{fig:4}~The GFS formula and the exact wavefunction constructed from GFS basis.
    (a) The blue dots represent the OBC eigenvalues with the system size $L_x{=}L_y{=}101$. For the chosen generic energy $E_0{=}0.819{-}1.108i$ (the red dot), the GFS projections onto the $\beta_x$ and $\rho_y$ planes are shown in (b) and (c), respectively. (d) The analytical wavefunction for the OBC eigenvalue $E_0$ fully agrees with the numerical wavefunction calculated from diagonalizing the Hamiltonian matrix. The Hamiltonian parameters in Eq.~(\ref{EQ_2DTBModel}) are set to be $t_x=t_y=1, t_{xy}=i/2$, and $u=-i$. }
\end{figure*}

We take the Hamiltonian in Eq.~(\ref{EQ_2DTBModel}) as an example, where the Hamiltonian only involves the nearest-neighbor hopping along the $x$ direction, thus $M_x=N_x=1$. 
In this model, the GFS condition in the layer-$y$ basis becomes 
\begin{equation}\label{EQ_ModelGBZCondition}
    |\beta_{x,1}(\rho_{y},E)| = |\beta_{x,2}(\rho_{y},E)|.
\end{equation}
The open-boundary eigenvalues with system size $L_x=L_y=101$ are represented by the blue dots in Fig.~\ref{fig:4}(a), which covers a finite region in the complex energy plane, i.e., $\dim{\sigma_{\text{OBC}}}=2$. 
We take a generic OBC eigenvalue $E_0=0.819-1.108 i$, as labeled by the red dot in Fig.~\ref{fig:4}(a). 
The corresponding GFS can be obtained according to the GFS condition in Eq.~(\ref{EQ_ModelGBZCondition}). 
The projection of GFS onto complex $\beta_x$ and $\rho_y$ planes are illustrated in Figs.~\ref{fig:4}(b) and (c), respectively. 
We emphasize that by applying the standing-wave condition in Eq.~\eqref{EQ_ModelGBZCondition} along with the bulk characteristic equation, the GFS in the thermodynamic limit ($L\rightarrow \infty$) can be obtained analytically. The resultant method used to derive the GFS curves shown in Figs.~\ref{fig:4}(b) and (c) is detailed in Appendix~\ref{App_ResGFS}, and the analytic method for obtaining the GFS in 2D continuum systems is presented in Section~\ref{SecVII}. 
The open-boundary conditions in the $x$ direction has been satisfied by the standing-wave condition in the $x$ direction. Therefore, there are OBCs in the $y$ directions leaved. For a finite-size lattice system of size $L_x\times L_y$, there are $2L_x$ lattice points on the two $y$-boundaries, which provide $2L_x$ boundary conditions. Meanwhile, the GFS is discretized due to discrete lattice translation symmetry. The discretized GFS should match with the remaining $2L_x$ degrees of freedom on the open boundaries in the $y$ direction. 
Here, we use the layer transfer matrix approach to calculate the GFS in a finite-size lattice system. The layer transfer matrix $\mathbb{T}(E)$, defined in Eq.~\eqref{EQ_TransMat}, is a $2L_x \times 2L_x$ matrix, whose eigenvalues yield the $2L_x$ values of $\rho_y$. These values represent the discretized GFS curves in the finite size systems. 
Consequently, these $2L_x$ solutions of $\rho_y$, represented by blue and red dots in Fig.~\ref{fig:4}(c), form two arcs in the complex $\rho_y$ plane. 
For a given solution of $\rho_y=\rho_{i}$, there are two corresponding $\beta_x$ solutions having the same amplitude, i.e., $|\beta_{i,1}(\rho_i)|=|\beta_{i,2}(\rho_i)|$, as denoted in Figs.~\ref{fig:4}(b) and (c). These two $\beta_x$ solutions are used to form a standing wave, which coupled with $\rho_{i}$. The OBC eigenstate of energy $E_0$ can be constructed by these standing-wave components given by the GFS. 
To summarize, for a given energy $E_0$, the standing-wave condition gives rise to the 1D GFS curve. These non-Bloch waves on the GFS can be used to compose the eigenstate of energy $E_0$, as shown in Eq.~(\ref{EQ_WavFucRecipe}), and the superposition coefficients are determined by the open-boundary conditions in the $y$ direction, as will be shown in the following subsection. 

\subsection{The 2D wavefunction construction from GFS bases under open boundary conditions}\label{SubSecC}

Here, we demonstrate that the ansatz wavefunction constructed from the GFS basis fully agrees with the result obtained from direct numerical diagonalization of the $L^2 \times L^2$ Hamiltonian matrix, as shown in Fig.~\ref{fig:4}(d), thereby validating our 2D GFS formula. 

Notably, the GFS framework classifies the $d$-dimensional systems into $d$ distinct cases based on the standing-wave construction, as illustrated in Fig.~\ref{fig:1}. 
In our model given by Eq.~\eqref{EQ_2DTBModel}, the standing waves can be enforced along either the $x$ or $y$ direction, placing it into $n=1$ case, where $n$ denotes the number of standing wave conditions. 
Our framework effectively reduces the complexity of an order-$L^2$ diagonalization (solving an $L^2\times L^2$ Hamiltonian matrix derived from the original Schrödinger eigenequation) to the complexity of an order-$L$ diagonalization (solving an $L\times L$ sub-extensive boundary matrix).
More generally, the system can be reduced to an order-$L^{d-n}$ diagonalization problem. 
When $n=d$, order-$L^0$ boundary matrix is needed, indicating that in this case boundary details are irrelevant in the thermodynamic limit. The 1D non-Bloch band theory falls into this category, where $n=d=1$. 
Otherwise, when $n<d$, solving an order-$L^{d-n}$ boundary matrix becomes necessary, which underscores the crucial role of boundary conditions in these scenarios. 

In the model Hamiltonian given by Eq.~(\ref{EQ_2DTBModel}), the GFS corresponding to $E_0$ is obtained and its projections onto complex $\beta_x$ and $\rho_y$ planes are shown in Figs.~\ref{fig:4}(b) and (c). For a given $\rho_y$, there exist two non-Bloch waves, $\beta_{x,1}(\rho_y)$ and $\beta_{x,2}(\rho_y)$, that have the same amplitude $|\beta_{x,1}(\rho_y)|=|\beta_{x,2}(\rho_y)|$. 
These non-Bloch waves combine to form standing waves that satisfy OBCs in the $x$ direction, expressed as $\varphi_{\rho_y}(x) = \beta^x_{x,1}(\rho_y) + B \beta^x_{x,2}(\rho_y)$, where $B$ is a superposition coefficient. 
The OBCs in the $x$ direction impose the conditions $\varphi_{\rho_y}(x=0)=\varphi_{\rho_y}(x=L_x+1)=0$ on the standing-wave component. 
Specifically, $\varphi_{\rho_y}(x=0)=0$ leads to $B=-1$, and $\varphi_{\rho_y}(x=L_x+1)=0$ discretizes the continuous GFS curve into discrete points, with the number of points matching the boundary degrees of freedom. See more details in Appendix~\ref{App_ResGFS}. 
An OBC wavefunction is the superposition of $2L_x$ such standing wave components, each coupled with the associated $\rho_y(\beta_x,E_0)$. 
Therefore, for this model, the standing-wave construction in Eq.~\eqref{EQ_WavFucRecipe} can be explicitly expressed as 
\begin{equation}\label{EQ_AnsatzWave}
    \Psi_{E_0}(\bm{r}) = \sum\nolimits_{i=1}^{2L_x} A_i \, \rho_i^y \, \varphi_i(x) = \sum\nolimits_{i=1}^{2L_x} A_i \, \rho_i^y \, [\beta^x_{i,1} - \beta^x_{i,2}],
\end{equation}
where the subscripts $x$ and $y$ for $\beta$ and $\rho$ are omitted for simplicity, and $\varphi_{i}(x)$ signifies the $i$-th standing-wave component associated with $\rho_i$. 

The $2L_x$ pairs of non-Bloch wave components, $\left( \rho_{i}; \beta_{i,1},\beta_{i,2} \right)|_{i=1,\dots,2 L_x}$, constitute the discretized GFS for $E_0$ and are represented by the red and blue dots in Figs.~\ref{fig:4}(b) and (c). The $2L_x$ superposition coefficients $\{A_{i=1,\dots, 2L_x}\}$ are exactly determined by the $2L_x$ open-boundary conditions in the $y$ direction. 
The open boundary conditions in the $y$ direction are defined by the specific configurations on the boundary layers at $y=1$ and $y=L_y$. 
More generally, these open boundary conditions can be encapsulated in the following equations: 
\begin{equation}\label{EQ_BoundConds}
\begin{split}
(\textbf{b}_1- E_0 \, \mathbb{I}_{L_x}) \, \psi_{1} + \tbf{b}_{+} \, \psi_{2} &= \textbf{0}; \\ 
\tbf{b}_{-} \, \psi_{L_y-1} + (\textbf{b}_{L_y}- E_0 \, \mathbb{I}_{L_x}) \, \psi_{L_y} &= \textbf{0}.
\end{split}
\end{equation}
Here, $\psi_i$ represents a column of wavefunction components at layer $y=i$, each of which includes $L_x$ components. Specifically, $\textbf{b}_1$ and $\textbf{b}_{L_y}$ are the $L_x\times L_x$ matrices, which are given by the specific boundary conditions at the layers $y=1$ and $y=L_y$. The $\tbf{b}_{+}$ ($\tbf{b}_{-}$) represents the adjacent matrix between boundary layer $y=1$ ($y=L_y$) and the bulk. In general, we can add the boundary disorders to modify the boundary conditions $\{ \textbf{b}_1, \textbf{b}_{L_x}, \textbf{b}_{\pm} \}$. The case with boundary disorders is discussed in Section~\ref{SecVI}. 
For notational clarity, we label the standing-wave component as
\begin{equation*}
    \varphi_{x,j} \equiv \beta^x_{1}(\rho_{j}) - \beta^x_{2}(\rho_{j}), 
\end{equation*}
where $x$ denotes the boundary lattice sites (ranging from $1$ to $L_x$), and $j$ indexes the $j$-th GFS basis (running from $1$ to 2$L_x$) for a system of size $L_x \times L_y$. 
Then, we define the following boundary matrix elements: 
\begin{equation*}
    \begin{split}
    & \mathcal{B}_{ij} = \sum_{x} \{[(\textbf{b}_1)_{i,x}-E_0 \delta_{i,x}] \rho_j + (\textbf{b}_+)_{i,x} \rho_{j}^2 \} \, \varphi_{x,j}; \\ 
    & \mathcal{C}_{ij} = \sum_{x} \{(\textbf{b}_-)_{i,x} \rho_j^{L_y-1} + [(\text{b}_{L_y})_{i,x}-E_0 \delta_{i,x}] \rho^{L_y}_j \} \, \varphi_{x,j}.
    \end{split}
\end{equation*}
The subscripts $i,x$ corresponds to the lattice sites and runs from $1$ to $L_x$, and the index $j$ of $\rho_j$ corresponds to discretized GFS components and runs from $1$ to $2L_x$. Therefore, $\mathcal{B}$ and $\mathcal{C}$ are two rectangular matrices with size of $L_x \times 2L_x$. 
With these notations, the boundary conditions in Eq.~\eqref{EQ_BoundConds} can be expressed in a matrix form: 
\begin{widetext}
\begin{equation}
\mathcal{M}_B(E_0) \, \textbf{A} = 
\begin{pmatrix}
    \mathcal{B}_{1,1} & \mathcal{B}_{1,2} & \dots & \mathcal{B}_{1,L_x} & \mathcal{B}_{1,L_x+1} & \dots & \dots & \mathcal{B}_{1,2L_x} \\
    \vdots & \vdots & \vdots & \vdots & \vdots & \vdots  & \vdots & \vdots \\
    \mathcal{B}_{L_x,1} &  \mathcal{B}_{L_x,2} & \dots &  \mathcal{B}_{L_x,L_x} &  \mathcal{B}_{L_x,L_x+1} & \dots & \dots &  \mathcal{B}_{L_x,2L_x} \\
    \mathcal{C}_{1,1} & \mathcal{C}_{1,2} & \dots & \mathcal{C}_{1,L_x} & \mathcal{C}_{1,L_x+1} & \dots & \dots & \mathcal{C}_{1,2L_x} \\
    \vdots & \vdots & \vdots & \vdots & \vdots & \vdots & \vdots & \vdots\\
    \mathcal{C}_{L_x,1} & \mathcal{C}_{L_x,2} & \dots & \mathcal{C}_{L_x,L_x} & \mathcal{C}_{L_x,L_x+1} & \dots & \dots & \mathcal{C}_{L_x,2L_x} \\
\end{pmatrix}
\begin{pmatrix}
    A_1 \\ \vdots \\ A_{L_x} \\ A_{L_x+1} \\ \vdots \\ A_{2L_x}
\end{pmatrix} = \textbf{0}.  
\end{equation}
\end{widetext}
The boundary matrix $\mathcal{M}_B(E_0)$ is a $2L_x\times 2L_x$ matrix that depends on the energy $E_0$. For the boundary matrix $\mathcal{M}_B$, the first $L_x$ rows correspond to the boundary conditions at $y=1$ (the lower boundary), while the last $L_x$ rows correspond to the boundary conditions at $y=L_y$ (the upper boundary). The $2L_x$ columns of the boundary matrix correspond to the $2L_x$ discretized GFS components.
The null space of the boundary matrix $\mathcal{M}_B(E_0)$ determines the superposition coefficients $\textbf{A}=\{A_1, \dots, A_i, \dots, A_{2L_x}\}^T$ in Eq.~\eqref{EQ_AnsatzWave}. With this, the ansatz wavefunction constructed from the GFS basis is fully determined. 
Additionally, the boundary matrix provides a criterion for determining open-boundary eigenvalues:
\begin{equation}
    \det[\mathcal{M}_B(E_0)] = 0
\end{equation}
if and only if ${E_0}$ belongs to open-boundary spectrum. 

As an illustration, we compute the superposition coefficients using the boundary matrix at energy $E_0 = 0.819 - 1.108i$, and construct the ansatz OBC wavefunction, as shown in Fig.~\ref{fig:4}(d). This wavefunction is analytical in the sense that its expression in Eq.~\eqref{EQ_AnsatzWave} is derived from the GFS basis, with the corresponding GFS curve obtained analytically via the resultant method (see Appendix~\ref{App_ResGFS} for details). Thus, the constructed wavefunction is fundamentally distinct from the numerical wavefunction obtained by direct diagonalization of the full Hamiltonian matrix.
To validate our approach, we compare the ansatz wavefunction (blue lines) with the numerically computed wavefunction (red dots) at three representative layers in Fig.~\ref{fig:4}(d). The perfect agreement between the two confirms the accuracy of the GFS-based formulation.

\subsection{The non-uniqueness of GFS in higher dimensions}\label{SubSecD}

Unlike in one dimension, the GFS in higher-dimensional systems may not be uniquely defined for a given Hamiltonian. This non-uniqueness arises from the wide variety of possible boundary conditions or geometries in higher dimensions, for instance, the infinite family of parallelogram-shaped boundaries in 2D.
When a system is sensitive to boundary geometry, as reflected in its spectrum or wavefunction, a GFS theory must explicitly account for these boundary-specific features. As a result, the GFS becomes inherently non-unique.
However, for certain special systems that can be decomposed into independent 1D subsystems, the GFS or GBZ can be constructed as a direct product of 1D solutions, rendering them uniquely defined in such cases.

We now provide an intuitive explanation for this non-uniqueness based on the standing-wave conditions.
In a $d$-dimensional system, if the number of imposed standing-wave conditions (denoted by $n$) is less than the spatial dimension $d$, i.e., $n < d$, the GBZ and GFS manifolds are not uniquely determined by the bulk Hamiltonian. This ambiguity arises from the freedom in choosing the directions along which standing waves are imposed.
For instance, in two dimensions with $n = 1$, one may impose the standing-wave condition along either the $x$- or $y$-direction. Each choice yields a complete non-Bloch basis capable of reconstructing the OBC wavefunction, but the resulting GBZ and GFS manifolds are generally incompatible and distinct. An explicit example illustrating this non-uniqueness is provided in Appendix~\ref{App_DiffGFS}.
In contrast, when $n = d$, the standing-wave conditions fully constrain the system, leaving only one consistent way to construct the non-Bloch basis. 
As a result, the GBZ and GFS manifolds are uniquely defined.

As an example, the Hamiltonian in Eq.~(\ref{EQ_2DTBModel}) can also be expressed in the layer-$x$ basis. Following a similar procedure, one can derive the non-Bloch layer transfer matrix along the $x$-direction, denoted by $\mathbb{T}(\beta_y, E)$, along with the corresponding bulk equation $f(\rho_x, \beta_y, E) = 0$, where $\rho_x$ represents the eigenvalues of the transfer matrix along $x$ direction. 
For this representation, the GFS condition becomes
\begin{equation}\label{EQ_LayerXGBZ}
    |\beta_{y,1}(\rho_x,E)| = |\beta_{y,2}(\rho_x,E)|. 
\end{equation}
Compared to the GFS condition in Eq.~(\ref{EQ_ModelGBZCondition}) under the layer-$y$ basis, this alternative leads to a distinct GFS curve.
In the layer-x basis, the projection of the GFS onto the $\rho_x$-plane forms arcs, while its projection onto the $\beta_y$-plane yields closed curves that satisfy the condition in Eq.~(\ref{EQ_LayerXGBZ}). Within this basis, the open-boundary wavefunction can be constructed as a superposition of standing waves along the y-direction, each coupled to a corresponding transfer eigenvalue $\rho_x$:
\begin{equation}\label{EQ_WavFucBasisX}
    \begin{split}
    \Psi_E(x,y) = \sum\nolimits_{\rho_x\in \text{GFS}} A(\rho_x) \, \rho^x_{x} \, \varphi_{\rho_x}(y),
    \end{split}
\end{equation}
where $\varphi_{\rho_x}(y)$ indicates the standing wave coupled with the transferring component $\rho_x$. 
The coefficients $A(\rho_x)$ are determined by the open-boundary conditions along the $x$-direction.
It is important to emphasize that although the GFS bases obtained via standing-wave conditions along the $x$- and $y$-directions differ, they ultimately yield the same OBC eigenvalues and eigenstates.

\subsection{Discussion on other scenarios in two dimensions}\label{SubSecE}

Here, we consider special 2D cases involving symmetry-protected bulk Hamiltonians or particular boundary geometries, which require fine-tuning and occupy a Hamiltonian parameter space of measure zero.
Based on the dimensionality of the GBZ and the coverage of the OBC spectrum, there are four possible scenarios: $\dim{\sigma_{\text{OBC}}}=2$ or $1$ and $\dim{\text{GBZ}}=3$ or $2$. 
The most generic 2D case, as previously shown, features a 3D GBZ manifold and a 2D (area-covering) OBC spectrum. The remaining cases, which are either forbidden by general constraints or only realized through fine-tuning, will be discussed in this section. 

We first consider the case where the OBC eigenvalues span a finite area (i.e., $\dim \sigma_{\text{OBC}} = 2$) and the GBZ is two-dimensional (i.e., $\dim \text{GBZ} = 2$), as illustrated in Fig.~\ref{fig:2}.
According to the dimensional relation in Eq.~(\ref{eq:GFSdimrelation}), the corresponding GFS in this case reduces to a set of isolated points, i.e., $\dim \text{GFS} = 0$.
The zero-dimensional GFS means that OBC eigenstates are formed by a discrete number of Bloch or non-Bloch waves and thus manifest as extended bulk waves or exponentially localized skin modes. This generally requires that the Hamiltonian respects certain spatial symmetry and that the open boundary geometry matches this symmetry. A representative example is as follows. 
For the example in Eq.~(\ref{EQ_2DTBModel}), we set the parameters $t_{xy}=0, t_{x}=1$, and $t_{y}=i$, such that the Hamiltonian is decoupled in the $x$ and $y$ directions 
\begin{equation}\label{EQ_ReducedGBZ}
    \mathcal{H}(\beta_x,\beta_y) = \beta_x+\beta_x^{-1} + i (\beta_y+\beta_y^{-1})
\end{equation}
with open boundary along the $x$ and $y$ directions, i.e., square boundary geometry. 
For a fixed complex energy, say $E_0=1+i$, the $\beta_x$ and $\beta_y$ can be solved as four points where $|\beta_x|=|\beta_y|=1$. In this case, the GFS consists of these four isolated Bloch points, reducing from 1D curves to 0D points. Correspondingly, the GFS conditions include two constraints, $|\beta_{x,1}(\beta_y,E)|=|\beta_{x,2}(\beta_y,E)|$ and $|\beta_{y,1}(\beta_x,E)|=|\beta_{y,2}(\beta_x,E)|$, indicating that standing-wave conditions can be simultaneously enforced along both the $x$- and $y$-directions. Since the GFS is 0D for each fixed energy, scanning over the full energy continuum generates a 2D GBZ manifold. 

The case with $\dim \sigma_{\text{OBC}}=1$ and $\dim{\text{GBZ}}=3$ is impossible for the following reason. According to the relation in Eq.~(\ref{eq:GFSdimrelation}), GFS is a 2D manifold, i.e., $\dim{\text{GFS}}=2$ in this case. 
However, a 3D GBZ manifold would suggest that only one standing wave condition can be enforced based on Eq.~(\ref{eq:GFSdimrelation}). 
Under these conditions, once the layer-$y$ basis is selected, the OBC eigenstates are constructed based on Eq.~(\ref{EQ_WavFucRecipe}). 
While the wavefunction construction here only requires non-Bloch solutions with the number of order-$L$ on the GFS, suggesting that the actual GFS is 1D curves, i.e., $\dim{\text{GFS}}=1$. 
Consequently, there are contradictions with $\dim{\text{GFS}}=2$ the case suggests, making this case impossible. 

The remaining possible case is $\dim \sigma_{\text{OBC}}=1$ and $\dim{\text{GBZ}}=2$. 
One representative example is the 2D Hatano-Nelson model, whose Hamiltonian is given by 
\begin{equation}
    \mathcal{H}(\beta_x,\beta_y) = t_x \beta_x + w_x \beta_x^{-1} + t_y \beta_y + w_y \beta_y^{-1}
\end{equation}
with the real valued $t_i,w_i>0$ and $t_i \neq w_i$ for $i=x,y$. The GBZ can be directly solved through an imaginary gauge transformation: $\beta_x\to \sqrt{w_x/t_x} \, \widetilde{\beta}_x$, and $\beta_y \to \sqrt{w_y/t_y} \, \widetilde{\beta}_y$. After this transformation, the 2D non-Hermitian Hamiltonian is mapped to a Hermitian Hamiltonian in the basis of $\widetilde{\beta}_x$ and $\widetilde{\beta}_y$: $\mathcal{H}(\widetilde{\beta}_x,\widetilde{\beta}_y) = \sqrt{t_x w_x} (\widetilde{\beta}_x + \widetilde{\beta}_x^{-1}) + \sqrt{t_y w_y} (\widetilde{\beta}_y + \widetilde{\beta}_y^{-1})$. 
For the Hermitian Hamiltonian $\mathcal{H}(\widetilde{\beta}_x,\widetilde{\beta}_y)$, the GBZ coincides with the BZ, and its GFS reduces to a 1D Fermi surface. The OBC eigenstates are Bloch wave standing waves along both $x$ and $y$ directions. 
Upon transforming this Hermitian Hamiltonian back to the non-Hermitian one, we can obtain the GBZ in the original non-Hermitian Hamiltonian as a 2D distorted torus with radius $|\beta_i|=\sqrt{w_i/t_i}$ for $i=x,y$. 
In this case, the OBC eigenstates are Bloch waves modified by an exponential prefactor, causing the wavefunctions to concentrate at the corners of open boundaries. 

We have now discussed other possible scenarios in two dimensions, all of which are compatible with and can be described by the established GFS framework.
Notably, in the two fine-tuned cases considered above, both the GBZ and GFS are uniquely determined, as the standing-wave conditions can be simultaneously imposed along both spatial directions, eliminating any redundancy in the GFS basis. 
In these cases, the OBC wavefunctions in both cases exhibit either extended Bloch waves or exponentially localized skin modes, due to the 0-dimensional GFS ($\dim \text{GFS}=0$), as shown in Fig.~\ref{fig:2}.

\section{Algebraic non-Hermitian skin effect in two dimensions}~\label{SecIV}

Using the above GFS framework, we can exactly construct open-boundary wavefunctions from the GFS basis. Based on this theoretical framework, this section presents a central result of this work: in dimensions higher than one, skin modes can exhibit quasi-long-range algebraic localization, with wavefunction amplitudes decaying as a power law away from the boundary. This behavior contrasts sharply with the conventional expectation of exponential localization in the non-Hermitian skin effect. 

The emergence of the algebraic skin effect hinges on the following key mechanism. In two dimensions, an OBC eigenstate consists of a continuum of non-Bloch wave components determined by the GFS curves. The wavefunction exhibits algebraic localization if and only if these components include an extended Bloch wave with a diverging localization length. The necessary and sufficient conditions for the appearance of the algebraic skin effect can be formally stated as follows:
(i)~$\dim(\text{GFS}) \geq 1$;
(ii)~the GFS curve contains real momentum components, i.e., it intersects the Brillouin zone with nonzero superposition weights. 
These conditions restrict the algebraic skin effect to systems with spatial dimension higher than one. In principle, they can be satisfied in both reciprocal and non-reciprocal systems, implying that the algebraic skin effect may arise in any non-Hermitian system beyond one dimension. 

Under these two conditions, we further deduce that in two or higher dimensions, reciprocal non-Hermitian skin effects, such as the geometry-dependent skin effect~\cite{Kai2022NC}, generally manifest as algebraic localization under generic boundary geometries. 
In the following, we illustrate this sufficient condition using a reciprocal skin effect model and present a non-reciprocal example that exhibits algebraic skin modes in Appendix~\ref{App_NRSEModel}. 

\begin{figure}[b]
    \begin{center}
    \includegraphics[width=1\linewidth]{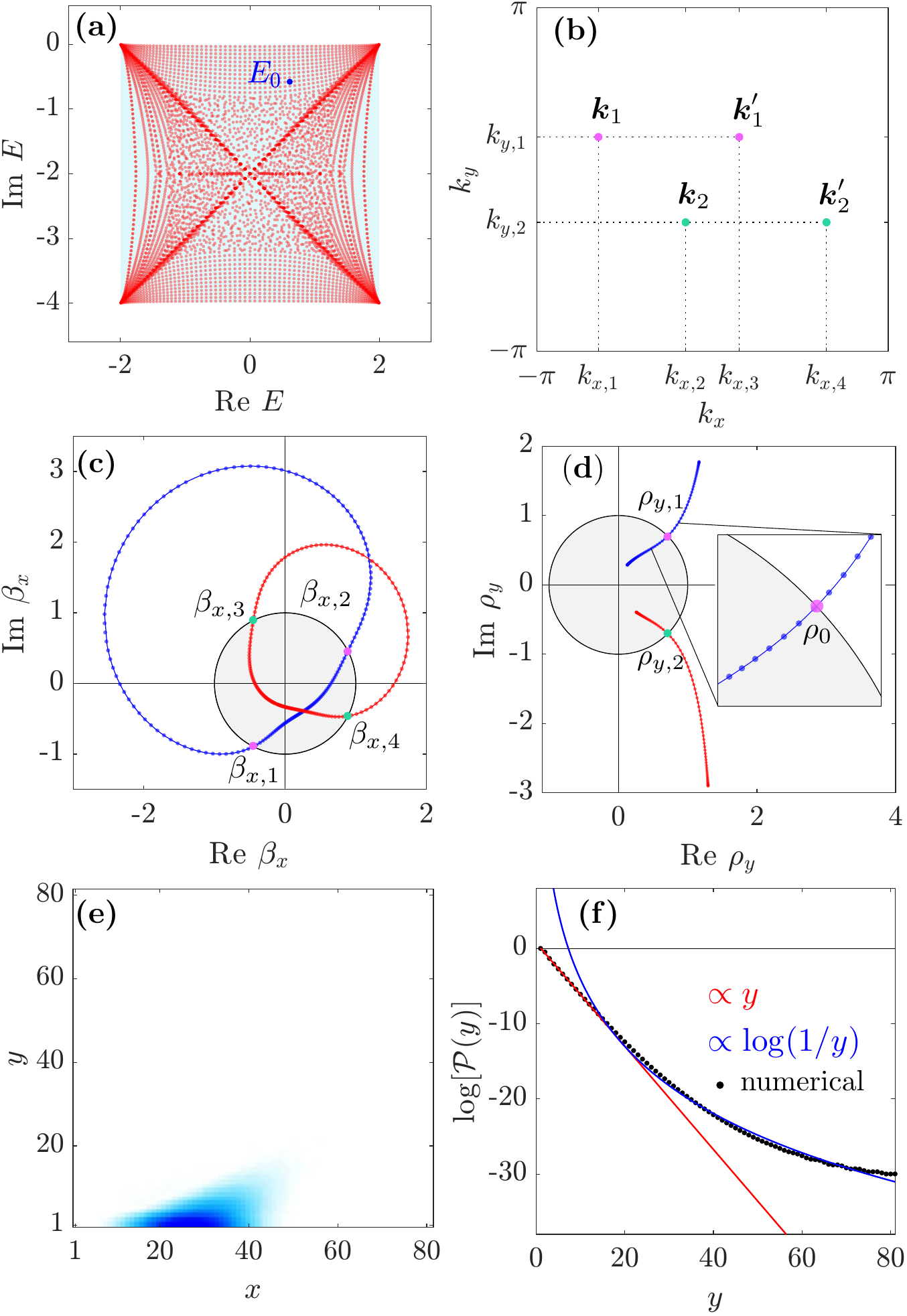}
    \par\end{center}
    \protect\caption{\label{fig:5}~The power-law localization of non-Hermitian skin effect. 
    (a) The comparison between the PBC spectrum (the blue region) and OBC eigenvalues (the red dots), with the system size of $L_x=L_y=81$. 
    (b) For the selected OBC eigenvalue $E_0=0.61-0.58 i$ (the blue dot in (a)), there are four corresponding Fermi points $\bm{k}_i$ distributed in Brillouin zone, where $\det{[h(\bm{k}_i)-E_0]}{=}0$. 
    The projections of the GFS of $E_0$ on the $\beta_x$ and $\rho_y$ planes are shown in (c) and (d), respectively. 
    The intersections between the GFS and the Brillouin zone (the black unit circle) are labeled by the purple and green dots, which correspond to the four Fermi points in (b). 
    (e) The partial wavefunction that is constructed from the GFS inside the unit circle. 
    (f) The layer density of the wavefunction exhibits power-law localization when it is away from the edge. } 
\end{figure}

For demonstration, we consider the reciprocal model given by Eq.~(\ref{EQ_2DTBModel}) with system parameters set to $t_x = 0$, $t_y = i$, $t_{xy} = 1$, and $u = -2i$. 
The non-Bloch Hamiltonian becomes
\begin{equation}\label{EQ_RecipHam}
    \mathcal{H}(\beta_x,\beta_y) = i (\beta_y+\beta_y^{-1}) + \beta_x\beta_y + (\beta_x\beta_y)^{-1} -2 i,
\end{equation}
which respects the reciprocity $H^{T}(\beta_x,\beta_y)=H(1/\beta_x,1/\beta_y)$. 
When the Hamiltonian is reciprocal, it exhibits the spectral property that all bulk eigenvalues under OBCs lie within the Bloch spectrum, i.e., $E_{\text{OBC}} \in \sigma_{\text{PBC}}$. A direct proof is provided in Appendix~\ref{App_FPsProof}. 
As a result, for a given OBC eigenvalue $E_0$ in reciprocal systems, there are always Bloch-wave solutions $\bm{k}_i$ obtained by solving $\det[h(k_x,k_y)-E_0]=0$ where $h(k_x,k_y)$ denotes the Bloch Hamiltonian. These Bloch wave solutions, termed Fermi points, necessarily lie on the GFS curves in reciprocal systems. 

For the Hamiltonian given by Eq.~(\ref{EQ_RecipHam}), the Bloch spectrum is represented by the blue region in Fig.~\ref{fig:5}(a), and open-boundary eigenvalues for a system size of $L_x=L_y=81$ are indicated by red dots in Fig.~\ref{fig:5}(a). 
We select a generic OBC eigenvalue $E_0 = 0.61 - 0.58 i$, denoted by the blue dot in Fig.~\ref{fig:5}(a). 
The distribution of the corresponding Fermi points in the Brillouin zone is illustrated in Fig.~\ref{fig:5}(b). 
As shown in Fig.~\ref{fig:5}(b), a pair of Fermi points, $\bm{k}_1$ and $\bm{k}^{\prime}_1$ (or $\bm{k}_2$ and $\bm{k}^{\prime}_2$), project to the same point $k_{y,1}$ ($k_{y,2}$) on the $k_y$ axes. 
Therefore, we can let Bloch waves $k_{x,1}$ and $k_{x,3}$ form a standing wave in the $x$ direction, which then propagates along the $y$ direction characterized by $k_{y,1}$. 
This forms one Bloch standing-wave component of the OBC wavefunction. 
Therefore, these Fermi points are components on the GFS. 
The discretized GFS for $E_0$ is obtained using the transfer matrix approach within the $x$-standing wave condition. The projections of GFS onto the complex $\beta_x$ and $\rho_y$ planes are shown in Figs.~\ref{fig:5}(c) and (d), respectively.
The presence of Fermi points for $E_0$ means that its GFS must intersect the Brillouin zone, represented by the black unit circle in Figs.~\ref{fig:5}(c) and (d). These intersections, marked by purple and green dots, correspond to the Fermi points in Fig.~\ref{fig:5}(b), specifically $\rho_{y,i} {=} e^{ik_{y,i}}$ and $\beta_{x,i}{=}e^{ik_{x,i}}$. 
In the thermodynamic limit, the GFS on the $\rho_y$ plane forms 1D continuous curves that crosses the unit circle. 
Consequently, in reciprocal non-Hermitian systems, a generic OBC eigenstate always includes Bloch standing-wave components characterized by $|\rho_y| = |\beta_x| = 1$. 
The intersections between GFS and the Brillouin zone are guaranteed by the reciprocity of the Hamiltonian. 
Next, we demonstrate that the presence of these Bloch wave components in the OBC wavefunction leads to a power-law localization. 

We focus on the intersection point $\rho_{y,1}$, which is relabeled by $\rho_0$ for notational clarity as shown in Fig.~\ref{fig:5}(d). 
In the continuum limit, the summation over $\rho_y$ points in the wavefunction of Eq.~(\ref{EQ_AnsatzWave}) becomes the integral over the $\rho_y$ curve. 
The contribution of wave components near the intersection $\rho_0$ is given by 
\begin{equation}\label{EQ_PLDecay}
    \Psi_{E_0,\rho_0}(x,y) = \int_{\delta \rho} \dd\rho \, A(\rho) \, \rho^y \, [\beta_1^x(\rho)-\beta_2^x(\rho)],
\end{equation}
where $\delta \rho$ represents a piece of $\rho_y$ that is a small deviation from the $\rho_0$. 
Here, we focus on the localization of wavefunction at the bottom edge of the system. The localization of the wavefunction at the bottom edge is primarily determined by the $\rho_y$ components inside the unit circle ($|\rho_y| \leq 1$), while the localization near the top edge is governed by their reciprocal counterparts outside the unit circle. 
We select the $\rho_y$ components inside the unit circle and set $|\rho_y|\leq 1$ to be the integral region. 
According to GFS condition in Eq.~(\ref{EQ_ModelGBZCondition}), we have $|\beta_1(\rho)| = |\beta_2(\rho)| = e^{\mu_x(\rho)}$. Consequently, the standing wave component in Eq.~(\ref{EQ_PLDecay}) becomes $\beta_1^x(\rho)-\beta_2^x(\rho) = e^{\mu_x(\rho) x} \sin{(\theta_\rho x)}$, where $\theta_\rho$ is the real-valued phase difference that depends on $\rho$. 
The intersection $\rho_0$ corresponds to $\beta_{x,1}$ and $\beta_{x,2}$ in the $\beta_x$ plane, where $|\beta_{x,1}| = |\beta_{x,2}|=1$. Near $\rho_0$ and within the unit circle, $|\beta_x(\rho)| \lesssim  1$ and $\mu_x(\rho) \lesssim 0$. The function $\mu_x(\rho)$ can be approximated as $\mu_x(\rho) = \mu_x(\rho_0) + \mu^{\prime}_x(\rho_0) \, \delta\rho + o(\delta\rho) \approx \mu_x(\rho_0) = 0$, where only the leading term is retained. 
Adopting this approximation, the Eq.~(\ref{EQ_PLDecay}) can be represented as 
\begin{equation}\label{EQ_ApproxWav}
    \begin{split}
    \Psi_{E_0,\rho_0}(x,y) \approx \int_{\delta \rho} \dd\rho \, A(\rho) \, \rho^y \, \sin{(\theta_\rho x)}.
    \end{split}
\end{equation}
We now define the layer density of the wavefunction to capture the decay behavior along the $y$ direction, expressed as
\begin{equation}\label{EQ_AveDensity}
    \mathcal{P}_{E_0,\rho_0}(y) := \int \dd x \, |\Psi_{E_0,\rho_0}(x,y)|^2.
\end{equation}
Substituting Eq.~(\ref{EQ_ApproxWav}) into Eq.~(\ref{EQ_AveDensity}), noting that the cross terms will vanish due to $\int \dd x \sin(\theta_{\rho_{1}}x) \, \sin(\theta_{\rho_2}x) = 0$ for $\rho_{1}\neq \rho_2$, we finally obtain that $\mathcal{P}_{E_0,\rho_0}(y) \approx \int_{\delta \rho} d|\rho| \, |A(\rho)|^2\,|\rho|^{2y}$. 
The layer density can be further expressed as 
\begin{equation}\label{EQ_ApproxAveDensity}
    \mathcal{P}_{E_0,\rho_0}(y) \approx \int_{-\infty}^{0} \dd\mu_y \, |A(\mu_y)|^2 e^{\mu_y {(2y+1)}},
\end{equation}
where $e^{\mu_y} = |\rho|$, $d|\rho| = e^{\mu_y} d\mu_y$, and $\mu_y\leq 0$ in this integral region. 
Here, $|A(\mu_y)|^2$ corresponds to the superposition coefficients. 
It can be expanded near the $\mu_y=0$ as $|A(\mu_y)|^2 = |A_0|^2 + |A_1|^2 \delta\mu_y + o(\delta\mu_y)$ with $A_0 \equiv A(\mu_y=0)$. 
Finally, the leading order for the localization law of the layer wavefunction density becomes 
\begin{equation}\label{EQ_PowerDecay}
    \mathcal{P}_{E_0,\rho_0}(y) \sim |A_0|^2 \, y^{-1} + o(y^{-1}),
\end{equation}
where $|A_0|^2 \neq 0$ means that the open-boundary wavefunction includes the Bloch standing-wave components, which holds true when the Hamiltonian respects reciprocity. See more detailed discussion in Appendix~\ref{App_GDSEareASE}. 
So far, we have demonstrated that the presence of Bloch standing-wave components leads to the power-law decay ${r^{{-}\alpha}}$ ($\alpha \geq 1$) for the open-boundary wavefunctions, where $r$ is the distance from the open boundaries and the exponent $\alpha$ is determined by the structure of the GFS near the unit circle. 

The power-law localization of the wavefunction is verified in Figs.~\ref{fig:5}(e) and (f). 
To clearly demonstrate the algebraic decay, we examine the OBC wavefunction components near the bottom edge, which are constructed from the GFS bases inside and on the unit circle, as indicated by the gray regions in Figs.~\ref{fig:5}(c) and (d). 
As shown in Fig.~\ref{fig:5}(e), this partial wavefunction is localized at the bottom edge due to a localization exponent $\mu_y \leq 0$. 
By reciprocity, the remaining components, constructed from non-Bloch basis elements outside the unit circle, are localized at the top edge. 
We calculate the layer density for this partial wavefunction, shown as black dots in Fig.~\ref{fig:5}(f), with the blue and red curves representing exponential and power-law fittings, respectively.
The results clearly reveal that the OBC wavefunction exhibits a power-law tail, decaying as $y^{-\alpha}$ away from the bottom edge.

For a finite-size system with size $L_x \times L_y$, there are $2L_x$ values of $\rho_y$ solutions. As shown in the inset of Fig.~\ref{fig:5}(d), it exhibits a minor gap between $\rho_y$ solutions and the unit circle. The gap scales with the length of $L_x$ and can be quantified as $\log{\Delta} \sim c/L_x$, where $c>0$ is a constant factor. This gap, determined solely by the length $L_x$ and independent of $L_y$, modifies the upper limit of the integral in Eq.~(\ref{EQ_ApproxAveDensity}) to $-c /L_x$. 
Consequently, the layer density with length of $L_x$ becomes: 
\begin{equation}
\mathcal{P}_{E_0,\rho_0,L_x}(y) \sim y^{-1} \, e^{- c \, y/L_x},
\end{equation}
which exponentially converges to a perfect power-law decay in Eq.~(\ref{EQ_PowerDecay}) as $y$ increases. 
The finite-size effects on decay behavior along the $y$ direction are determined by the system length $L_x$ in the $x$ direction. This follows from the fact that a finite-size wavefunction is composed of $2L_x$ non-Bloch wave components and is independent of the length in the $y$ direction. 

\section{The robust algebraic localization with boundary disorders}~\label{SecV}

\begin{figure}[t]
    \begin{center}
    \includegraphics[width=1\linewidth]{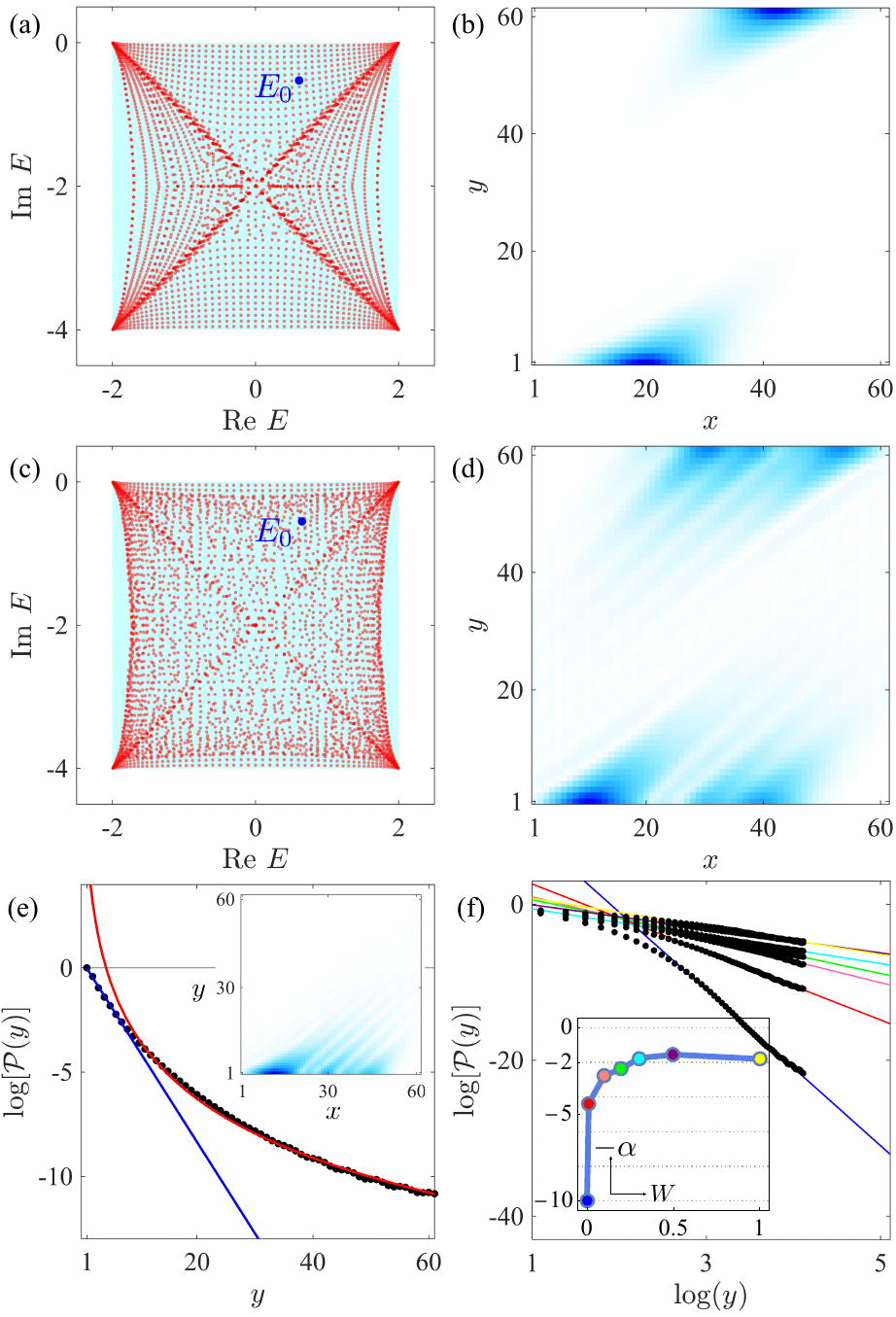}
    \par\end{center}
    \protect\caption{\label{fig:6}~The robustness of algebraic skin effect with random boundary disorder. 
    The spectrum and wavefunction corresponding to $E_0=0.61-0.52i$ without and with the presence of boundary disorder are shown in (a)(b) and (c)(d), respectively. To simulate the boundary disorder, we use onsite potentials with strengths randomly distributed in the range of $[0,W]$ at the layers $y=1$ and $y=L_y$. (c) and (d) are results for $W=1/100$. 
    In the inset of (e), we plot the partial wavefunction constructed from GFS basis inside the unit circle. The black dots in (e) represent layer density decaying along the $y$ direction, with blue and red curves showing linear and power-law fittings ($y^{-\alpha}$ with $\alpha \geq 1$), respectively. (f) shows a log-log plot for different strengths of boundary disorder $W$. The relation between disorder strength $W$ and power-law decay index $\alpha$ is shown in the inset of (f). Disorder strengths are calculated for $W=0$ (blue color), $0.01$ (red), $0.1$ (pink), $0.2$ (green), $0.3$ (cyan), $0.5$ (purple) and $W=1$ (yellow). }
\end{figure}

The algebraic decay behavior in the higher-dimensional skin effect is robust, even in the presence of random open-boundary disorders. 
In this section, we numerically verify the robustness of the algebraic skin effect against boundary disorder. 
The main results are illustrated in Fig.~\ref{fig:6}. 
As shown in Fig.\ref{fig:6}(f), even weak boundary disorder, by breaking reciprocity, rapidly enhances the power-law decay. As the disorder strength increases from zero, the decay exponent quickly saturates to a steady value (Fig.\ref{fig:6}(f)), indicating that the algebraic decay is not only amplified but also stabilized by the disorder.
These results confirm that the algebraic skin effect persists and remains robust even under strong boundary disorder. 

We use the reciprocal example to demonstrate the robustness of the algebraic skin effect, whose Hamiltonian is given by Eq.~\eqref{EQ_RecipHam}. 
The Hamiltonian under square OBC geometry respects reciprocity. Once the random boundary disorder is introduced, the reciprocity is broken at the boundary. According to our GFS framework, the GFS curve is solely determined by the bulk Hamiltonian, and the specific boundary conditions determine the superposition coefficients of these GFS basis. 
Therefore, even with the random boundary disorder, the GFS curve is the same as that without boundary disorder. 
In Section~\ref{SecIV}, we have demonstrated that the GFS curve for this reciprocal Hamiltonian intersects the Brillouin zone. 
In addition, the random boundary disorder cannot diminish the Bloch standing-wave components. 
These two reasons guarantee the robustness of the existence of algebraic skin effect. 

In the numerical simulation, we set the boundary disorder to be on-site potentials with strengths randomly distributed in the range of $[0,W]$ at the boundary layers $y=1$ and $y=L_y$. 
For the reciprocal Hamiltonian in Eq.~\eqref{EQ_RecipHam}, the OBC spectra without and with boundary disorder of strength $W=1/100$ are shown in Figs.~\ref{fig:6}(a) and (c), respectively, with the system size $L_x=L_y=61$. 
The spectrum density exhibits extreme sensitivity to the weak boundary disorder, which is due to the reciprocity breaking induced by the random boundary disorder. 
Correspondingly, the shapes of wavefunctions of OBC eigenvalue $E_0\approx 0.61-0.52i$ are largely modified by the boundary disorder, as shown in Figs.~\ref{fig:6}(b) and (d), respectively. 
Despite the large modification in the wavefunction, its power-law decay behavior remain robust even with the presence of boundary disorder. 
In Fig.~\ref{fig:6}(e), we plot the layer density (the black dots) for the partial wavefunction constructed by the GFS basis inside the unit circle, which follows a similar approach as Fig.~\ref{fig:5}(e) but has different superposition coefficients modified by the boundary disorder. 
The blue line and the red curve in Fig.~\ref{fig:6}(e) refers to the exponential and power-law fittings respectively, which clearly shows the power-law decay of the layer density along the $y$ direction, i.e., $\mathcal{P}_{E_0}(y) \propto y^{-\alpha}$ where the $\alpha$ is the power-law decay index. 
A smaller value of $\alpha$ indicates a slower power-law decay away from the boundary. 
In Fig.~\ref{fig:6}(f), we show the log-log plot of the layer density for different strengths $W$ of boundary disorder. Disorder strengths are calculated for $W=0, 0.01, 0.1, 0.2, 0.3, 0.5$, and $1$, marked in different colors. The straight lines in the log-log plot are linear fittings indicating the power law decay behavior. 
Notably, as the boundary disorder strength increases from zero, which corresponds to the blue fitting line and blue dot in the inset of Fig.~\ref{fig:6}(f), the power-law exponent $\alpha$ saturates exponentially to a steady value, as shown in the inset of Fig.~\ref{fig:6}(f). The simulation results indicate the role of boundary disorder. The presence of boundary disorder breaks reciprocity, making the GFS components mix sufficiently. Consequently, the power-law decay behavior is stabilized by the boundary disorder, which demonstrates the robustness of the algebraic skin effect even with the presence of boundary disorder. 

\section{The GFS formula for generic higher dimensional lattice systems}~\label{SecVI}

Thus far, we have established the GFS formula in two dimensions, based on the standing-wave construction illustrated in Fig.~\ref{fig:1}, which allows us to construct the OBC wavefunctions. Building on this foundation, we derived the quasi-long-range algebraic decay characteristic of the 2D non-Hermitian skin effect. We now extend these conclusions to generic higher dimensions. 

In two dimensions, we show that, unlike in one dimension, the number of standing wave conditions can be either one or two, leading to a 3D or 2D GBZ manifold, respectively. 
In $d$ dimensions, the complex momenta span a $2d$-dimensional real space, and the number of standing wave conditions can be within the range of ${1\leq n \leq d}$. 
Correspondingly, the GBZ dimension is determined to be ${2d-n}$. 
Therefore, for a $d$-dimensional lattice system, the dimension of the GBZ manifold is in the range: 
\begin{equation}\label{EQ_GBZRange}
    {d \leq \dim{\text{GBZ}} \leq 2d-1}. 
\end{equation}
The GFS is submanifold embedded in the GBZ for a fixed open-boundary eigenenergy, and they follow the dimensionality relation: $\dim{\text{GFS}} = \dim{\text{GBZ}} - \dim{E}$. 
Notably, when $d=1$, this inequality in Eq.~(\ref{EQ_GBZRange}) constrains the GBZ manifold to one dimension, and the open-boundary spectrum is restricted to forming some arcs in the complex energy plane, namely $\dim{E}=1$. Consequently, for a given OBC eigenvalue, the GFS consists of some points and has the dimensionality of $\dim{\text{GFS}} = \dim{\text{GBZ}} - \dim{E} = 0$. 
When $d\geq 2$, the case of GFS becomes more complicated due to the fact that the OBC eigenvalues can cover areas in the complex energy plane. In conclusion, the dimension of the GFS in $d\geq 2$ dimensions falls within the range: 
\begin{equation}\label{EQ_GFSRange}
    \begin{split}
    {d-2} \leq \dim{\text{GFS}} \leq {2d-3}, \quad 
    \text{for} \,\, {d \geq 2}. 
    \end{split}
\end{equation}
For example, in $d=2$, the open-boundary spectrum covers a finite area in the complex energy plane, and the GBZ dimension could be either 2D or 3D. Correspondingly, the GFS is composed of 0D points or 1D curves. All these scenarios in two dimensions have been discussed in our previous discussions. 
The dimensions of GBZ and GFS in $d$-dimensional systems are summarized in Table~\ref{table1}. 

\begin{table}[t]
    \caption{Based on the number of standing wave conditions (in the horizontal direction) and the case of the open-boundary spectrum (in the vertical direction), the dimensions of GBZ and GFS in $d$-dimensional lattice systems are summarized in the following table.} 
    \begin{center}
    \begin{tabular}{c|c|c|c|c}
    \hline\hline
    \;\;\;\;\;\;\;\;\;\;\;\;& \;\;\;${n=1}$\;\;\; & \;\;\;${1\leq n\leq d}$\;\;\; & \;\;\;${n=d}$\;\;\; & \;\;\;\;\;\;\;\;\;\;\;\; \\
    \hline
    arc spectrum  &  &  & $d$ & ${\dim{\text{GBZ}}}$ \\ 
    $({d = 1})$ &  &  & ${d-1}$ & ${\dim{\text{GFS}}}$ \\ 
    \hline
    area spectrum & ${2d-1}$ & ${2d-n}$ & $d$ & ${\dim{\text{GBZ}}}$ \\
    $({d \geq 2})$ & ${2d-3}$ & ${2d-n-2}$ & ${d-2}$ & ${\dim{\text{GFS}}}$ \\
    \hline\hline
    \end{tabular}
    \end{center}
    \label{table1}
\end{table}

According to the dimension counting in Eq.~(\ref{EQ_GBZRange}) and Eq.~(\ref{EQ_GFSRange}), in three dimensions (${d=3}$), the GBZ dimension could be $3$, $4$, or $5$, accordingly, the GFS dimension ranges from $1$ to $3$. 
For most generic case, $\dim{\text{GBZ}}=5$ and $\dim{\text{GFS}}=3$. We now provide an example for this generic case. 

For a three-dimensional non-Hermitian Hamiltonian, its open-boundary eigenvalues are typically complex-valued, and their number scales with the system size. 
In the most generic cases, these eigenvalues cover finite areas in the complex energy plane, that is, $\dim{E}=2$. For this case, we can extend the transfer matrix approach to three dimensions to obtain the GFS formula and construct the analytic open-boundary wavefunctions. 
To demonstrate this generalization, we consider an example of a 3D non-Bloch Hamiltonian:
\begin{equation}\label{EQ_3DModelHam}
\begin{split}
    \mathcal{H}(\beta_x,\beta_y,\beta_z) &= \sum_{i=x,y,z} t_i \, (\beta_i+\beta_i^{-1}) \\ & + 
    t_{xyz}\,(\beta_x\beta_y\beta_z+\beta^{-1}_x\beta^{-1}_y\beta^{-1}_z),
\end{split}
\end{equation}
where $t_{i=x,y,z}$ are the nearest neighbor hopping strengths along $x,y$, and $z$ directions, respectively. The term $t_{xyz}\neq 0$ represents hopping along the cube diagonal direction, effectively coupling the lower-dimensional systems in the $x,y$, and $z$ directions. 
One can first choose the transfer matrix basis. Without loss of generality, the Hamiltonian can be written in the planar-$xy$ basis and transferred along the $z$ direction. 
We assume the system size to be $L^3$, with $L$ being the system length. Similar to Eq.~(\ref{EQ_TransMat}), a planar transfer matrix of dimension $2L^2 \times 2 L^2$ can be obtained. The eigenvalues of this transfer matrix, denoted as $\rho_z$, are allowed by the bulk Hamiltonian. In general, these $2L^2$ non-Bloch solutions of $\rho_z$ cover a finite region in the complex $\rho_z$ plane. 
For each fixed $\rho_{z_0}$, the 3D non-Bloch Hamiltonian Eq.~(\ref{EQ_3DModelHam}) reduces to a 2D subsystem, $\mathcal{H}(\beta_x,\beta_y,\rho_{z_0})$. One can treat this 2D subsystem as a 2D system, and solve it using the layer transfer matrix approach once again in two dimensions. 
According to the established transfer matrix method for a 2D Hamiltonian, a 1D GFS curve in the $(\beta_x,\rho_y)$ space can be obtained, which varies with the choice of $\rho_{z_0}$. 
For each fixed $\rho_{z_0}$, 1D GFS curves in the $(\beta_x,\rho_y)$ for the 2D subsystem can be solved. As $\rho_z$ scans through the 2D region of the transfer-matrix eigenvalues, a 3D GFS for the fixed energy is obtained in this 3D system, indicating $\dim{\text{GFS}}=3$. Consequently, the GBZ for this Hamiltonian has the dimension $\dim{\text{GBZ}} = \dim{\text{GFS}} + \dim{E} = 5$. 

For each fixed $\rho_{z_0}$ and $\rho_{y_0}$, the Hamiltonian reduces to 1D subsystem. Consequently, the standing wave condition along the $x$ direction can be enforced, which gives rise to the GFS condition for this 3D system: 
\begin{equation}\label{EQ_3DGBZCondition}
    |\beta_{x,1}(\rho_{y},\rho_{z},E)| = |\beta_{x,2}(\rho_{y},\rho_{z},E)|.
\end{equation}
Depending on the transfer matrix basis, the GFS condition will change and is not unique, similar to the 2D cases discussed in subsection~\ref{SubSecD}. 
Similarly, the 3D OBC wavefunction can be constructed as follows: Let the non-Bloch waves first form standing waves along the $x$ direction; then transfer these standing waves along the $y$ direction to form planar components; each planar component is further transferred along the $z$ direction to generate a standing wave component. 
Therefore, the OBC wavefunctions for this model can be constructed with these standing wave components: 
\begin{equation}
    \Psi_{E_0}(x,y,z) = \sum_{i=1}^{2 L^2} \sum_{j=1}^{2 L} A_{ij} \, \rho_{z,i}^z \, \rho_{y,j}^y \, (\beta^x_{x,1} - \beta^x_{x,2}),
\end{equation}
where the tensor $A_{ij}$ represents the superposition coefficients, and $\beta_{x,1}$ and $\beta_{x,2}$ depend on the values of $\rho_{z,i}$ and $\rho_{y,j}$ and satisfies the GFS condition in Eq.~(\ref{EQ_3DGBZCondition}). 
We take the Hamiltonian in Eq.~(\ref{EQ_3DModelHam}) with the lattice size of $L=12$ and set the parameters to $t_x=t_y=1/2$, $t_z=1$, and $t_{xyz}=i/2$. 
The analytic wavefunction, constructed from the GFS basis, fully aligns with the numerical wavefunction obtained by diagonalizing the Hamiltonian matrix. 
See Appendix~\ref{App_3DGBZFomula} for verification details. 
This verifies the 3D generalization of the 2D GFS formula. As an illustration, this case features a 3D GFS submanifold embedded in a 5D GBZ manifold. 

\section{The GFS formula in 2D continuous systems}~\label{SecVII}

To show the applicability of the GFS formula we developed in lattice models, we now extend it to a two-dimensional continuous system. We demonstrate that the GFS formula can be achieved without the help of the transfer matrix approach. 

The strategy is as follows: we first extend the GFS formula to continuous models. Specifically, the GFS curve can be analytically obtained by imposing the standing-wave condition along the $x$ direction. With this choice, the GFS becomes a collection of standing wave bases along the $x$ direction. 
It's worth noting that, due to the quantization of the standing wave numbers under OBCs in the $x$ direction, the GFS is inherently an infinite discrete set, even in continuous systems, rather than a continuous set. 
Using the GFS basis, we can construct the OBC wavefunction in the continuous system with undetermined coefficients. 
By imposing OBCs in the $y$ direction, we obtain an infinite-dimensional boundary matrix, whose null space determines these coefficients. 
The existence of a null space (or equivalently, a zero determinant) in this boundary matrix serves as a reliable criterion for identifying the OBC energy spectrum of the continuous system. 
Finally, we confirm that the numerical wavefunction coincides with the analytic wavefunction constructed from GFS basis. 

\subsection{The GFS formula in reciprocal continuous systems}

To elaborate on the above approach, we consider the following concrete partial differential equation: 
\begin{equation}
    (-\partial_x^2 - A \, \partial_y^2 - B \, \partial_x\partial_y) \, \psi_E(x,y) = E \psi_E(x,y),
\end{equation}
where $A$ and $B$ are generally complex values, thus the differential operator is non-Hermitian. 
This operator can be expressed in complex-valued momentum space as
\begin{equation}\label{eq:ContinuousRSE}
    \mathcal{H}(q_x,q_y) = q_x^2 + A \, q_y^2 + B \, q_x q_y,
\end{equation}
where two complex momenta $q_x=k_x+i\kappa_x$ and $q_y=k_y+i\kappa_y$. 
The linear terms in $q$ in the Hamiltonian are forbidden due to reciprocity: $\mathcal{H}(q_x,q_y){=}\mathcal{H}^T(-q_x,-q_y)$. 
Finally, Eq.~\eqref{eq:ContinuousRSE} provides a general expression for reciprocal continuous systems, neglecting higher-order terms. 
It's worth noting that although the non-Bloch framework has been applied to 1D continuous systems~\cite{YMHu2023continuous}, extending it to two-dimensional non-Hermitian continuous systems remains challenging, due to the lack of a theoretical framework describing the reciprocal non-Hermitian skin effect. 
In the following, we demonstrate that the GFS formula can be applied to 2D continuous systems. Using the GFS formula, the complex-wave structure of OBC wavefunction and OBC spectrum in the continuous system can be determined. 

As the generalization of the GFS formula to continuous system, we first apply the standing-wave condition along the $x$ direction. 
For a given energy $E_0$, there are two branches of complex $q_{x,1}(q_y,E_0)$ and $q_{x,2}(q_y,E_0)$ by solving $\mathcal{H}(q_x,q_y)-E_0=0$. 
The standing-wave condition in the $x$ direction implies that two wave numbers, $k_x$ and $k_x + \delta k_x$, share the same localization length $\kappa_x$. Therefore, the $x$-direction standing-wave condition imposes a real constraint on the complex momentum $q_x=k_x+i\kappa_x$:
\begin{equation}\label{eq:ContinuousSWCs}
    \Im q_{x,1}(q_y,E_0) = \Im q_{x,2}(q_y,E_0). 
\end{equation}
Within the two real constraints from the characteristic equation $f(q_x, q_y, E) = \mathcal{H}(q_x, q_y) - E = 0$ and one real constraint from the standing-wave condition in Eq.~\eqref{eq:ContinuousSWCs}, the 4D momentum space ($k_x,k_y,\kappa_x,\kappa_y$) reduces to $4 - 2 - 1 = 1$, yielding a 1D GFS curve.

\begin{figure*}[t]
    \begin{center}
    \includegraphics[width=\linewidth]{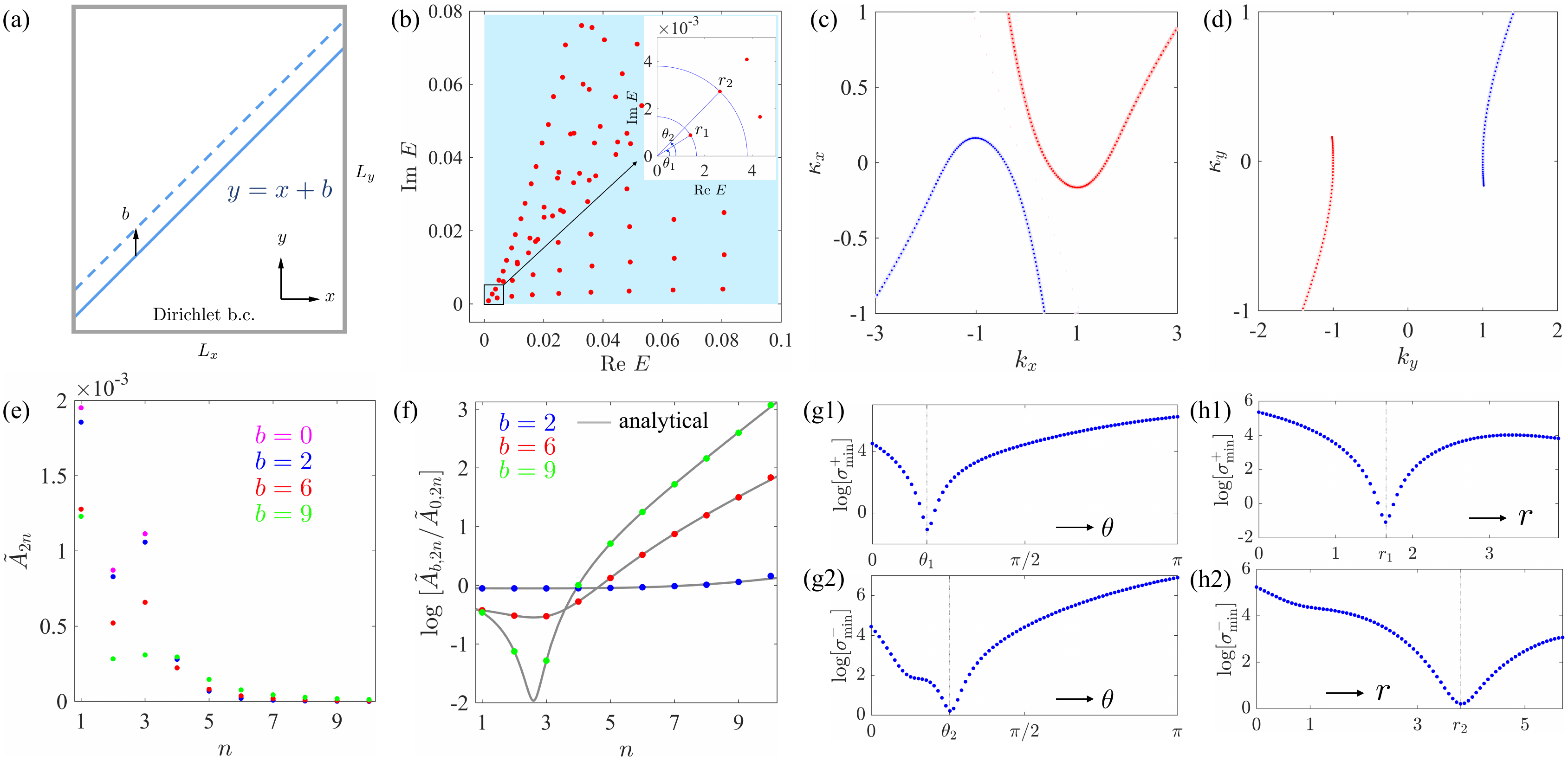}
    \par\end{center}
    \protect\caption{\label{fig:7}~(a)~Illustration of Dirichlet boundary conditions (also known as open boundary conditions) in a rectangular geometry with system size $L_x = 100$ and $L_y = 120$ for the continuous Hamiltonian given by Eq.~\eqref{eq:ContinuousRSE}. The blue solid and dashed lines represent the wavefunction line cuts. 
    (b)~The bluish region shows the PBC spectrum, while the red dots represent the first eighty eigenvalues under open boundary conditions. 
    For the selected OBC eigenvalue $E_0=1/3+i$, the analytic expression for its generalized Fermi surface (GFS) curves can be obtained using the resultant method. The projections of the GFS curve onto the complex-$q_x$ and complex-$q_y$ planes are shown in (c) and (d), respectively. The red and blue dots indicate that the GFS curves are parameterized by the $x$-standing wave number $n \in \mathbb{Z}^{+}$. 
    For the numerical OBC wavefunction of $E_0$, we extract its components along different line cuts with different values of $b$, as illustrated in (a), and then apply the Fourier transform in Eq.~\eqref{eq:FourierTransWave}. The first ten Fourier coefficients (from $n=1$ to $n=10$) for different line cuts $b$ are displayed in (e). 
    (f)~We compare the logarithms of the shift functions $\log[\tilde{A}_{b,2n}/\tilde{A}_{0,2n}]$ obtained from the numerical wavefunction (dots) with the analytical expression (black curves). The excellent agreement between the two confirms the validity of our GFS formula. 
    In (g1) and (g2), we calculate the logarithm of the amplitude of the minimal eigenvalue for the even (odd)-parity boundary matrix, denoted as $\sigma^+_{\text{min}}$ ($\sigma^-_{\text{min}}$), as the energy moves along the inner (outer) arc trajectory shown in the inset of (b). At angle $\theta_1$ ($\theta_2$), the trajectory crosses the first OBC eigenvalue of the system, where $\log[\sigma^+_{\text{min}}]$ ($\log[\sigma^-_{\text{min}}]$) reaches its minimum.
    In (h1) and (h2), we also compute the even (odd)-parity boundary matrix as the energy follows the radial trajectories shown in the inset of (b). At distance $r_1$ ($r_2$), the trajectory crosses the first(second) OBC eigenvalue, where $\log[\sigma^+_{\text{min}}]$ ($\log[\sigma^-_{\text{min}}]$) also reaches its minimum. 
    }
\end{figure*}

To obtain the analytic expression of this 1D GFS curve, we utilize the resultant method~\cite{ZSaGBZPRL}. 
The resultant formula works as follows: consider two constraints on three variables $(x_1, x_2, x_3)$, given by $f_1(x_1, x_2, x_3) = 0$ and $f_2(x_1, x_2, x_3) = 0$. The common solutions of $f_1$ and $f_2$ in the $(x_1, x_2)$ space are given by the zero locus of their resultant with respect to $x_3$: $g(x_1, x_2) \equiv \operatorname{Res}_{x_3}[f_1, f_2] = 0$. 
Let the complex energy be $E = E_r + i E_i$, where $E_r$ and $E_i$ represent the real and imaginary parts of the energy, respectively. 
Thus, the projection of the GFS on the complex-$q_x$ plane can be obtained through the following two resultants:
\begin{align}\label{eq:resultant}
    &g(q_x,\delta k_x,E) {=}\operatorname{Res}_{q_y}[f(q_x,q_y,E),f(q_x+\delta k_x,q_y,E)] \nonumber \\
    &R_x(k_x,\kappa_x,E_r,E_i) {=} \operatorname{Res}_{\delta k_x}[\Re g,\Im g],
\end{align}
where $q_x,q_y,E\in \mathbb{C}$ and $\delta,k_x,\kappa_x,E_r,E_i \in \mathbb{R}$. 
Here, $\Re g$ and $\Im g$ denote the real and imaginary parts of the complex function $g$, respectively. 
Finally, we obtain the real function $R_x$. For each given energy $E = E_r + i E_i$, the equation $R_x(k_x, \kappa_x, E_r, E_i) = 0$ determines the projection of the 1D GFS curve in the $k_x{-}\kappa_x$ plane. 
As $E$ spans a 2D region in the complex-energy plane, the equation $R_x(k_x, \kappa_x, E_r, E_i) = 0$ generates a 3D GBZ manifold. The dimensionality relation follows $\dim{\text{GBZ}}=\dim{\text{GFS}}+\dim{\text{E}}$ given by Eq.~\eqref{eq:GFSdimrelation}. 
Specifically, we set the parameters in Eq.~\eqref{eq:ContinuousRSE} to $A = 1 + i$ and $B = 2$. With these parameters, the continuous model relates to the lattice tight-binding model given by Eq.~\eqref{EQ_RecipHam}, where the skin modes are localized at bottom and top edges. 
We select the energy $E_0 = 1/3 + i$, the projection of the GFS curve on the $k_x {-} \kappa_x$ plane is illustrated in Fig.~\ref{fig:7}(c). 
Next, the $q_y$-projection of the GFS curve for energy $E_0$ can be obtained using the resultant method and shown in Fig.~\ref{fig:7}(d). 
For each $q_y$, there are two $q_x$ solutions with the same imaginary part but different real parts, which can form standing waves in the $x$ direction. 

When we impose the zero boundary conditions (or open-boundary condition) on the continuous system with size $0\leq x \leq L_x$ and $0\leq y \leq L_y$, the standing wave conditions in the $x$ direction quantize the GFS bases into an infinite discrete set, as illustrated in Figs.~\ref{fig:7}(c)(d). 
Now we demonstrate the quantization of the GFS bases under the $x$-direction standing-wave condition. 
The OBC wavefunction can be constructed from the GFS basis as: 
\begin{align}
    \psi_E(x,y) &{=} \int_\text{GFS} \dd q_y A(q_y) e^{i q_y y} \left(e^{iq_{x,1}(q_y) x} + C \, e^{iq_{x,2}(q_y) x}\right) \nonumber \\
    & {=} \int_\text{GFS} \dd q_y A(q_y) e^{iq_y y} e^{-\kappa_x x} e^{i k^+_{x} x} \sin{(k^{-}_x x)}, \label{eq:continSWs}
\end{align}
where $C$ is the superposition coefficient for two exponential waves characterized by $q_{x,1}$ and $q_{x,2}$ in the $x$ direction, and $k^{\pm}_x {=} (k_{x,1}\pm k_{x,2})/2$.
Imposing the left open boundary conditions, $\psi_E(x=0,y)=0$, gives $C=-1$. 
According to the standing wave condition in Eq.~\eqref{eq:ContinuousSWCs}, $\kappa_x = \kappa_{x,1} = \kappa_{x,2}$. 
This means two counter-propagating waves with wavevectors $k_{x,1}$ and $k_{x,2}$ share the same decay length $\kappa_x$ and form standing waves in the $x$ direction. 
These standing waves are coupled to corresponding exponential waves $q_y$ in the $y$ direction, leading to the final form of the OBC wavefunction given in Eq.~\eqref{eq:continSWs}. 
By imposing the right open boundary conditions, $\psi_E(x{=}L_x,y){=}0$, we obtain the constraint $\sin{(k_x^{-} L_x)}{=}0$, which leads to the quantization of the GFS bases: 
\begin{equation}\label{eq:quantizationGFS}
    k_{x,1}(q_y,E) - k_{x,2}(q_y,E) = 2 n \pi/L_x, \quad n \in \mathbb{Z}. 
\end{equation}
This quantization condition causes the GFS curve to become discrete dots, as illustrated in Figs.~\ref{fig:7}(c)(d). 

The Hamiltonian respects inversion (reciprocity) symmetry, therefore, its eigenfunctions can be labeled by the odd or even parity. 
For this reason, we offset our lattice coordinates to the range $-L_x/2 \leq x \leq L_x/2$ and $-L_y/2 \leq y \leq L_y/2$.
Based on the resultant method from Eq.~\eqref{eq:resultant} and the quantization condition from Eq.~\eqref{eq:quantizationGFS}, the GFS curve can be parameterized by the standing wave numbers $n\in \mathbb{Z}$. 
The final expression is as follows:
\begin{equation}\label{eq:GFSexpression}
\begin{split}
    &\kappa_{x,n}=\sqrt{-E_i/2+\sqrt{E_i^2+ \left[ E_r-(n\pi/L_x)^2\right]^2}/2}; \\ 
    &k_{y,n} = \operatorname{sgn}\left[E_r-(n\pi/L_x)^2 \right] \sqrt{E_i+\kappa_{x,n}^2}; \\
    &\kappa_{y,n}=-\kappa_{x,n}; \quad k^+_{x,n}=-k_{y,n}; \quad k^-_{x,n}=n\pi/L_x,
\end{split}
\end{equation}
where the notation $\operatorname{sgn}[x]$ denotes the sign of $x$. 
For any given energy $E=E_r+iE_i$, running $n$ from $1$ to $+\infty$ can generate the GFS curve. 
Consequently, the OBC eigenfunction with even or odd parity ($\psi^+_E$ or $\psi^-_E$) can be discretized as: 
\begin{align}\label{eq:paritymodes}
    \psi^{\pm}_E(x,y) = & \sum_{n=1}^{+\infty} A_n \varphi^{\pm}_n(x,y)= \sum_{n=1}^{+\infty} A_n \sin{\left(\frac{n\pi}{L_x}x+\frac{n\pi}{2}\right)} \nonumber \\ 
    & \times \left[e^{\alpha_n (x-y)}\mp (-1)^n e^{-\alpha_n (x-y)}\right],
\end{align}
where $\alpha_n = \kappa_{x,n} + i k_{y,n}$, $A_n$ represents the superposition coefficient, and $\varphi_n^{\pm}$ denotes the standing wave component. 
For even and odd parity, it can be directly verified that $\psi^{\pm}_E(x,y)=\pm\psi^{\pm}_E(-x,-y)$. 
The superposition coefficients $A_n$ can be determined by the open boundary conditions in the $y$ direction. 

The open boundary conditions in the $y$ direction require that $\psi_E(x,y=\pm L_y/2)=0$. 
We express the boundary conditions in terms of standing wave components, this implies that the projection of edge wavefunction $\psi^{\pm}_E(x,y=\pm L_y/2)$ onto each standing wave component $\varphi_n^{\pm}$ must vanish, i.e., $\langle \varphi^{\pm}_n|\psi^{\pm}_E\rangle=0$. 
It can be further expressed as:
\begin{align}\label{eq:StandingWaveBMatrix}
    & \sum_{m=1}^{+\infty} A_m \int_{-L_x/2}^{L_x/2} \dd x \, \varphi^{\pm \ast}_n(x,L_y/2) \varphi^{\pm}_m(x,L_y/2) \nonumber \\
    & = \sum_{m=1}^{+\infty} \mathcal{M}^{\pm}_{nm} A_m = 0; \quad \forall n\in\mathbb{Z}^+,
\end{align}
where $\mathbb{Z}^+$ represents the positive integer. 
In this formalism, only boundary conditions at edge $y=L_y/2$ are used, as the other boundary conditions at $y=-L_y/2$ become redundant for wavefunctions with even or odd parity. 
Therefore, the boundary matrix is given by $\mathcal{M}_{nm}^{\pm}$ with infinite matrix elements. Meanwhile, the null space of the boundary matrix determines the superposition coefficients $\{A_m\}$. 

\subsection{Numerical verifications of wavefunctions and eigenvalues}

So far, we have extended the GFS formula [Eq.~\eqref{eq:ContinuousSWCs}] to continuous systems and constructed the analytical form of the OBC wavefunction [Eq.~\eqref{eq:continSWs} or Eq.~\eqref{eq:paritymodes}], which represents the standing waves along the $x$-direction, coupled with exponential waves along the $y$-direction. 
We now perform the numerical verification of the wavefunctions and spectrum to confirm the validity of this GFS formula in continuous systems. 

To verify the OBC wavefunctions, we take several line cuts along the diagonal direction, as shown in Fig.~\ref{fig:7}(a). These cuts are defined by $y = x + b$, where $b$ represents the offset from the origin along the $y$-axis. Without loss of generality, we focus on the wavefunction of the energy $E_0=1/3+i$, which exhibits odd parity. 
Using the odd-parity wavefunction from Eq.~\eqref{eq:paritymodes}, the expression for the line-cut component simplifies to: 
\begin{widetext}
\begin{equation}\label{eq:slicedwaves}
    \psi^{-}_E(x,x+b) = \sum_{m=1}^{+\infty} 2A_{2m} \cosh{(\alpha_{2m} \, b)} (-1)^m \sin{\frac{2m\pi x}{L_x}} \nonumber + 2A_{2m-1} \sinh{(\alpha_{2m-1} \, b)} (-1)^m \cos{\frac{(2m-1)\pi x}{L_x}},
\end{equation}
\end{widetext}
where $\alpha_{n}=\kappa_{x,n}+ik_{y,n}$ is determined by the GFS formula and has been given by Eq.~\eqref{eq:GFSexpression} explicitly. 
To extract the superposition coefficients of the standing wave components, we perform the following Fourier transform on the odd-parity wavefunction:
\begin{equation}\label{eq:FourierTransWave}
    \tilde{A}_{b,2n} = \frac{(-1)^n}{L_x}\int_{-L_x/2}^{L_x/2} \dd x \, \sin{(\frac{2n\pi x}{L_x})} \, \psi_E^{-}(x,x+b),
\end{equation}
where $\tilde{A}_{b,2n}\equiv A_{2n} \cosh{(\alpha_{2n} \, b)}$ and $\alpha_{2n}=\kappa_{x,2n}+ik_{y,2n}$ is given by Eq.~\eqref{eq:GFSexpression}. 
The numerical wavefunctions are evaluated using the finite difference method, allowing us to compute the numerical coefficients $\tilde{A}_{2n}$ for line cuts at $b = 0, 2, 6, 9$. The first ten coefficients are presented in Fig.~\ref{fig:7}(e). 
Additionally, the ratio of Fourier coefficients between two different line cuts, $b_1$ and $b_2$, is given by $\cosh(\alpha_{2n} b_2) / \cosh(\alpha_{2n} b_1)$, where $\alpha_{2n} = \kappa_{x,2n} + ik_{y,2n}$, as determined by the GFS formula. 
In Fig.~\ref{fig:7}(f), we compare the ratio of analytical coefficients, $\tilde{A}_{b,2n}/\tilde{A}_{0,2n}$, with the corresponding numerical ratio of Fourier transform coefficients between different line cuts, as illustrated by the black curves and colored dots, respectively. 
The perfect agreement between the two confirms the validity of the OBC wavefunction obtained from the GFS formula in reciprocal continuous systems. 

To verify the OBC eigenvalues, we apply the boundary matrix criterion from Eq.~\eqref{eq:StandingWaveBMatrix}. Whenever the boundary matrix $\mathcal{M}^{\pm}$ has a null space as we vary $E$ on the complex plane , the corresponding energy is identified as an eigenvalue. The superscript $\pm$ denotes the parity of the eigenvalue, with either an even or odd parity wavefunction.
In continuous systems, there are infinitely many eigenvalues, and the first $80$ eigenvalues (sorted by their amplitudes) are illustrated in Fig.~\ref{fig:7}(b), calculated numerically using the finite difference method. 
The wavefunctions of the first and second eigenvalues exhibit even and odd parity, respectively. Therefore, we use the corresponding boundary matrices, $\mathcal{M}^+$ and $\mathcal{M}^-$, to probe these two eigenvalues.
The boundary matrix element $M_{nm}^{\pm}$ is derived from the overlap of standing-wave components, as shown in Eq.~\eqref{eq:StandingWaveBMatrix}. Rather than calculating the null space of the infinite boundary matrix, we impose a cutoff and consider only the first 30 standing-wave bases, i.e., $n$ and $m$ running from 1 to 30. This results in a $30 \times 30$ boundary matrix, from which we calculate their smallest eigenvalue, denoted as $\sigma^{\pm}_{\text{min}}$. 
As shown in the insets of Figs.~\ref{fig:7}(g)(h), we calculate the logarithm of $\sigma^{\pm}_{\text{min}}$. 
The result demonstrates that $\sigma^{+}_{\text{min}}$ reaches its minimum as the energy approaches the first eigenvalue along the arc/ray trajectory, as shown in Figs~.\ref{fig:7}(g1) and (h1). 
Meanwhile, the smallest eigenvalue of the odd-parity boundary matrix, $\sigma^{-}_{\text{min}}$, reaches its minimum when the energy trajectory crosses the second eigenvalue, as shown in Figs~.\ref{fig:7}(g2) and (h2). 
This demonstrates that even with a cutoff, the boundary matrix method effectively identifies both the eigenvalues and their parities in continuous systems. 
The reason for this effectiveness lies in the fact that we order the boundary matrix basis according to the standing wave number. As a result, the initial standing-wave bases correspond to relatively long wavelengths, enabling the cutoff boundary matrix to accurately capture the low-energy parts. 
For higher eigenvalues, however, a larger cutoff of the boundary matrix is required. 
So far, we have verified the validity of the GFS framework in the 2D continuum systems. 

\section{Discussion and Conclusion}~\label{SecVIII}

We clarify the distinctions and connections between our findings and earlier related works. 
In particular, we emphasize the differences between the algebraic non-Hermitian skin effect introduced in this study and the 1D scale-free critical skin effect reported in Refs.~\cite{LiLH2020NC,LLHScaleFree2021,Kawabata2023PRX}. 
We also discuss the relation between our GFS framework, which is constructed based on standing-wave conditions, and the amoeba-based approach proposed in Ref.~\cite{HYWang2022}. 

In this work, the emergence of the algebraic skin effect requires: 
(i)~$\dim{\text{GBZ}}>d$, with $d$ denoting the real-space dimension;
(ii)~A one-dimensional GFS basis that includes Bloch standing-wave components. 
These two conditions immediately rules out the 1D systems. Therefore, the algebraic skin effect in our framework is forbidden in one dimension. 
In addition, these two conditions does not require the system to be in a critical phase in parameter space, encompassing a broad class of higher-dimensional systems that meet these conditions. For example, the algebraic non-Hermitian skin effect generally exists in 2D reciprocal non-Hermitian systems for almost all OBC energies. 

In contrast, the power-law localization of the skin effect discussed in Ref.~\cite{Kawabata2023PRX} arises at a phase transition point characterized by exceptional points, marking the transition from a phase without skin effect to one with exponentially localized skin effect. 
In addition, the critical skin effect proposed in Ref.~\cite{LiLH2020NC} is scale-free and exhibits a localization behavior in which the exponential decay exponent scales with the system size. 
Due to their underlying mechanisms, these scale-free skin effects can occur in arbitrary dimensions, including in one-dimensional systems as illustrated in Refs.~\cite{Kawabata2023PRX,LiLH2020NC}. Moreover, they generally require fine-tuning and occupy a zero-measure subset of the system’s parameter space. Therefore, these phenomena are clearly distinct from the algebraic skin effect proposed in this work.

The amoeba-based formulation proposed in Ref.~\cite{HYWang2022} provides a boundary-free framework that captures universal bulk information independent of boundary conditions. Within this framework, each OBC wavefunction is associated with a unique exponential localization factor. As a result, this framework effectively captures exponential non-Hermitian skin effects in arbitrary dimensions. 
In contrast, our framework develops a GFS framework based on standing-wave construction. We find that under open boundary conditions, for example, in two dimensions, each wavefunction is composed of a continuum of non-Bloch wave components with distinct localization lengths. When these components include diverging localization lengths (i.e., Bloch-wave components), the resulting OBC wavefunctions exhibit quasi-long-range algebraic decay in space. This enables our framework to capture the algebraic non-Hermitian skin effect in two and higher dimensions. Furthermore, in the GFS framework, boundary conditions play a crucial role in determining both the OBC wavefunctions and the energy spectrum. 

In summary, this work establishes the GFS framework based on the standing-wave construction, which applies to non-Hermitian systems with parallelogram-shaped open boundary geometries. 
We find that in a $d$-dimensional non-Hermitian system, in general, the GBZ manifold's dimensionality must fall into the range from $d$ to $2d-1$, denoted by ${d \leq \dim\text{GBZ} \leq 2d-1}$. In 1D, this inequality is trivial because the upper and lower bounds converge, forcing the GBZ's dimensionality to match with that of the physical space, in agreement with existing knowledge on the exponential non-Hermitian skin effect. However, in 2D and above, this inequality indicates that there is no obligation for the GBZ's dimensionality to concur with the physical space's dimensionality, which gives rise to a new class of non-Hermitian skin effects, specifically, the algebraic non-Hermitian skin effect.

Contrary to conventional wisdom, which associates the non-Hermitian skin effect with exponential localization at open boundaries, the algebraic localization commonly exists in higher-dimensional systems reflects quasi-long-range spatial correlations. 
Due to its power-law decay, this universal algebraic skin effect lies beyond the scope of existing theoretical frameworks, highlighting its uniqueness and foundational significance in non-Hermitian physics. 
Because algebraic localization is more nonlocal than exponential localization, it is expected to give rise to novel physical consequences and provide a basis for future research. 
One such example is the bi-impurity-induced ultra spectral sensitivity proposed in Ref.~\cite{Chang2024Ultra}. 
Additionally, we became aware of a proposal for the photonic metamaterial realization of the algebraic skin effect in Ref.~\cite{KunDing2025}.


\begin{acknowledgments}
    We thank Chen Fang and Zhesen Yang for valuable discussions. This work was supported in part by the Office of Naval Research (MURI N00014-20-1-2479). 
\end{acknowledgments}

\appendix

\section{An analytic resultant method for GFS in the thermodynamic limit}~\label{App_ResGFS}

In Section~\ref{SecIII}, we utilize the layer transfer matrix approach to calculate the discretized generalized Fermi surface (GFS) for a finite-size lattice system. 
Here, we introduce an analytic approach, using the resultant method, to obtain continuous GFS curves in the thermodynamic limit. 
This analytic method avoids the need to diagonalize a layer transfer matrix, reducing the computational complexity to $\mathcal{O}(1)$, independent of system size. 

The key aspect of the GFS framework is the standing-wave construction in arbitrary spatial dimension. 
As illustrated in Fig.~\ref{fig:1}, in $d$-dimensional systems, there are $d$ classes as the number of standing wave conditions $n$ ranges from $1$ to $d$. For example, in two dimensions, two distinct cases exist: $n = 1$ and $n = 2$. 
For the case of $n=1$, the standing-wave condition can be enforced along either the $x$ or $y$ direction, expressed as: 
\begin{align}
    &|\beta_{x,M_x}(\rho_y,E_0)| = |\beta_{x,M_x+1}(\rho_y,E_0)| \quad \text{or} \label{EQA_StandingWaveX} \\ 
    &|\beta_{y,M_y}(\rho_x,E_0)| = |\beta_{y,M_y+1}(\rho_x,E_0)|, \label{EQA_StandingWaveY}
\end{align}
where $\beta$ denotes the non-Bloch variable along the standing-wave direction, and $\rho$ corresponds to the transfer direction. 
Notably, the two standing-wave conditions, Eqs.~\eqref{EQA_StandingWaveX} and \eqref{EQA_StandingWaveY}, yield distinct sets of GFS bases that are mutually incompatible.
In the case of $n=d=2$, the standing wave conditions can be imposed along all directions simultaneously, expressed as:
\begin{equation}\label{EQA_StandingWaveXY}
    \begin{split}
    &|\beta_{x,M_x}(\beta_y,E_0)| = |\beta_{x,M_x+1}(\beta_y,E_0)| \quad \text{and} \\
    & |\beta_{y,M_y}(\beta_x,E_0)| = |\beta_{y,M_y+1}(\beta_x,E_0)|. 
    \end{split}
\end{equation}
The notations $\beta_x$ and $\beta_y$ in Eq.~\eqref{EQA_StandingWaveXY} indicate that both the $x$ and $y$ directions are standing-wave directions. 
These two standing wave conditions are imposed simultaneously and determine the GFS. 
In general, when $n < d$, multiple choices of standing-wave directions are possible, resulting in distinct sets of GFS bases and thus non-unique GFS curves. In contrast, for $n = d$, the standing-wave conditions fully constrain the system, uniquely determining the GFS basis. 
When the standing wave conditions are fixed, the corresponding GFS in the thermodynamic limit can be derived analytically. Below, we use the example used in Fig.~\ref{fig:4} to demonstrate the analytic method. 

The non-Bloch Hamiltonian for Hamiltonian in Eq.~\eqref{EQ_2DTBModel} is expressed as
\begin{equation}\label{EQA_NonBlochHam}
    \begin{split}
    \mathcal{H}(\beta_x,\beta_y) & = t_x (\beta_x+\beta_x^{-1}) + t_y (\beta_y+\beta_y^{-1}) \\ 
    & + t_{xy} \left[ \beta_x\beta_y+(\beta_x\beta_y)^{-1} \right] + u,
    \end{split}
\end{equation}
where the Hamiltonian parameters are set to $(t_x,t_y,t_{xy},u) = (1,1,i/2,-i)$. 
In the main text, it is known that this Hamiltonian belongs to the case $n=1<d$. 
Therefore, the standing wave condition is given by either Eq.~\eqref{EQA_StandingWaveX} or Eq.~\eqref{EQA_StandingWaveY}. 
Without loss of generality, we impose the standing wave condition along the $x$ direction. 
Therefore, for a given energy $E_0$, the complex momenta are under constrains of the bulk characteristic equation $f_{E_0}(\beta_x,\rho_y)=0$ and the standing wave condition Eq.~\eqref{EQA_StandingWaveX}. 
These conditions actually provide three real constrains, which restrict the four-dimensional momentum variables into a 1D curve, i.e., GFS curve. 
We utilize the resultant method to obtain the GFS curve, which serves as the thermodynamic limit version of the discretized GFS obtained from the layer transfer matrix. 

\begin{figure}[h]
    \begin{center}
    \includegraphics[width=1\linewidth]{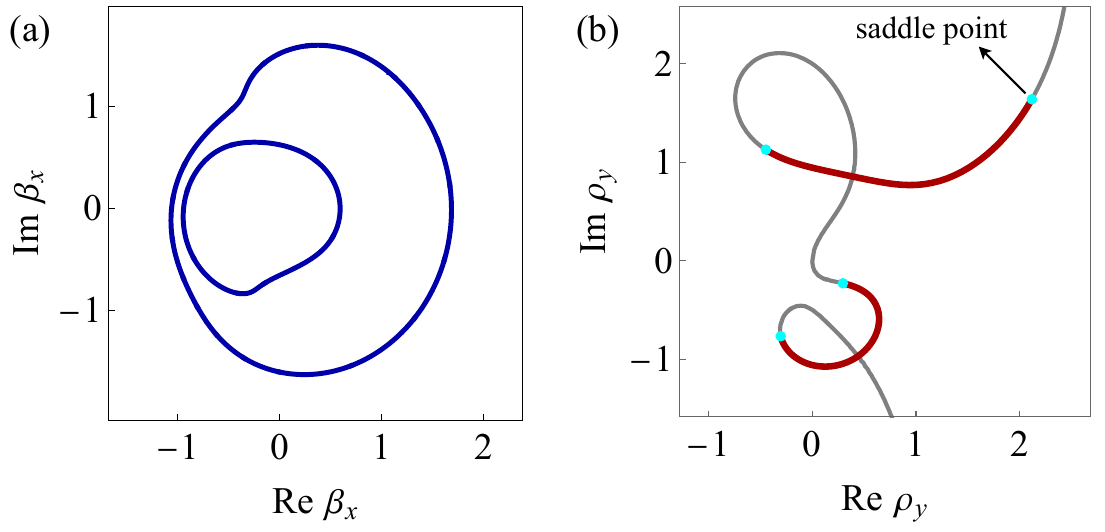}
    \par\end{center}
    \protect\caption{\label{fig:A1}~The GFS curves in the thermodynamic limit are analytically solved by resultant method and plotted in complex $\beta_x$ (a) and $\rho_y$ (b) planes. 
    In (b), the two $\rho$ arcs are terminated at four saddle points (the cyan dots) defined by Eq.~\eqref{EQA_SaddlePoints}. 
    The Hamiltonian follows Eq.~\eqref{EQA_NonBlochHam}, using the same parameters as in Fig.~\ref{fig:4}. The GFS curves correspond to the energy $E_0{=}0.819{-}1.108i$. }
\end{figure}

The resultant formula works as follows: consider two constraints on three variables $(x_1, x_2, x_3)$, given by $f_1(x_1, x_2, x_3) = 0$ and $f_2(x_1, x_2, x_3) = 0$. The common solutions of $f_1$ and $f_2$ in the $(x_1, x_2)$ space are given by the zero locus of their resultant with respect to $x_3$: $g(x_1, x_2) \equiv \operatorname{Res}_{x_3}[f_1, f_2] = 0$. 
We denote the characteristic equation as $f_{E_0}(\beta_x,\rho_y)=\det[\mathcal{H}(\beta_x,\rho_y)-E_0]=0$.
Because for a given $\rho_y$, there are two $\beta_x$ solutions with the same amplitude but different phases, a second equation should be satisfied: $f_{E_0}(\beta_x e^{i\theta},\rho_y)=\det[\mathcal{H}(\beta_x,\rho_y)-E_0]=0$, where the phase factor $\theta$ is real. 
Using the Weierstrass substitution, we have $e^{i\theta}\equiv (1+2i\tau-\tau^2)/(1+\tau^2)$ and $\tau$ is a real variable. 
Now we have four real constraints coming from these two complex equations, and five real variables: the real and imaginary parts of $\beta_x$ and $\rho_y$, and the real variable $\tau$. 
Using these notations, we have the following resultant equations to obtain the GFS in the $\beta_x$ plane: 
\begin{align}
    &g_x(\Re{\beta}_x,\Im{\beta}_x,\tau) \nonumber \\
    &= \text{Res}_{\beta_y}[f_{E_0}(\beta_x,\rho_y),f_{E_0}(\beta_x \, \frac{1+2i\tau-\tau^2}{1+\tau^2},\rho_y)] \label{EQA_Resx1Eq1} \\ 
    &\text{GFS}_x(\Re{\beta}_x,\Im{\beta}_x) =  \text{Res}_{\tau}[\Re(g_x),\Im(g_x)].
\end{align}
Note that Eq.~\eqref{EQA_Resx1Eq1} is a complex function since $f_{E_0}$ is complex, and $\text{GFS}_x(\Re{\beta}_x,\Im{\beta}_x)$ is a real function of variables $\Re{\beta}_x$ and $\Im{\beta}_x$. 
Eventually, $\text{GFS}_x(\Re{\beta}_x,\Im{\beta}_x)$ gives rise to the GFS projection onto the complex $\beta_x$ plane.
The zero locus of $\text{GFS}_x(\Re{\beta}_x,\Im{\beta}_x)$ is plotted by the red curves in Fig.~\ref{fig:A1}(a). 
Likewise, the GFS projection onto $\rho_y$ plane can be obtained using the following resultants: 
\begin{align}
    & g_y(\Re{\rho}_y,\Im{\rho}_y,\tau) \nonumber \\ 
    &= \text{Res}_{\beta_x}[f_{E_0}(\beta_x,\rho_y),f_{E_0}(\beta_x \, \frac{1+2i\tau-\tau^2}{1+\tau^2},\rho_y)] \label{R1_Resx1Eq1} \\ 
    &\text{GFS}_y(\Re{\rho}_y,\Im{\rho}_y) =  \text{Res}_{\tau}[\Re(g_y),\Im(g_y)].
\end{align}
The zero-locus equation $\text{GFS}_y(\Re{\rho}_y,\Im{\rho}_y)=0$ is plotted by the gray curve in Fig.~\ref{fig:A1}(b). 
It's important to note that the gray analytic curve contains the GFS projection curve, but it is not identical to it; rather, they share the same underlying algebraic equation. 
The GFS projection curve in the $\rho_y$ plane need to satisfy additional requirement. 
Since the projection of the GFS onto the $\rho_y$ plane forms arcs, its endpoints (cyan dots in Fig.~\ref{fig:A1}(b)) are determined by the saddle-point condition: 
\begin{equation}
    \partial \rho_y/\partial\beta_x=0.
\end{equation}
Considering the saddle-point condition above, the endpoints of $\rho_y$ curves (or arcs) can be determined by the following resultant: 
\begin{align}\label{EQA_SaddlePoints}
    g_s(\rho_y) = \text{Res}_{\beta_x}[f_{E_0}(\beta_x,\rho_y),-\frac{\partial_{\beta_x}f_{E_0}(\beta_x,\rho_y)}{\partial_{\rho_y}f_{E_0}(\beta_x,\rho_y)}].
\end{align}
The second term in the resultant is the saddle-point condition. 
The $\rho_y$ solutions of $g_s(\rho_y)=0$ determine the endpoints of the $\rho_y$ curves, indicated by the cyan dots in Fig.~\ref{fig:A1}(b).  
Thus far, we have analytically obtained the GFS curves in the thermodynamic limits, based on the standing-wave condition in the $x$ direction, which are shown by the red and blue curves in Fig.~\ref{fig:A1}. 

For a finite size lattice system, the GFS curve becomes discretized, with the number of discrete points matching the number of boundary conditions in the $y$ direction. 
For the model in Eq.~\eqref{EQA_NonBlochHam}, assuming a system size of $L_x\times L_y$, there are $2L_x$ boundary conditions in the $y$ direction. 
In Section~\ref{SecIII}, we use the layer transfer matrix to discretize the GFS curve. 
In the following, we demonstrate that the GFS becomes discretized due to the quantization of standing waves along the $x$ direction, consistent with the results obtained from the layer transfer matrix approach.  

The GFS provides the basis to construct the OBC wavefunction, expressed as: 
\begin{align}
    \psi_{E_0}(x,y) & = \sum_{\rho \in \text{GFS}} A(\rho) \, \rho^y \varphi_{\rho}(x)  \nonumber \\
    & = \sum_{\rho \in \text{GFS}} A(\rho) \, \rho^y [ \beta^x_{1}(\rho) + B(\rho) \beta^x_{2}(\rho)] ,
\end{align}
where the subscripts $x,y$ of $\beta$ and $\rho$ are neglected for simplicity. 
Notation $\varphi_{\rho}(x)$ represents the standing-wave component characterized by $\rho$. 
For each $\rho$, the corresponding standing wave component $\varphi_{\rho}(x)$ is required to satisfy the open boundary conditions in the $x$ direction. 
Therefore, we have
\begin{equation}\label{EQA_SWCond}
    \varphi_{\rho}(x=0) = 0; \quad 
    \varphi_{\rho}(x=L_x+1) = 0.
\end{equation}
For the first condition, we further obtain: $B(\rho)+1=0$, and then $B(\rho)=-1$. Consequently, the standing-wave component has the form of $\varphi_{\rho}(x) = \beta_1^x(\rho) - \beta_2^x(\rho) = |\beta_1(\rho)|^x (e^{i \theta_1 x}-e^{i \theta_2 x}) = |\beta_1(\rho)|^x \, 2i \, e^{i \theta_+ x} \, \sin( \theta_- x)$. 
Here, $\theta_{i}$ is the phase of non-Bloch variable $\beta_{i}$ and $\theta_\pm = (\theta_1\pm \theta_2)/2$. 
The second condition in Eq.~\eqref{EQA_SWCond} gives rise to $\sin[(L_x+1)\theta_-] = 0$. 
It further leads to 
\begin{equation}
    \theta_-(\rho) = \frac{\theta_1(\rho)-\theta_2(\rho)}{2} = \frac{\pi m}{L_x+1}; \quad m \in \mathbb{Z}. 
\end{equation}
Here, $\theta_-$, which represents the phase difference between the $\beta$ solutions, depends on the value of $\rho$. 
This equation imposes a quantization condition for standing waves along the $x$ direction, discretizing the GFS curves to match the lattice boundary degrees of freedom. 
We emphasize that, by using the layer transfer matrix approach, the resulting discretized GFS automatically satisfies this quantization condition. 

\section{Non-uniqueness of GFS bases}~\label{App_DiffGFS}

In this section, we illustrate with an example that, for the cases of $n<d$, as shown in Fig.~\ref{fig:1}, the GFS varies depending on the choice of standing-wave basis for a given OBC eigenenergy and eigenstate. 
This non-uniqueness in the GFS bases aligns with the flexibility in selecting the standing-wave direction. 
In contrast, the GFS becomes uniquely determined in the case of $n=d$ in $d$ dimensions. 

\begin{figure}[h]
    \begin{center}
    \includegraphics[width=1\linewidth]{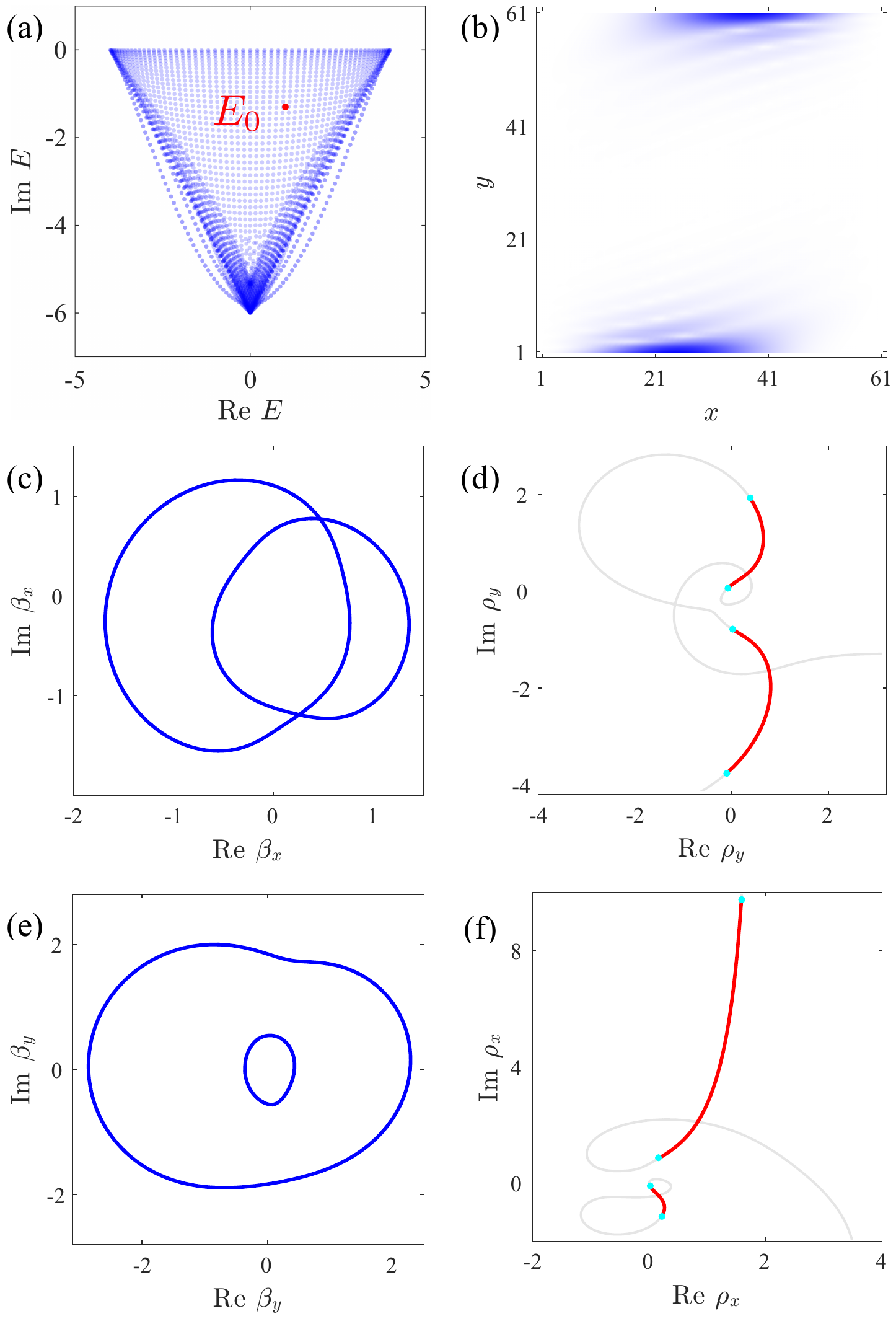}
    \par\end{center}
    \protect\caption{\label{fig:A2}~(a)~presents the open boundary eigenvalues of the Hamiltonian in Eq.~\eqref{EQA_NonBlochHam} for a system size of $L_x = L_y = 61$, with system parameters set to $(t_x,t_y,t_{xy},u)=(1,1,i/2,-i)$. The selected OBC eigenvalue, $E_0 = 1 - 1.3i$, is denoted by a red dot. (b) illustrates the wavefunction corresponding to this eigenvalue $E_0$. For this same eigenvalue and eigenstate, the distinct GFSs are shown under the $x$-direction and $y$-direction standing wave bases in (c)(d) and (e)(f), respectively. }
\end{figure}

The model Hamiltonian in Eq.~\eqref{EQA_NonBlochHam} is used, and the system parameters are set to ($t_x,t_y,t_{xy},u$)=($1,1,i/2,-i$). 
In this case, the standing waves can be enforced along either the $x$ or $y$ direction. 
Different standing-wave bases lead to distinct GFS curves. 
The OBC eigenvalues of the Hamiltonian, with the system size $L_x=L_y=61$, are represented by the blue dots in Fig.~\ref{fig:A2}(a). The OBC wavefunction with energy $E_0$ (red dot in (a)) is shown in Fig.~\ref{fig:A2}(b). 
Within the standing wave condition in the $x$ direction, the GFS in the complex $\beta_x$ and $\rho_y$ planes are obtained in (c) and (d), respectively. 
The analytic resultant method is discussed in Appendix~\ref{App_ResGFS}. 
Notably, the GFS in the $\rho_y$ space are two red arcs ended at the saddle points (cyan dots), as illustrated in Fig.~\ref{fig:A2}(d). 

If we choose the standing wave basis in the $y$ direction, the standing wave condition switches from Eq.~\eqref{EQA_StandingWaveX} to Eq.~\eqref{EQA_StandingWaveY}. 
This condition, along with the characteristic equation, determines the alternative set of basis functions, specifically the GFS curves under the $y$-direction standing wave basis, as shown in Figs.~\ref{fig:A2}(e) and (f). 
Here, we use the notations $\beta_y$ and $\rho_x$, indicating that the standing waves are formed in the $y$ direction. 
From Figs.~\ref{fig:A2}(c)(d) and (e)(f), it is evident that the GFSs differ based on the choice of standing-wave basis, despite having the same eigenvalues and eigenstates. This flexibility in choosing the standing-wave basis arises when $n<d$, where standing waves cannot simultaneously form along all directions. 

To solve for the GFS under the $y$-direction standing wave basis, the complex momenta need to satisfy the following two complex equations:
\begin{align}
    &f_{E_0}(\beta_x,\beta_y)=\det[\mathcal{H}(\beta_x,\beta_y)-E_0]=0; \\ 
    &f_{E_0}(\beta_x,\beta_y e^{i\theta})=\det[\mathcal{H}(\beta_x,\beta_ye^{i\theta})-E_0]=0.
\end{align}
Following the same procedure discussed in the Appendix~\ref{App_ResGFS}, the GFS can then be determined using the resultant method and the saddle point condition. 

\section{A non-reciprocal model with algebraic non-Hermitian skin effect}~\label{App_NRSEModel}

\begin{figure}[h]
    \begin{center}
    \includegraphics[width=1\linewidth]{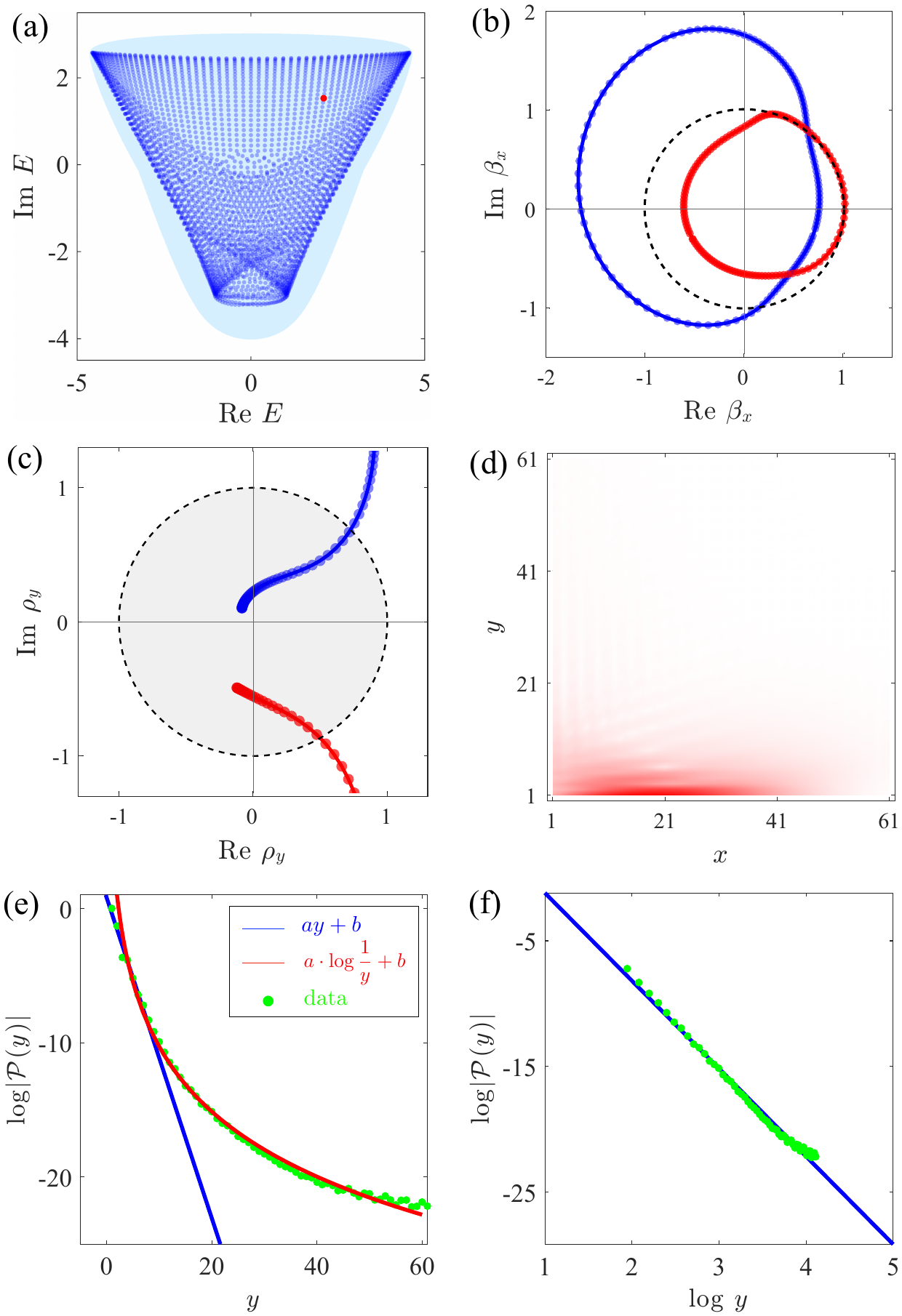}
    \par\end{center}
    \protect\caption{\label{fig:A3}~Algebraic skin modes in a non-reciprocal model. (a) shows the periodic boundary spectrum (the light blue region) and open boundary spectrum (the blue dots) for the non-reciprocal model Hamiltonian given by Eq.~\eqref{EQC_NRSEModel}. 
    (b) and (c) represent the projections of the GFS curves onto the complex $\beta_x$ plane and the $\rho_y$ plane, respectively, corresponding to the OBC eigenvalue $E_0=2+1.5i$ [the red dot in (a)]. In both (b) and (c), the black dashed circle is a unit circle for reference. To demonstrate the quasi-long-range algebraic decay, (d) shows the partial wavefunction composed of the non-Bloch wave components with $\rho_y$ inside the unit circle [the gray region in (c)]. (e) plots the layer density defined in Eq.~\eqref{EQC_LayerDensity}, with the data (green dots) fitted using a linear function (blue line) and a power-law function (red line). Furthermore, the straight line fitting in the log-log plot in (f) clearly indicates the power-law decay of the layer density. }
\end{figure}

Here, we provide a non-reciprocal tight-binding model exhibiting the algebraic skin effect. The Hamiltonian in the non-Bloch form is given by: 
\begin{equation}\label{EQC_NRSEModel}
    \mathcal{H}(\beta_x,\beta_y) = 2 \beta_x + \beta_x^{-1} + \frac{3}{2} \beta_y + \beta_y^{-1} + \frac{1}{2} \beta_x\beta_y + (\beta_x\beta_y)^{-1},
\end{equation}
which incorporates non-reciprocal hopping strengths along the $x$, $y$, and $x+y$ diagonal directions, respectively. 
The eigenvalues of the Hamiltonian within a square lattice of length $L_x=L_y=61$ are shown by the blue dots in Fig.~\ref{fig:A3}(a). 
We select the eigenvalue $E_0=2+1.5i$ (red dot), and plot its GFS in the complex $\beta_x$ and $\rho_y$ planes in Figs.~\ref{fig:A3}(b) and (c). 
Here, we use the basis of standing waves in the $x$ direction. 
Using the resultant method, we can obtain the 1D GFS curves, as shown in Figs.~\ref{fig:A3}(b) and (c). 
Note that the discrete dots in the GFS arise from the finite lattice degrees of freedom. 
As the lattice size approaches the thermodynamic limit, these discrete dots form the continuous curves shown in Figs.~\ref{fig:A3}(b) and (c). 

Within the GFS basis, we can construct the open-boundary wavefunction of energy $E_0$, its form can be written as: 
\begin{equation}\label{EQC_wavefunc}
    \psi_{E_0}(x,y) = \sum_{\rho\in\text{GFS}} A(\rho)\, \rho^y \, [\beta^x_{1}(\rho)+B\beta^x_{2}(\rho)].
\end{equation}
Imposing the open-boundary conditions in the $x$ direction on the wavefunction in Eq.~\eqref{EQC_wavefunc}, the coefficient $B=-1$, which means that the standing waves have the sin-form functions in the $x$ direction. 
Imposing the open-boundary conditions in the $y$ direction on the wavefunction in Eq.~\eqref{EQC_wavefunc}, we obtain a boundary matrix, and the null space gives rise to the coefficients $A(\rho)$. 
Finally, the OBC wavefunction $\psi_{E_0}(x,y)$ from GFS basis is obtained, and it fully agrees with the direct diagonalization of the Hamiltonian matrix. 

To verify the quasi-long-range algebraic decay behavior, we extract the partial wavefunction of $\psi_{E_0}(x,y)$ that is constituted by the complex-wave components $\rho$ inside the unit circle, as illustrated by the gray region in Fig.~\ref{fig:A3}(c). Since the corresponding coefficients of these complex-wave components have been solved, the corresponding partial wavefunction, denoted by $\psi_{\text{in}}(x,y)$, can be obtained and shown in Fig.~\ref{fig:A3}(d). 
We further calculate its layer density:
\begin{equation}\label{EQC_LayerDensity}
    \mathcal{P}_{\text{in}}(y) := \int \dd x \, |\psi_{\text{in}}(x,y)|^2. 
\end{equation}
The layer density is shown in Fig.~\ref{fig:A3}(e). 
It's evident that the layer density of wavefunction exhibits quasi-long-range algebraic decay, instead of commonly believed short-range exponential decay. 
To further confirm its algebraic decay, we show the log-log plot for the layer density in Fig.~\ref{fig:A3}(f). 
In the fitting, we discard the first 6 layers near the boundary, and the straight-line fitting on the log–log plot is often called the signature of a power law. 
We have thus far demonstrated that the quasi-long-range algebraic decay is not exclusive to reciprocal systems and widely distributed in higher-dimensional systems with $\dim \text{GBZ}>d$. 

\section{The existence of Fermi points ensured by reciprocity}~\label{App_FPsProof}

Here, we prove that the existence of Fermi points for open-boundary bulk eigenvalues is guaranteed by the reciprocity of the Hamiltonian. 

For a given Bloch Hamiltonian $h(k_x,k_y)$, we first define the spectral winding number in the $x$ direction with fixed $k_y$
\begin{equation}\label{App_SpecWinding1D}
    w(E_0, k_y) = \frac{1}{2\pi} \int_{-\pi}^{\pi} d k_x \, \partial_{k_x} \arg\det[h(k_x,k_y)-E_0],
\end{equation}
where $E_0$ represents a generic OBC eigenvalue, considered as the reference energy here. 

The reciprocity of the Hamiltonian ensures that the OBC bulk eigenvalues must be included in its PBC spectrum. Here, we provide a proof for this statement. 
Suppose an open-boundary bulk eigenvalue $E_0 \notin \sigma_{\text{PBC}}$, and the Hamiltonian respects reciprocity. 
It is known that the OBC bulk spectrum is obtained from the collapse of the PBC spectrum. Thus, the OBC bulk spectrum cannot exceed the outer boundary of the PBC spectrum. 
If we assume $E_0 \notin \sigma_{\text{PBC}}$, the only possible case is that $E_0$ is located within an inner gap of the PBC spectrum, such as some inner holes of the PBC continuum spectrum. 
Below, we prove that this case is forbidden by the reciprocity of the Hamiltonian. Consequently, open-boundary eigenvalues must belong to the PBC spectrum of $h(k_x,k_y)$, and for a given open-boundary eigenvalue $E_0$, there exist Fermi points ensured by reciprocity. 

Due to the reciprocity of the Hamiltonian, the spectral winding number in Eq.~(\ref{App_SpecWinding1D}) satisfies $w(E_0, k_y)=-w(E_0, -k_y)$. 
At the high symmetry points ${k^{\ast}_y=0,\pi}$, we have $k^{\ast}_y=-k^{\ast}_y$. Therefore, for a generic OBC eigenvalue $E_0$, the spectral winding number satisfies $w(E_0, k^{\ast}_y)=-w(E_0, k^{\ast}_y)=0$. 
It means that the PBC spectrum of $h(k_x,k_y^{\ast})$ collapses into arcs in the complex-energy plane~\cite{Kai2020}. 
Similarly, the spectrum of $h(k^{\ast}_x,k_y)$ also collapses into arcs at the high symmetry points ${k_x^{\ast}=0,\pi}$. 
Assuming $E_0$ is in the inner gap of the PBC spectrum, there is a $k_y^0$ such that the spectrum of $h(k_x,k_y^0)$ has a nonzero spectral winding number with respect to $E_0$. 
When we smoothly transition $k_y$ from $k_y^0$ to $k_y^{\ast}$ (for example, $k^{\ast}_y=0$), the corresponding PBC spectrum of the 1D subsystem of $k_x$ must transition smoothly. 
Therefore, it is impossible to squeeze the spectrum from a loop encircling energy $E_0$ (at $k_y=k_y^0$) into an arc (at $k_y=k_y^{\ast}$) without passing through $E_0$. 
Consequently, the PBC spectrum must sweep through $E_0$ at some $k_{y,i}$. These $k_{y,i}$ correspond to the Fermi points of $E_0$. Therefore, open-boundary eigenvalue $E_0$ must belong to the periodic-boundary spectrum of $h(k_x,k_y)$ and the existence of Fermi points is ensured by reciprocity.  

\begin{figure}[t]
    \begin{center}
    \includegraphics[width=1\linewidth]{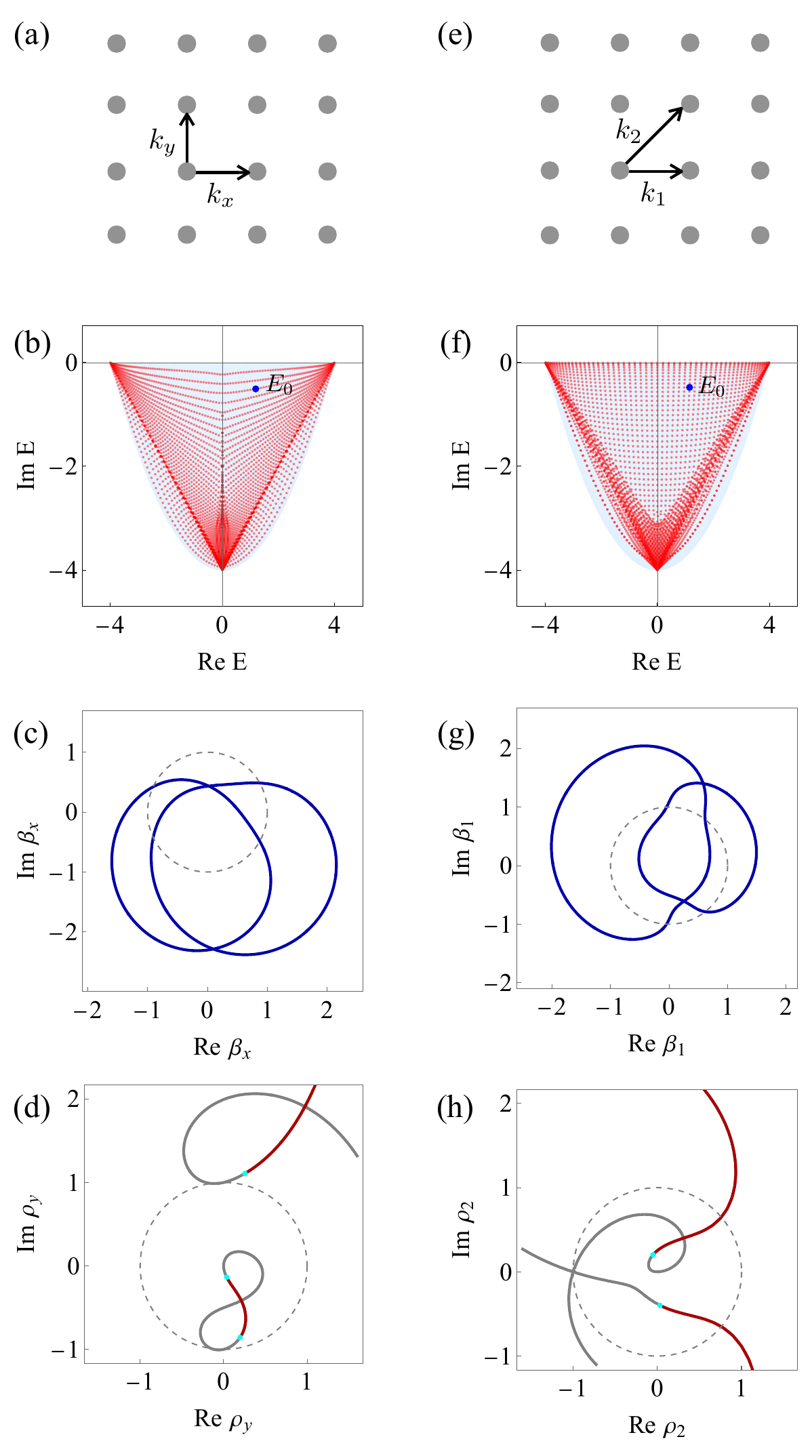}
    \par\end{center}
    \protect\caption{\label{fig:A4}~Generalized Fermi surfaces of the same bulk Hamiltonian for square [(a)-(d)] and parallelogram [(e)-(h)] geometries. 
    (a) and (e) illustrate the different momentum bases. In (b) and (f), the bluish regions indicate the periodic boundary spectrum, while the red dots represent the open boundary eigenvalues of the Hamiltonians in Eq.~\eqref{EQE_BasisHam1} and Eq.~\eqref{EQE_BasisHam2}, respectively. The blue and red curves in (c) and (f) represent the GFS in the complex-$\beta_x$ and $\rho_y$ planes, where the black dashed circle is a unit circle for reference. 
    In contrast, (g) and (h) show the GFS of the Hamiltonian in Eq.~\eqref{EQE_BasisHam2} after a basis transformation.}
\end{figure}

\section{The reciprocal systems ensure the existence of algebraic skin effect}~\label{App_GDSEareASE}

In this section, we demonstrate that reciprocal systems, such as those exhibiting the geometry-dependent skin effect, ensure the occurrence of the algebraic skin effect, which consistently appears under generic open boundary geometries. 
Using a reciprocal Hamiltonian, we illustrate that the GFS is gapped from the Brillouin zone under square geometries, whereas it intersects with the Brillouin zone under other parallelogram geometries, indicating the reappearance of the algebraic skin effect. 

The tight-binding Hamiltonian in momentum space reads 
\begin{equation*}
\begin{split}
    \mathcal{H}(k_x,k_y) & = 2 t_x \cos{k_x} {+} 2 t_y \cos{k_y} {+} 2 t_{xy} \cos{(k_x{+}k_y)} {+} u.
\end{split}
\end{equation*}
The system parameters are set to $(t_x,t_y,t_{xy},u)=(1,1,i,-2i)$. 
When we choose the ($k_x,k_y$) basis as shown in Fig.~\ref{fig:A4}(a), the corresponding open boundary geometry is a square geometry, whose boundaries are parallel to these two wavevectors. 
With open boundary conditions, the momenta generally take complex values; hence, the non-Bloch form of the Hamiltonian is written as: 
\begin{equation}\label{EQE_BasisHam1}
    \begin{split}
    \mathcal{H}(\beta_x,\beta_y) & = t_x (\beta_x+\beta_x^{-1}) + t_y (\beta_y+\beta_y^{-1}) \\
    & + t_{xy} (\beta_x\beta_y+\beta_x^{-1}\beta_y^{-1}) + u,
    \end{split}
\end{equation}
where $\beta_x=e^{ik_x}$ and $\beta_y=e^{ik_y}$. 
The periodic-boundary and open-boundary spectra are shown in Fig.~\ref{fig:A4}(b), which are represented by the bluish region and the red dots, respectively. The system size is $L_x=L_y=61$. 
Using the resultant method discussed in Appendix~\ref{App_ResGFS}, the analytic GFS curves under square geometry can be solved in Figs.~\ref{fig:A4}(c) and (d). 
It shows that the GFS in the $\rho_y$ plane has a gap from the Brillouin zone (the dashed unit circle), indicating the absence of algebraic decay along the $y$ direction. 

Now, we perform a basis transform [which belongs to the special linear group SL(2,$\mathbb{Z}$)] as shown from Fig.~\ref{fig:A4}(a) to Fig.~\ref{fig:A4}(e), expressed as: 
\begin{equation*}
    \begin{pmatrix}
    k_x \\ k_y 
    \end{pmatrix} =
    \begin{pmatrix}
    1 & 0 \\ -1 & 1
    \end{pmatrix}
    \begin{pmatrix}
    k_1 \\ k_2
    \end{pmatrix}
\end{equation*}
After applying this transformation, we can express the Hamiltonian in terms of the new momentum basis ($k_1,k_2$) as: 
\begin{equation*}
    \tilde{H}(k_1,k_2) = 2 t_x \cos{k_1} {+} 2 t_y \cos{(k_1{-}k_2)} {+} 2 t_{xy} \cos{k_2} {+} u.
\end{equation*}
Within the ($k_1,k_2$) basis, the corresponding OBC boundary geometry is a parallelogram whose boundaries are parallel to these two wavevectors. 
The corresponding non-Bloch Hamiltonian can be written as:
\begin{equation}\label{EQE_BasisHam2}
    \begin{split}
    \tilde{H}(\beta_1,\beta_2) &= t_x (\beta_1+\beta_1^{-1}) + t_y (\beta_1 \beta^{-1}_2+ \beta_2 \beta^{-1}_1) \\ 
    & + t_{xy} (\beta_2+\beta_2^{-1}) + u,
    \end{split}
\end{equation}
where $\beta_1=e^{ik_1}$ and $\beta_2=e^{ik_2}$. 
Remarkably, as illustrated in Fig.~\ref{fig:A4}(f), under this basis transformation, the periodic boundary spectrum (bluish region) remains invariant since the Brillouin zone is unaffected by the transformation. However, the density of the open boundary spectrum (red dots) changes significantly due to the variation in the GFSs under different boundary geometries. 
The GFS curves in the parallelogram geometry, using the ($k_1,k_2$) basis, can be analytically solved via the resultant method, as shown in Figs.~\ref{fig:A4}(g) and (h). 
Notably, the GFS in the $k_2$ direction [represented by the GFS curves in the $\rho_2$ space in Fig.~\ref{fig:A4}(h)] intersects the Brillouin zone, indicating the presence of algebraic decay in the wavefunction along the $k_2$ direction. 
The intersection between the GFS and the Brillouin zone along a specific direction is guaranteed by the presence of Fermi points, which is further ensured by reciprocity [see discussion in Appendix~\ref{App_FPsProof}]. 
Therefore, reciprocal systems always show the algebraic skin effect under specific open boundary geometries. 

\begin{figure}[t]
    \begin{center}
    \includegraphics[width=1\linewidth]{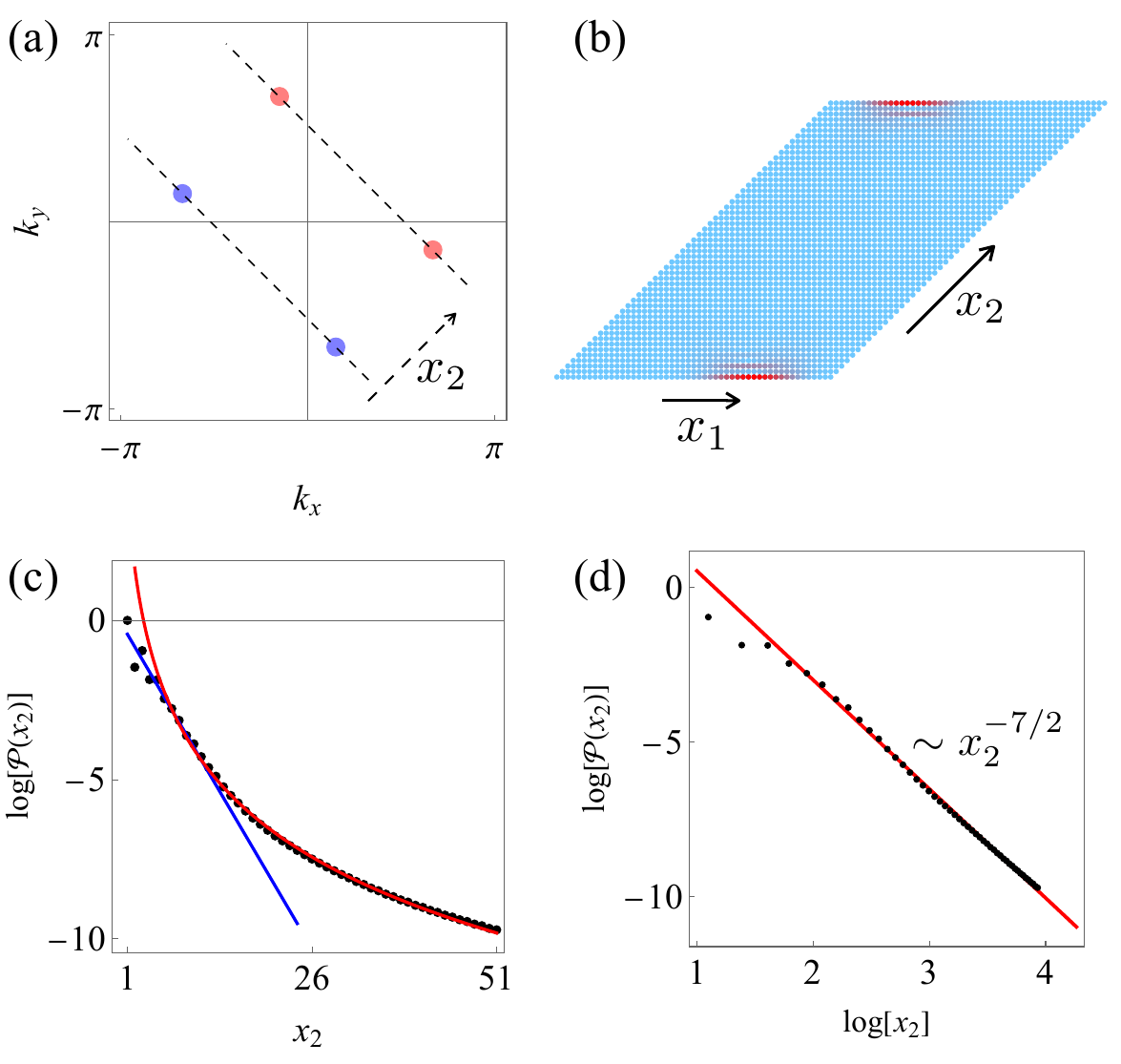}
    \par\end{center}
    \protect\caption{\label{fig:A5}~(a) Distribution of Fermi points for $E_0 = 0.819 - 1.108 i$ in the Brillouin zone. (b) The spatial distribution of wavefunction of $E_0$ under a parallelogram boundary geometry. (c) The layer density in the parallelogram geometry exhibits algebraic localization along the $x_2$ direction. (d) Log-log plot of layer density further confirms the power-law localization with a decay exponent of approximately $x^{-3.5}$. }
\end{figure}

\section{Algebraic skin effect under different boundary geometries}

In this section, we use the model in Fig.~\ref{fig:4} to demonstrate that the quasi-long-range algebraic localization occurs when an appropriate boundary geometry is selected. 

We calculate the Fermi points corresponding to $E_0 = 0.819 - 1.108 i$, which are the solutions $\textbf{k}$ to characteristic equation $\det[H(\textbf{k})-E_0]=0$. Here, $H(\textbf{k})$ represents the Bloch Hamiltonian representation of Eq.~\eqref{EQ_2DTBModel}. 
In the Brillouin zone, there are four Fermi points, as shown in Fig.~\ref{fig:A5}(a).
According to the distribution of Fermi points, we consider a parallelogram shaped open boundary geometry where a pair of Fermi points (in the same color) project onto the same location along the slant edge (along $\hat{x}_2=\hat{x}+\hat{y}$ direction), as illustrated in Fig.~\ref{fig:A5}(b). 
The other edge of the parallelogram is along the $\hat{x}_1=\hat{x}$ direction. 
Next, we compute the layer density by integrating the density along the $x_1$-edge and examine its decay behavior along the slant $x_2$-edge. 
Fig.~\ref{fig:A5}(c) clearly shows the algebraic localization with a characteristic extended long tail. The log-log plot in Fig.~\ref{fig:A5}(d) further confirms the power-law localization. 
These numerical verifications of layer density demonstrate that in reciprocal non-Hermitian systems, the algebraic non-Hermitian skin effect can always occur when the open boundary geometry is appropriately selected. 

\section{The GFS formula in three-dimensional systems}~\label{App_3DGBZFomula}

\begin{figure}[t]
    \begin{center}
    \includegraphics[width=1\linewidth]{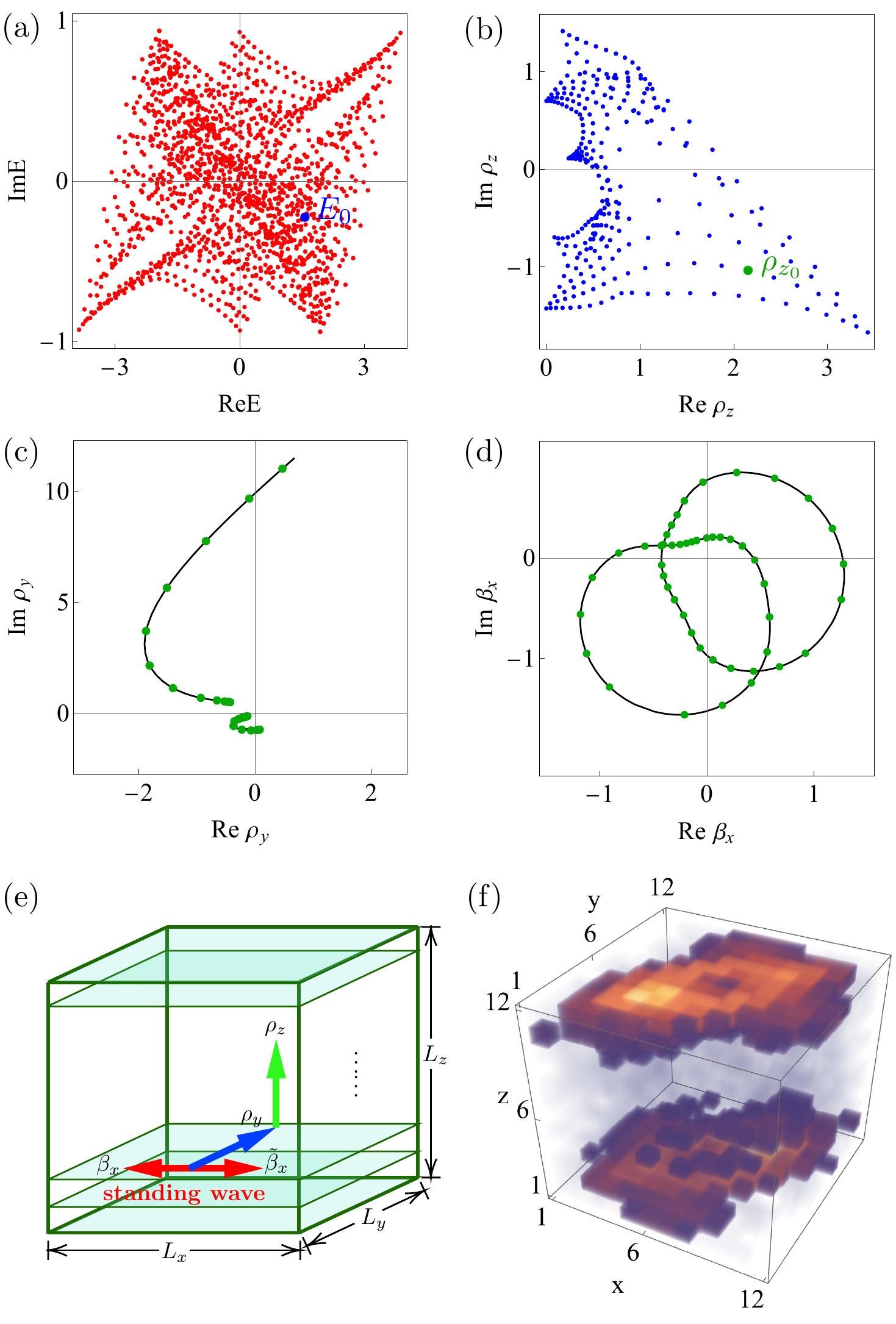}
    \par\end{center}
    \protect\caption{\label{fig:A6}~Numerical verification of the 3D GFS formula is presented using the tight-binding model Hamiltonian in Eq.~(\ref{EQ_3DModelRealHam}) [or Eq.~(\ref{EQ_3DModelHam})]. (a) shows the OBC eigenvalues as red dots, with the selected eigenvalue $E_0$ marked by a green dot. (b)-(d) illustrate the GFS for the OBC eigenvalue $E_0$. (b) shows the transfer matrix eigenvalues $\rho_z$, with the selected value $\rho_{z_0}$ indicated by a green dot. For $\rho_{z_0}$, the 3D Hamiltonian simplifies to a 2D subsystem, and the sub GFS for this subsystem is determined; its projections onto $\rho_y$ and $\beta_x$ planes are illustrated in (c) and (d), respectively. (e) visualizes the formation of a standing wave component for the 3D OBC eigen wavefunction. The standing wave is formed along the $x$ direction by two equal-amplitude non-Bloch waves, where $|\beta_x|=|\widetilde{\beta}_x|$. (f) shows the constructed 3D wavefunction of energy $E_0$ using the GFS basis, which fully agrees with the numerical wavefunction obtained by diagonalizing the Hamiltonian matrix. The system parameters specified in Eq.~(\ref{EQ_3DModelRealHam}) are $t_x=t_y=1/2$, $t_z=1$, $t_{xyz}=i/2$, and the system size is $L_x=L_y=L_z=12$. }
\end{figure}

Here, we provide a detailed numerical verification for GFS formula of the 3D tight-binding model given by Eq.~(\ref{EQ_3DModelHam}). 
The verification includes two steps. 
First, we choose an appropriate planar basis and compute the GFS under this basis. In the most generic cases, $\dim{\text{GFS}}=3$ and $\dim{\text{GBZ}}=5$. 
Second, for a specific OBC eigenvalue, we construct the corresponding OBC eigenstate using the GFS basis and demonstrate that it matches perfectly with the numerical wavefunction obtained from direct diagonalization of the full Hamiltonian matrix.

The real space Hamiltonian for model given by Eq.~(\ref{EQ_3DModelHam}) is written as 
\begin{align}\label{EQ_3DModelRealHam}
    H = \sum_{x,y,z} & t_x (c_{x+1,y,z}^{\dagger}c_{x,y,z} {+} \text{h.c.}) + t_y (c_{x,y+1,z}^{\dagger}c_{x,y,z} {+} \text{h.c.}) \nonumber \\
    & + t_z (c_{x,y,z+1}^{\dagger}c_{x,y,z} + \text{h.c.}) \nonumber \\ 
    & + t_{xyz} (c_{x+1,y+1,z+1}^{\dagger}c_{x,y,z} + \text{h.c.}),
\end{align}
where $t_{x,y,z}$ are the nearest neighbor hopping strengths along $x,y$, and $z$ directions, respectively. The term $t_{xyz}$ represents hopping along the cube diagonal direction. The Hamiltonian is reciprocal and non-Hermitian when the hopping strengths are complex values. 
Without loss of generality, we choose the $xy$-planar basis, and write down the Hamiltonian within this basis
\begin{equation}
    H = \sum_z \sum_{l=0,\pm 1} \textbf{c}^{\dagger}_{z}\, \textbf{h}_{l} \, \textbf{c}_{z+l},
\end{equation}
where $\textbf{c}^{\dagger}_z$ represents a planar creation operator that encompasses all creation operators at the $z$ plane, specifically $\textbf{c}^{\dagger}_z = \{c^{\dagger}_{1,1,z}, \dots, c^{\dagger}_{L_x,1,z}, \dots, c^{\dagger}_{1,L_y,z}, \dots, c^{\dagger}_{L_x,L_y,z}\}$. Here, the lattice size is taken to $L_x \times L_y \times L_z$. 
Therefore, the length of vector $\textbf{c}^{\dagger}_z$ is $L_x\times L_y$, and under this basis $\textbf{h}_l$ represents a matrix of dimension $L_x\times L_y$. 
The subscript $l$ denotes the hopping range along the $z$ direction. 
The bulk eigenequation can be transformed into a recursion equation along the $z$ direction
\begin{equation}
    \tbf{h}_{-1} \, \psi_{z-1} + (\textbf{h}_0- E \, \mathbb{I}_{L_x\times L_y}) \, \psi_{z} + \tbf{h}_{1} \, \psi_{z+1} = 0. 
\end{equation}
Here, the planar component $\psi_z$ can be represented as $\psi_z = (\Psi_{1,1,z},\dots, \Psi_{L_x,1,z},\dots, \Psi_{1,L_y,z},\dots, \Psi_{L_x,L_y,z})^T$, with $\Psi_{x,y,z}$ denoting the wavefunction component at lattice $(x,y,z)$. 
Finally, the transfer matrix can be obtained by writing the recursion equation into a matrix form
\begin{align}
    \begin{pmatrix}
        \psi_{z+1} \\ \psi_{z}
    \end{pmatrix} &= \mathbb{T}(E) \begin{pmatrix}
        \psi_{z} \\ \psi_{z-1}
    \end{pmatrix} \\
    &\nonumber=
    \begin{pmatrix}
        \textbf{h}_1^{-1}(E\,\mathbb{I}_{L_x\times L_y}-\tbf{h}_0) & -\tbf{h}_1^{-1}\tbf{h}_{-1}\\ 
        \mathbb{I}_{L_x\times L_y} & 0 
    \end{pmatrix}
    \begin{pmatrix}
        \psi_{z} \\ \psi_{z-1}
    \end{pmatrix}.
\end{align}
The transfer matrix $\mathbb{T}(E)$ has the dimension of $2L_x\times L_y$. 

We set the Hamiltonian parameters to $t_x=t_y=1/2$, $t_z=1$, $t_{xyz}=i/2$, and take the system size as $L_x=L_y=L_z=L=12$. The corresponding OBC eigenvalues are plotted by the red dots in Fig.~\ref{fig:A6}(a). 
We pick up a generic OBC eigenvalue $E_0=1.55391 - 0.22258 i$ (the blue dot) and calculate its GFS. According to our formula discussed in Sec.~\ref{SecV}, the bulk allowable $\rho_z$ values can be solved as the eigenvalues of the transfer matrix $\mathbb{T}(E_0)$. These $2L^2$ values of $\rho_z$ are plotted by the blue dots in Fig.~\ref{fig:A6}(b), which spans a finite area in the complex $\rho_z$ plane. Hence, we have $\dim{\rho_z}=2$. For each fixed $\rho_z$, the 3D Hamiltonian reduces to a 2D subsystem, which can be solved using the layer transfer matrix approach once again. Here, we show the results that for a fixed $\rho_{z_0}= 2.15655 - 1.03812 i$, the sub GFS for the 2D subsystem is solved and its projections onto $\rho_y$ and $\beta_x$ planes are illustrated in Figs.~\ref{fig:A6}(c) and (d), respectively. Here, for the 2D subsystem, layer-$y$ basis has been selected. For each fixed $\rho_z$, the GFS for the corresponding 2D subsystem forms 1D curves in ($\beta_x,\rho_y$) space. Meanwhile, $\rho_z$ spans a finite region. Consequently, GFS corresponding to a specific energy $E_0$ manifests as a 3D manifold. As $E_0$ sweeps through the entire OBC continuum spectrum, the collection of GFSs constitutes the GBZ, which is a 5D manifold. Therefore, through Figs.~\ref{fig:A6}(a)-(d), we numerically verified the dimensions for GFS and GBZ using the 3D tight-binding model in Eq.~(\ref{EQ_3DModelRealHam}). 

Now we examine the OBC wavefunctions using the GFS basis. 
According to Eq.~(\ref{eq:GBZ_cont}), the number of standing wave conditions is $2d-\dim{\text{GBZ}} = 1$. It means that one can enforce the standing wave condition along either one direction. Based on the basis we choose above, the standing waves are formed in the $x$ direction. An illustration of a standing wave component of an OBC eigen wavefunction is shown in Figs.~\ref{fig:A6}(e). The standing wave is first formed along the $x$ direction, and then transfer along the $y$ and $z$ directions. Therefore, the OBC wavefunction can be constructed as follows
\begin{equation}\label{EQ_3DConstructedWave}
    \Psi_{E_0}(x,y,z) = \sum_{i=1}^{2 L^2} \sum_{j=1}^{2 L} A_{ij} \, \rho_{z,i}^z \, \rho_{y,j}^y \, (\beta^x_{x,1} - \beta^x_{x,2}),
\end{equation}
where $\rho_z$, $\rho_y$, and $\beta_x$ are taken on the 3D GFS of energy $E_0$. The superposition coefficients $A_{ij}$ can be determined by applying the open boundary conditions (or termed zero boundary conditions). 
We obtain the coefficients $A_{ij}$ by solving the boundary matrix, and plot the constructed wavefunction of Eq.~(\ref{EQ_3DConstructedWave}) in Figs.~\ref{fig:A6}(f), which fully coincides with the numerical wavefunction by diagonalizing the Hamiltonian matrix. This agreement in OBC wavefunction validates the 3D GFS formula presented in Sec.~\ref{SecV}.


%

\end{document}